\documentclass[%
        10pt,
        nofootinbib,
 amsmath,amssymb,
 aps,
        prc,
        floatfix,
        tightenlines
]{revtex4-1}
\usepackage{graphicx}
\usepackage{dcolumn}
\usepackage{bm}
\usepackage{bbm}
\usepackage{booktabs}
\usepackage{physics}
\usepackage{hyperref}
\usepackage[mathlines]{lineno}
\usepackage{xcolor}
\usepackage{footmisc}
\usepackage{amsmath}
\usepackage{subcaption} 

\hypersetup{breaklinks=true,colorlinks=true,linkcolor=blue,citecolor=blue,
filecolor=magenta,urlcolor=blue}

\newcommand{\alphavec}{\bm{\alpha}}
\newcommand{\xvec}{\vb{x}}
\newcommand{\pibar}{\bar{\pi}}
\newcommand{\etabar}{\bar{\eta}}
\newcommand{\abar}{\bar{a}}
\newcommand{\bbar}{\bar{b}}
\newcommand{\poisson}[2]{\qty{#1, #2}_{\text{PB}}}

\newcommand{\HBRST}{H_{\text{BRST}}}

\newcommand{\DD}[1]{\mathcal{D}{#1}\,}
\newcommand{\potential}[2]{V(#1 - #2)}
\newcommand{\PAV}{\text{PAV}}
\newcommand{\VAP}{\text{VAP}}
\newcommand{\proj}{\text{proj}}
\newcommand{\lambdaop}{\widehat\lambda}
\newcommand{\Gammaop}{\widehat\Gamma}
\newcommand{\Phat}{\widehat P}
\newcommand{\Bop}{\widehat B}

\newcommand{\Hint}{H_{\text{int}}}

\newcommand{\Hho}{H_{\text{ho}}}

\newcommand{\com}{{\text{CoM}}}
\newcommand{\WS}{{\text{WS}}}

\newcommand{\beq}{\begin{equation}}
\newcommand{\eeq}{\end{equation}}

\newcommand{\rhotilde}{\tilde{\rho}}

\newcommand{\Nt}{N_t}

\begin{document}

\title{BRST quantization for the restoration of broken symmetries: a pedagogical example}

\author{Pranav Sharma}
\email{sharma.1098@osu.edu}
\affiliation{Department of Physics, The Ohio State University, Columbus, OH 43210, USA}

\author{R.~J. Furnstahl}
\email{furnstahl.1@osu.edu}
\affiliation{Department of Physics, The Ohio State University, Columbus, OH 43210, USA}

\date{\today}
\begin{abstract}
We present a pedagogical showcase of BRST quantization for restoring symmetries in nuclear many-body systems as a reformulation and potential alternative to conventional projection methods.
The formalism is illustrated using translational invariance for a simple system of two interacting masses in one dimension,
but with an eye toward generalizing to more particles, 
higher dimensions, and other symmetries.
We explain the considerations underlying particular choices within the BRST construction, and 
develop both Hamiltonian and path integral formulations to provide guidance 
for the variety of many-body and effective field theory contexts where gauge fixing for symmetry restoration might be useful. 
For the demonstration system we show how to diagonalize within the extended BRST phase space, how variation after projection is recovered for product reference states, how the corresponding gauge-fixed functional integral is constructed, and how collective zero modes are isolated and controlled.
Throughout, we keep in mind extensions of BRST symmetry to various approximation schemes as a guide for consistent symmetry restoration.

\end{abstract}
\maketitle

    \section{Introduction}

    Atomic nuclei are self-bound, which creates computational problems for many-body solution methods that build on a reference state, such as a Hartree-Fock Slater determinant in coupled cluster or the in-medium similarity renormalization group.
    These problems are also present in techniques that are based on background field methods with a mean-field reference, such as in density functional theory (DFT).
    The common issue is that there are symmetries associated with the collective dynamics of nuclei that are not respected by the reference state.
    These symmetries include collective translations, rotations, and phase (conjugate to particle number), which are broken because the reference state is centered at a definite point in space, oriented in a definite direction, or set at a definite phase.
    The consequence is that observables calculated in such an approach do not respect the symmetries of the system as a whole~\cite{Blaizot:1985,ringschuck}, leading to theoretical errors (and conceptual issues for  DFT~\cite{Furnstahl:2019lue,Engel:2006qu,Giraud:2007pe,Giraud:2008zz,Schunck:2019book,Sheikh:2002zz}).
    These symmetry-breaking errors are typically corrected by explicitly projecting out components of the wavefunction with good quantum numbers corresponding to the symmetry --- the symmetry is then said to be restored~\cite{Sheikh:2019qdz,Schunck:2019book,ringschuck,brink2005nuclear,Blaizot:1985,Duguet:2010cv, Yao2020}.
    Performing a projection generally results in better ground states~\cite{Duguet:2010cv, Yao2020}, which can be understood from a variational perspective~\cite{ringschuck}.
    However, it can be computationally demanding, particularly if the projection precedes the variation~\cite{Sheikh:2019qdz}.
    
An alternative to explicit projection is the introduction of a collective coordinate~\cite{LIPKIN1960272}.
Such a coordinate introduces a redundancy into the description of the system, which can be understood in a field theory context as the promotion of a global symmetry to a local (in time) symmetry.
This allows us to treat the symmetry as a gauge symmetry, so the theory becomes a gauge theory.
The redundancy inherent in gauge theories when calculating observable quantities is well known from the example of electrodynamics, where 
there is a continuum of field configurations $\phi(\xvec, t), \vb{A}(\xvec, t)$, called \emph{gauge orbits}, that yield the same measurable $\vb{E}(\xvec, t)$ and $\vb{B}(\xvec, t)$ configurations.
\begin{figure}[hb]
    \centering
    \includegraphics[width=0.4\linewidth]{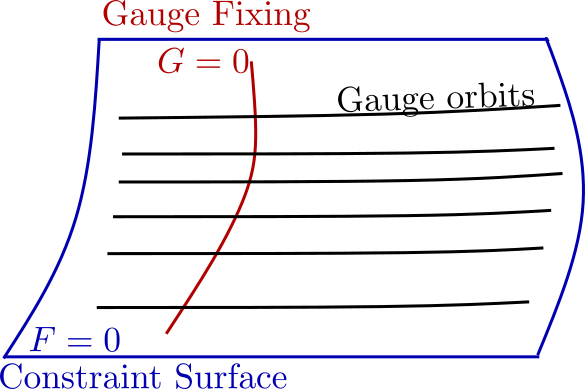}
    \caption{There are multiple physically equivalent trajectories (gauge orbits) in the phase space of a gauge theory. A gauge fixing condition (red line) picks just one point along each set of gauge orbits, 
    defining a ``gauge slice''. Figure adapted from~\cite{Henneaux:1992ig}.}
    \label{fig:gauge_orbits_fixing}
\end{figure}
The introduction of the collective coordinate can be offset by removing a degree of freedom (DoF) elsewhere; this is 
done in the case of translational symmetry by using Jacobi coordinates.
However, treating 
the collective coordinate as an additional DoF allows us to gauge the symmetries that are being broken, so we can use well-developed gauge theory methods for quantization in order to address the redundancy.

A technique called BRST quantization is the primary framework for accomplishing quantization of gauge theories such as those that define the Standard Model~\cite{Peskin:1995ev, Weinberg:1996II}.
There is much prior literature on adapting BRST methods to many-body problems~\cite{Bes:1990}, including several pedagogical treatments of one-body systems~\cite{Nemeschansky:1987xb,Bes:2002ajp, Bes:1990}.
However, as the BRST program has not been taken up as an alternative to conventional projection, it is our hope that a treatment that reflects the modern understanding and challenges of explicit projection will make more appealing the potential advantages of the BRST formalism.
Here we revisit this approach with a two-body system that has a direct many-body generalization to illustrate the key features, using implementations of both Hamiltonian and path integral  formalisms, and with an eye toward consistent approximations and effective field theory expansions.

The key feature of the program that we illustrate here is gauge fixing, which is a procedure by which a theory with redundancy is made to satisfy a condition once per gauge orbit  of a particular field configuration (as illustrated schematically in Fig.~\ref{fig:gauge_orbits_fixing}).
Fixing a gauge in a manner that does not spoil the gauge invariance of the measurable features of the theory is essential, since the gauge symmetry gives rise to Ward identities that ensure that conservation laws hold quantum mechanically.
This means that the effects of the gauge-fixing condition must be canceled in the calculation of any physical observable (which is necessarily gauge invariant).
The BRST framework uses Grassmann ``ghost'' variables in the implementation of the gauge fixing.
Ghost variables anticommute with each other, which is instrumental in the formulation of a systematic way to add gauge fixing terms to a Hamiltonian or action while having the effects on observables cancel out. 
Not only is this procedure systematically applicable to different symmetries (both abelian and non-abelian) with built-in compatibility with approximation schemes, but there is also a tremendous amount of freedom in the added terms that can
be exploited. 
This freedom has yet to be fully explored in the context of many-body symmetry restoration; we hope to help catalyze such explorations.

The application of BRST quantization to translational symmetry breaking for a two-body system allows us to focus on how the symmetry-breaking effects cancel in the calculation of observables.
This cancellation mechanism is the same whether the Hamiltonian is exact or approximate, so the simple illustration here serves as a guide for how the mechanism should work in many-body systems.
This methodology is distinct from explicit projection as derived from the generator coordinate method as we are changing the functionals used to calculate observables so that they cancel the contributions from collective symmetries, rather than solely modifying wavefunctions in a separate calculation.
This facilitates symmetry restoration that is consistent with approximations%
\footnote{Note that in an exact calculation we would expect to get the same answer as explicit projection.}
and EFT power counting.

We organize the pedagogical showcase of the BRST formalism as follows.
In Sec.~\ref{sec:setup}, we introduce the model system and discuss the addition of a collective coordinate corresponding to the symmetry broken by the reference state.
We introduce the formalism for BRST quantization in Sec.~\ref{sec:brst_ped_formalism} and apply it
in Sec.~\ref{sec:Hamiltonian_osc} to analyze the model system using the traditional Hamiltonian/operator-based form of quantum mechanics.
Variational approaches with product ans\"atze are considered in Sec.~\ref{sec:PAV_and_VAP}, with a comparison of conventional 
projection-after-variation (PAV) and variation-after-projection (VAP) to a naive prescription using an intrinsic Hamiltonian from Sec.~\ref{sec:Hamiltonian_osc}, before showing how a consistent BRST approach can recover the full VAP calculation but also motivate alternatives. 
This leads into the BRST formulation with path integrals in Sec.~\ref{sec:pi_osc}, which sets up for the use of the effective action formulation of density functional theory~\cite{Furnstahl:2019lue}.
Building on the development of the formalism in both frameworks, zero modes in perturbative calculations are discussed in Sec.~\ref{sec:zero_modes}.
Summary points and a guide to generalizations are given in Sec.~\ref{sec:summary}.
Three appendices are included. 
The first connects the gauge machinery that is core to our exposition with the familiar setting of classical electromagnetism.
The second appendix provides an alternate path integral prescription for a gauge-fixed action to describe the simple model addressed here.
The third appendix shows a way of constructing a variational ansatz with gauge-fixing as a potential source of inspiration.
Throughout, we take $\hbar = 1$ and denote operators with a hat only when there is a possibility of confusion.
        
\section{Setup: a simple model}\label{sec:setup}
        
Our demonstration model is a pair of identical particles%
\footnote{We assume the particles have different ``flavors'', so the spatial wavefunctions are the same for fermions and bosons.}  
of mass $m$ in one dimension interacting via a translationally invariant potential, e.g., a harmonic potential parametrized by spring constant $k$ or a contact interaction. 
The simple and familiar nature of this system makes it an ideal candidate for a pedagogical introduction to using BRST symmetry for symmetry restoration.
The essential ideas of the BRST program can be illustrated cleanly on this system and it enables a direct path to many-body systems, including exactly solvable models for future validation (e.g., as in \cite{Engel:2006qu}). 
The example of an abelian symmetry avoids some 
complications without restricting the generalization to the non-abelian case later.
This differs from prior pedagogical treatments~\cite{Bes:2002ajp, Nemeschansky:1987xb}, which focused more on the non-abelian cases. 
However, the overarching idea of how the BRST program achieves symmetry restoration is not altered at a conceptual level by nonzero structure constants of the (broken) symmetry group, so we choose to omit such distractions and focus on illustrating the motivations for the formalism and the mechanism of the restoration.

We also provide details for \textit{both} a Hamiltonian treatment of the formalism and a path-integral treatment, in contrast to prior works that focused solely on a Hamiltonian treatment of a simple system~\cite{Bes:1990} or solely on an expository example of a many-body system in the path integral framework~\cite{Alessandrini:1978qd}.
We present both perspectives so that the reader can follow the development of the ideas in a controlled setting with either formalism, with connections between the two discussed explicitly.
The attention to both treatments is motivated by the broad applicability of the BRST program for the restoration of broken symmetries.
The path integral treatment is of interest for the calculation of EDFs via effective actions, while the Hamiltonian treatment is of particular interest for ab-initio calculations.

As typical when considering gauge symmetries and BRST symmetry~\cite{Weinberg:1996II, Henneaux:1992ig}, we first revisit the classical description of the system to reassess the procedure of quantization.
The classical Lagrangian for the system is
\begin{align}
        L = \frac{m}{2}(\dot{x}_1^2 + \dot{x}_2^2) -
            \potential{x_2}{x_1}.
\end{align}
This system has a ``zero mode'' corresponding to the collective motion of both particles along the $x$ axis.
This contribution to the total energy does not depend on the interaction $V$ at all.
If viewed from the rest frame of the center of mass (CoM), the system has a spectrum that reflects the interaction;
this spectrum from the collective rest frame is the \textit{intrinsic spectrum}.
Solving for this spectrum is a very general problem, 
although it is generally much more complicated with a realistic potential and many particles.
The two-body problem with a simple potential can be solved in a simple manner by going to CoM coordinates but following an alternative path here enables us to exhibit the more general issues.
        
\begin{figure}[tbh!]
    \centering
    \begin{subfigure}[t]{0.49\textwidth}
        \centering
        \includegraphics[width=0.99\textwidth]{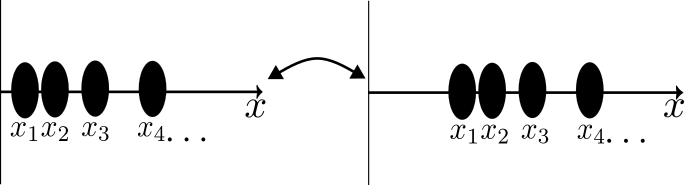}
    \end{subfigure}
        \hfill
    \begin{subfigure}[t]{0.49\textwidth}
        \centering
        \includegraphics[width=0.99\textwidth]{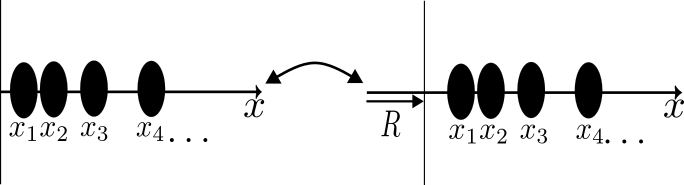}
    \end{subfigure}%
        \caption{a) The original system of particles has a global translational symmetry: the system can be uniformly translated to any location on the $x$ axis and have the same intrinsic dynamics.
    (b) The system can now be moved arbitrarily at each moment in time, with an adjustment in $R$ keeping it fixed as if it is at the same point on the line.}   
   \label{fig:global-local-translations}        
\end{figure}

To begin the BRST procedure, we must make our global translational symmetry a local translational symmetry by introducing a collective coordinate.
This is most simply accomplished by performing the symmetry transformation and then treating the parameter of the transformation as a dynamical coordinate.
For a system in one dimension, our symmetry transformation is a translation of the whole system on its axis, 
namely
\begin{align}
    x_i \to x_i + R\qc \dot{x}_i \to \dot{x}_i + \dot{R}
    ,
\end{align}
    so we treat $R$ as our new coordinate, meaning that our Lagrangian for the two-body system is 
\begin{align}
    L \longrightarrow \frac{m}{2}((\dot{x}_1 + \dot{R})^2 + (\dot{x}_2 + \dot{R})^2) - 
    \potential{x_2}{x_1} .
\end{align}
The addition of the extra coordinate gives us a redundancy in our description, and that redundancy is precisely what we get in a gauge symmetry!
The original description of the system was invariant under global shifts of the particles: the intrinsic spectrum was the same when the CoM was centered at any point on the $x$ axis (see left panel of Fig~\ref{fig:global-local-translations}).
Now, with the collective coordinate $R$, we have a stronger local symmetry: the CoM of the system can be shifted arbitrarily as a function of time with the collective coordinate changing in response to keep the same physics (right panel of Fig~\ref{fig:global-local-translations}).%
\footnote{Note that this can also be thought of as viewing the system from arbitrary locally changing frames of reference~\cite{Bes:1990, Henneaux:1992ig}.}
The redundancy is manifested when trying to invert the relationship between velocity and momentum for each coordinate in order to pass to the Hamiltonian formalism:
\begin{align}
    p_1 &= \pdv{L}{\dot{x}_1} = m(\dot{x}_1 + \dot{R}),\label{eq:p1}\\
    p_2 &= \pdv{L}{\dot{x}_2} = m(\dot{x}_2 + \dot{R}),\label{eq:p2}\\
    p_R &= \pdv{L}{\dot{R}} = m(\dot{x}_1 + \dot{R} + \dot{x}_2 + \dot{R}) = p_1 + p_2
    \label{eq:pR}
.\end{align}
The three momenta are not independent, which
means we cannot invert the relationship as we would normally do (i.e., the matrix from Eqs.~\eqref{eq:p1}, \eqref{eq:p2}, and \eqref{eq:pR} relating $(p_1, p_2, p_R)$ to $(\dot{x}_1, \dot{x}_2, \dot{R})$ is singular).

\begin{figure}[tbh!]
    \centering
    \begin{subfigure}[t]{0.49\textwidth}
        \centering
        \includegraphics[width=0.99\textwidth]{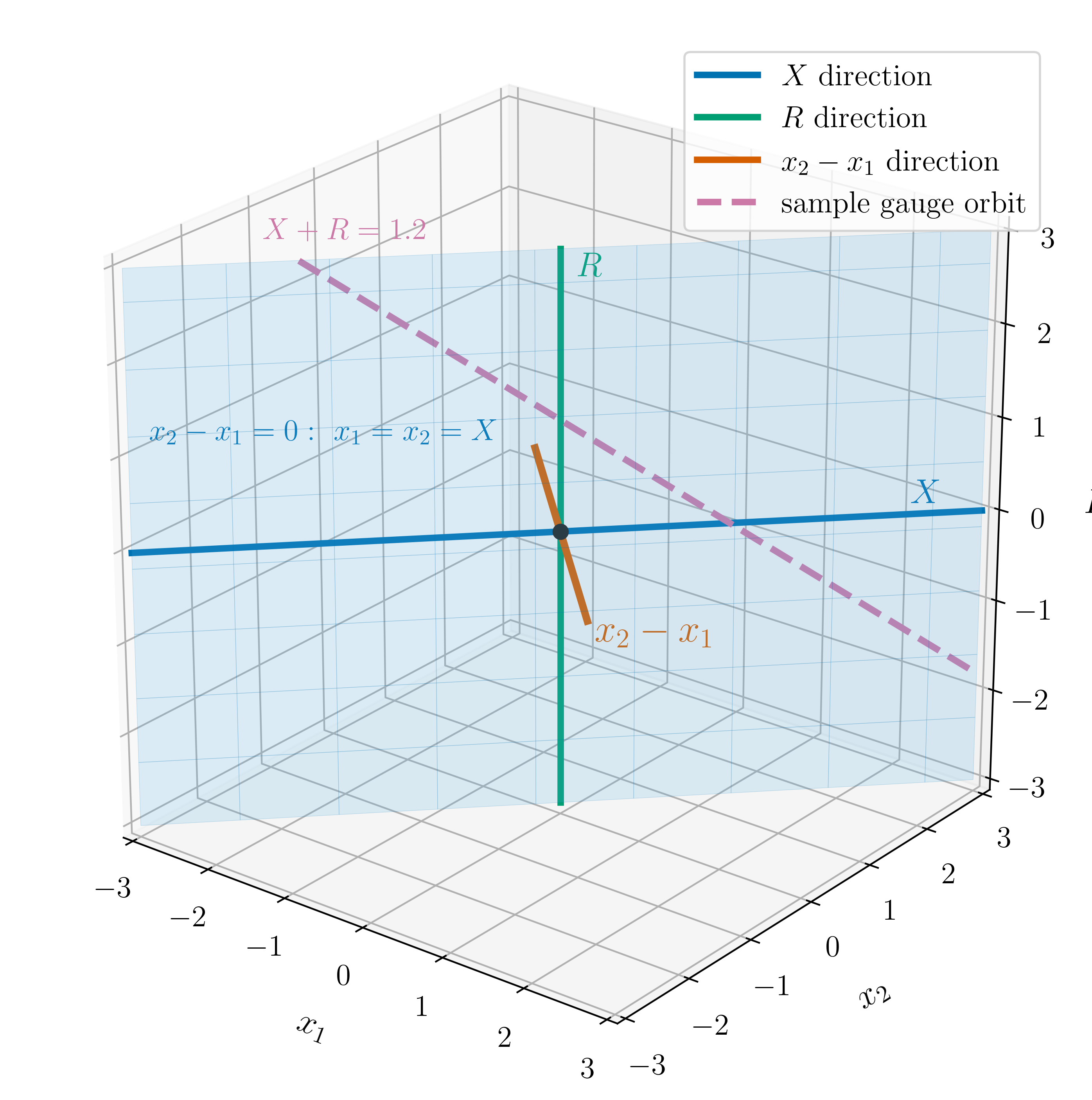}
    \end{subfigure}
        \hfill
    \begin{subfigure}[t]{0.49\textwidth}
        \centering
        \includegraphics[width=0.99\textwidth]{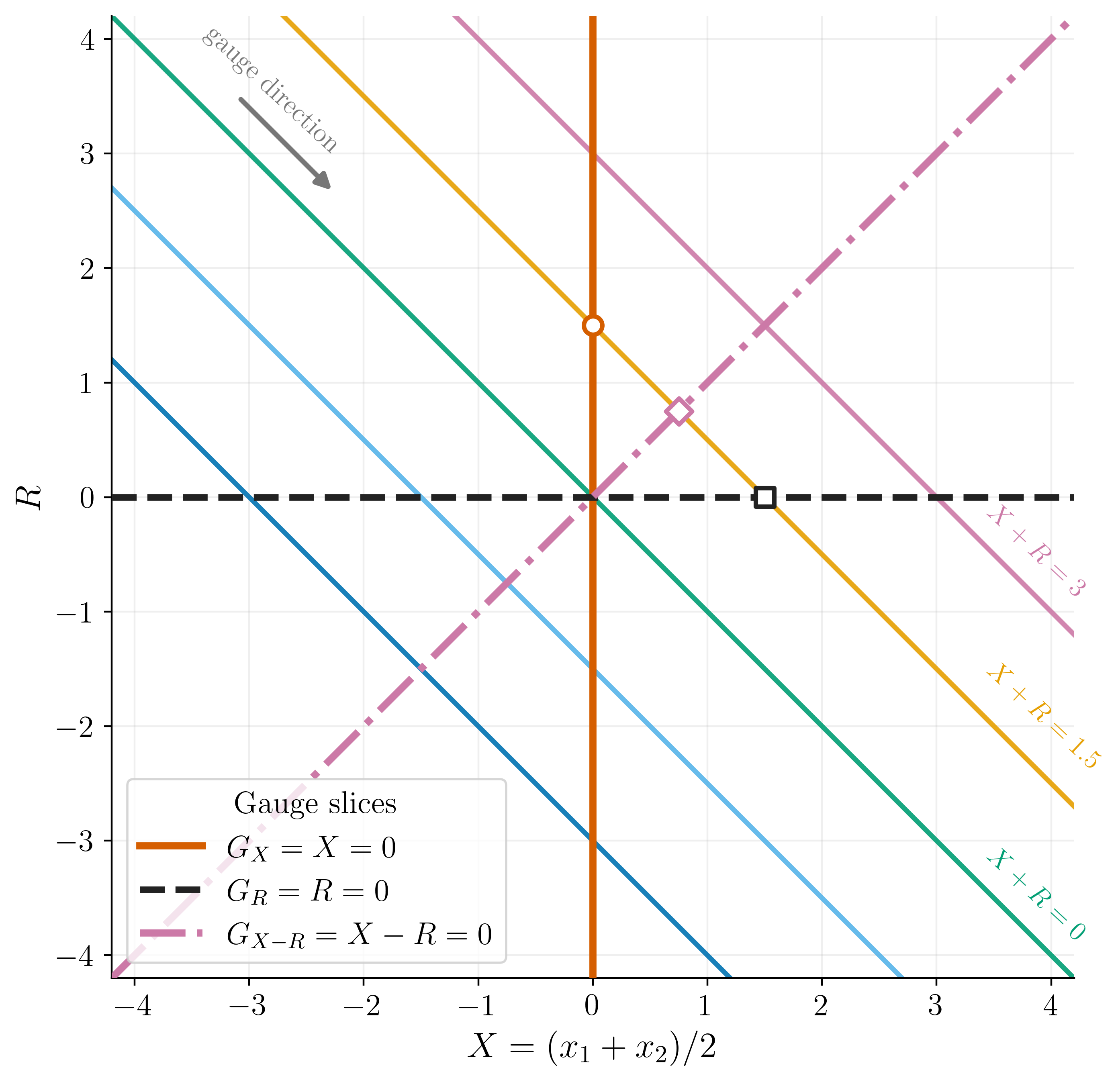}                
    \end{subfigure}%
    \caption{(a) The coordinate portion of the extended phase space for the two-body system, with the plane defined by the collective coordinate $R$ and the CoM of the original DoFs $X = (x_1+x_2)/2$ highlighted. Physically equivalent configurations of the system generated by the constraint Eq.~\eqref{eq:constraint_equation} (also known as gauge orbits) lie on such a plane.
    (b) The CoM and collective coordinate plane with several options for gauge-fixing illustrated (c.f.\ the red line in Fig.~\ref{fig:gauge_orbits_fixing}). The choice $X = 0$ corresponds to the gauge-fixing choice used to exactly diagonalize the Hamiltonian in Sec.~\ref{sec:Hamiltonian_osc}. The choice $R = 0$ reduces us back to the original problem.}   
   \label{fig:phase_space_translational}        
\end{figure}

The relationship between the momenta is called a first-class primary constraint in the language of constrained quantum mechanics, from which this formalism originates.
Such constraints always generate gauge transformations, and this is always going to be the nature of a constraint added by this manner of introducing collective coordinates.%
\footnote{First-class constraints are the only type we will need to consider here and so we will usually just refer to them as constraints for simplicity.
More details on first-class and second-class constraints can be found in chapter~1 of Ref.~\cite{Henneaux:1992ig}.
}
Constraints are derived as functions of phase space variables that are zero for physical trajectories of the system; we denote this one as 
\begin{align}
    F(x_i, p_i, R, p_R) = p_R - p_1 - p_2 .
    \label{eq:constraint_equation}
\end{align}
 In this case, the symmetry operation that we are treating as a gauge transformation is a collective motion of the whole system along the $x$ axis.
 This mathematically simple constraint generates the same translations in each original coordinate and the opposite translation in the collective coordinate (moving in the plane defined in the left panel of Fig.~\ref{fig:phase_space_translational}) --- this is nothing more than the familiar fact that momenta generate translations in their conjugate coordinates.
Through the Poisson bracket (PB), this constraint generates transformations corresponding to the symmetry that we want to gauge: for any function $f(x_i, R)$ of the coordinates in our system, a translation by $\epsilon$ is given by
 \begin{align}
     \var{f(x_i, R) }
        = \poisson{\epsilon F}{f}      
   =  \epsilon\pdv{f}{x_1} + \epsilon\pdv{f}{x_2} - \epsilon\pdv{f}{R}         
.\end{align}
This constraint has zero Poisson bracket with an interaction that is translationally invariant, so the Poisson bracket should also be zero with a translationally invariant $N$-particle Hamiltonian that consists of a kinetic term and such an interaction.
As such, there is no need to modify the Poisson bracket from this constraint. 

In general, respecting the constraints of a system may require a modification of the Poisson bracket, as the standard commutation relations are formulated without knowledge of the constraints.
There are two commonly used techniques for accomplishing this.
One way is to add a further term to the Poisson bracket that depends on the constraints.
This is precisely the Dirac bracket, as constructed in Section 1.3 of Ref.~\cite{Henneaux:1992ig}.
The use of the Dirac bracket in quantized systems is well discussed in field theory books~\cite{Weinberg:1996II, gitman_tyutin}.
The other way is to enlarge the phase space to include variables of opposite statistics.
The Poisson brackets taken over the enlarged set of variables then yield an enlarged symmetry that can stay respected even when a term that breaks the original gauged symmetry is added.
This is BRST symmetry.
Without some form of gauge-fixing, time evolution of the canonical degrees of freedom is ill-defined as there are large classes of arbitrary functions that can be added to each coordinate without changing any observables.\footnote{This is also the case in electrodynamics: a gauge condition must be chosen when working with the electromagnetic potentials, otherwise the EoMs for a particular system do not have well-defined solutions}
On the other hand, the gauge symmetry in the classical (quantum) theory gives rise to Noether currents (Ward identities) which are essential to preserve.
Introducing anti-commuting degrees of freedom provides a way to address both of these requirements, with the nilpotence that is implied by anti-commutation providing a means for the cancellation of gauge-fixing in final results, while still performing calculations that pick a particular gauge slice (and thus are not gauge-symmetric).
This works quantum mechanically as well, one needs only to replace the Poisson bracket with a commutator and include a factor of $i\hbar$.

The translational symmetry of our system is already local in time because we have elevated $R$ to the role of a dynamical coordinate.
Gauge invariant quantities here will then be intrinsic quantities: those which do not depend on the location of the center of mass.
More generally, gauge invariant quantities will be those which are invariant under the symmetry that we introduce a local coordinate for.

To go to the Hamiltonian, we perform the usual Legendre transformation~\cite{goldstein_classical} except that we use Eq.~\eqref{eq:pR} for $p_R$ and eliminate $\dot{x}_1$ and $\dot{x}_2$ using Eqs.~\eqref{eq:p1} and \eqref{eq:p2}, keeping the dependence on $\dot{R}$ (that is, the Hamiltonian is a function of $R$ and $\dot{R}$ rather than $R$ and $p_R$).\footnote{Strictly speaking, this is called a Routhian~\cite{goldstein_classical}.}
We find 
\begin{align}
 H &= p_1\dot{x}_1 + p_2\dot{x}_2 + (p_1 + p_2)\dot{R} - \frac{m}{2}(\dot{x}_1 + \dot{R})^2 - \frac{m}{2}(\dot{x}_2 + \dot{R})^2 + \potential{x_2}{x_1}
 \notag\\
   &= p_1\qty(\frac{p_1}{m} - \dot{R}) + p_2\qty(\frac{p_2}{m} - \dot{R}) + (p_1 + p_2)\dot{R} - \frac{p_1^2}{2m} - \frac{p_2}{2m}
   +\potential{x_2}{x_1}
   \notag\\
   &= \frac{p_1^2}{2m} + \frac{p_2^2}{2m}
   +\potential{x_2}{x_1}
.\end{align}
Note that the collective coordinate $R$ is implicitly present in the definitions of the momenta, but explicit dependence on it has disappeared in the Hamiltonian.
In principle, we could make the collective momentum manifest again by introducing a Lagrange multiplier term of the form $\lambda(p_R - p_1 - p_2)$ and treating $\lambda$ as a variational parameter.
However, the Lagrange multiplier only enforces constraints at the level of expectation values; the states that are variationally obtained with the use of Lagrange multipliers still break symmetries and need projection.%
\footnote{For example, the chemical potential is a Lagrange multiplier that fixes the expectation value of the number operator, but the BCS ground state that is obtained from the fixing of the chemical potential has an indefinite number of particles and a definite phase, ostensibly violating the phase symmetry of the system.} 
We can do better via the BRST program, which treats the Lagrange multiplier as a dynamical variable that assists in the cancellation of effects of a symmetry-breaking potential in the wavefunction, with the help of additional degrees of freedom.

\section[BRST Formalism]{\texorpdfstring{BRST Formalism\protect\footnote{I\lowercase{f the reader has experience with this formalism, or is simply eager to get to the calculations, they are invited to skip to the glossary at the end of the section.}}}{BRST Formalism\protect\footnote{If the reader has experience with this formalism, or is simply eager to get to the calculations, they are invited to skip to the glossary at the end of the section.}}}
\label{sec:brst_ped_formalism}

As noted in Sec.~\ref{sec:setup}, first-class constraints generate gauge transformations, and introducing a collective coordinate by promoting a collective transformation parameter to a coordinate will always leads to a first-class constraint.
This means that any constrained system has a gauge symmetry.
We can thus treat the symmetry associated with the collective dynamics as a gauge symmetry.
If a system has multiple constraints, then multiple different gauge transformations can be performed, one for every constraint.
So if we gauge translation in three dimensions, we have three constraints; 
if we also gauged rotations, then we would have six constraints in total. 
The BRST formalism makes it straightforward to treat multiple symmetry breakings simultaneously.

A general gauge transformation on a function $f$ of the degrees of freedom in a constrained system will be of the form 
\begin{align}
    \var{f} = \poisson{\epsilon_a F_a}{f} ,
\end{align}
where $\epsilon_a$ are the c-number parameters of the transformation and $a$ runs over all constraints $F_a$.
A BRST transformation is, on the original DoFs, a gauge transformation that uses Grassmann numbers instead of c-numbers as parameters.
Grassmann numbers are anticommuting numbers: $\eta_1 \eta_2 = -\eta_2\eta_1$, meaning that any Grassmann number (and any linear combination of Grassmann numbers) squares to zero: $\eta_1\eta_1 = -\eta_1\eta_1 = 0$.
This makes a BRST transformation \textit{nilpotent}.%
\footnote{See the discussion below Eq.~\eqref{eq:Q_BRST} for a comment on nilpotence in the non-abelian case.}
Nilpotence is exploitable to help isolate the gauge-invariant parts of a calculation.
The generator of a BRST transformation is called the BRST charge and is denoted $Q$.
The abelian BRST charge is defined minimally\footnote{The BRST formalism does not strictly \textit{need} Lagrange multipliers or other ``helper'' functions in order to work~\cite{Henneaux:1992ig, Alessandrini:1978qd}, though they can be convenient as illustrated in \cite{Bes:1990, Bes:2002ajp} and discussed in footnote \ref{fn:lagrange_multiplier}.}
as 
\begin{align}
    [Q]_{\text{minimal}} = -\eta_a F_a
    \label{eq:Qminimal}
.\end{align}
In a similar manner to how the gauge parameter $R$ was elevated to a canonical variable, we now treat every Grassmann number $\eta_a$ as a canonical variable (called a ghost), so each one needs a conjugate momentum variable $\pi_a$ (a ``ghost momentum'').
The minus sign in Eq.~\eqref{eq:Qminimal} is conventional, motivated by the anticommuting property of Grassmann numbers.
The Grassmann numbers are chosen as real,
by default, 
so the charge is hermitian upon quantization.
For each constraint, we add 
one pair
of Grassmann variables. 
This means that restoring translational symmetry in three dimensions would require three pairs of Grassmann variables (for any number of particles), restoring rotational symmetry in addition would require six total pairs, and so on.

Some special attention must be paid to Lagrange multipliers if they are present in the theory, as is the case in electrodynamics (see Appendix~\ref{app:constraints_em}) or in a pairing system with a chemical potential. 
We will use Lagrange multipliers for exact diagonalization; they should be thought of as present unless stated otherwise.
The Lagrange multiplier is required to transform under the gauge transformation in order to preserve gauge invariance of the Lagrangian. 
This means that the charge as stated requires modification. 
An examination of the Lagrangian
\begin{align}
    L = p_1\dot{x}_1 + p_2\dot{x}_2 - H - \lambda(p_R - p_1 - p_2)
\end{align}
reveals that under an infinitesimal gauge transformation $x_j \to x_j + \epsilon$, the first two terms generate a factor 
\begin{align}
    (p_1 + p_2)\dot{\epsilon}
.\end{align}
This can be canceled out if the Lagrange multiplier transforms as 
\begin{align}
    \lambda \to \lambda - \dot{\epsilon}
,\end{align}
and a term $p_{R}\dot{R}$ %
is added to the Lagrangian.
Thus, the inclusion of a Lagrange multiplier makes the dependence on the collective coordinate manifest, as the Lagrangian now becomes 
\begin{align}
 L =   p_1\dot{x}_1 + p_2\dot{x}_2 + p_R\dot{R} - H - \lambda(p_R - p_1 - p_2) .
\end{align}

Under a gauge transformation, the Lagrange multiplier transforms with the derivative of the gauge parameter rather than the gauge parameter itself.
This difference must be reflected in the BRST transformation by a term that has the derivative of a ghost $\dot{\eta}$.
Aside from this extra time derivative, we see that the gauge transformation on $\lambda$ acts like a translation, so we require a generator of translations in $\lambda$ to encode this transformation law in a generator for gauge or BRST transformations.
We therefore introduce a momentum conjugate to the Lagrange multiplier, $B$, with a Poisson bracket that becomes a commutator when quantized,
\begin{align}
    \poisson{\lambda}{B} = 1 \underset{\text{QM}}{\longrightarrow} \comm{\lambdaop}{\Bop} = i
.\end{align}
The BRST charge would then be modified by adding a term%
\footnote{Here, for brevity, we assume that we have a Lagrange multiplier for every symmetry broken.
Generalizing to a treatment where this is not the case is only a matter of adding a separate summation index. 
So there would be one index running over every symmetry that is being gauged through a collective coordinate, and one running over every gauged symmetry that we have introduced a Lagrange multiplier for.}
$\dot{\eta}_{a}B_{a}$ to $[Q]_{\text{minimal}}$, however
the inclusion of a time derivative is problematic, because there is no a priori Hamiltonian for the ghosts,\footnote{We will eventually arrive at a Hamiltonian for ghosts, however, that is only obtained from the charge being constructed as it is here, so such a Hamiltonian cannot be used in the definition of the charge.} so we do not have a way of computing $\dot{\eta}$.
To handle this, we introduce a second set of ghosts $\qty(\etabar, \pibar)$ such that $\pibar = \dot{\eta}$.
Treating the time derivative of a ghost as another canonical variable is not entirely out of the ordinary within this framework: the additional term in $Q$ can be seen as a separate constraint and would thus warrant its own set of ghosts according to how the charge was constructed above.
However, the fact that $\lambda$ transforms with the time derivative of the parameter means that the ghost cannot be the same as those used for the other DoFs.
The only other option for a Grassmann gauge parameter that yields the correct transformation law for this additional constraint is the canonical conjugate of a regular ghost.
So, with a Lagrange multiplier, the minimally defined abelian BRST charge is extended to be 
\begin{align}
    Q = -\eta_a F_a + \pibar_a B_a\label{eq:Q_BRST}
.\end{align}
Now, upon quantization, the anticommuting nature of ghosts means that they obey Fermi statistics.
However, they are also spin-0 modes, and we do not see spin-zero fermionic variables in any asymptotic physical state.
Ghost momenta can be thought of as removing ghosts in a manner similar to how momentum operators lower powers of $x$ in states in coordinate representation. 
However, $\pibar$ for the Lagrange multiplier plays the role of the ghost, which means that its conjugate, $\etabar$, can be thought of as removing $\pibar$. 
The reversal of the roles for the ghosts associated with the BRST transformation of $\lambda$ means that the barred ghost variables act really as \textit{anti}ghosts, hence the bar notation. 
There is a concept of ghost number~\cite{Henneaux:1992ig} which counts ghosts and antighosts in exactly the same way as lepton number, baryon number, etc. 
It can be shown from very general cohomological arguments that the canonical pair of Grassmann variables associated with a Lagrange multiplier constraint must be antighosts. 

Following this same line of argument, one can see that the same transformation law for the Lagrange multiplier (that is, transforming with the time derivative of the gauge parameter) emerges for other abelian symmetries, such as rotations in a plane or phase rotations. 
In the case of a non-abelian symmetry, the same line of reasoning yields a transformation law that is quite similar to what was seen here, but modified with an extra term that is proportional to the structure constants of the symmetry group.
In all cases, however, the ghosts associated with the Lagrange multiplier transformation must be antighosts.
As such, the form of $Q$ in Eq.~\eqref{eq:Q_BRST}
is quite generally applicable to abelian symmetries, and is modified only additively in the case of a non-abelian symmetry, as shown in section 9.4 of~\cite{Henneaux:1992ig}.

It is now helpful to introduce some terminology for using BRST symmetry to do calculations (summarized in Sec.~\ref{subsec:glossary}).
A \textbf{BRST-exact}%
\footnote{The nomenclature ``exact'' is borrowed from differential geometry, where a differential form is called exact if it can be written as the exterior derivative of something else.}
operator is an operator that results from a transformation on another operator by the BRST charge $Q$.
So given any general operator $\mathcal{G}$, a BRST-exact operator $\mathcal{A}$ is formed as 
\begin{align}
    \mathcal{A} = \comm{Q}{\mathcal{G}}_{\pm}
    \label{eq:BRST_exact_op}
,\end{align}
with the $\pm$ appearing according to 
whether $\mathcal{G}$ is a bosonic ($-$) or fermionic ($+$) operator.
Note that because $Q$ is nilpotent, we get zero if $\mathcal{G}$ itself is BRST-exact.
In a similar vein, a BRST-exact state is one resulting from a state that has been transformed by the BRST charge.
Thus, for any state $\ket{\phi}$ of nonzero norm, 
\begin{align}
    \ket{\chi} = Q\ket{\phi}
\end{align}
is a BRST-exact state.
Note that again by the nilpotence enforced by $Q$ being fermionic, the norm of a BRST-exact state is zero:
\begin{align}
    \braket{\chi} = \ev{Q Q}{\phi} = 0
.\end{align}
A \textbf{BRST-closed} operator $\mathcal{O}$ is one that (anti-)commutes with the charge:
\begin{align}
    \comm{Q}{\mathcal{O}}_{\pm} = 0
.\end{align}
A BRST-closed state $\ket{\psi}$ is similarly one that is annihilated by the charge:
\begin{align}
    Q\ket{\psi} = 0
    \label{eq:BRST_closed_state}
.\end{align}
Note that BRST-exact states are a subset of BRST-closed states.

Because of the nilpotence of the BRST transformation, we find the key result that BRST-exact parts of BRST-closed matrix elements
are automatically eliminated as part of the calculation with the extra DoFs (using the notation of states and operators from Eqs.\eqref{eq:BRST_exact_op}--\eqref{eq:BRST_closed_state}):
\begin{align}
    \qty{\bra{\psi'} + \bra{\chi'}}\qty(\mathcal{O} + \mathcal{A})\qty{\ket{\psi} + \ket{\chi}} 
    &= \qty{\bra{\psi'} +\bra{\phi'}Q  }\qty(\mathcal{O} + Q\mathcal{G} \pm \mathcal{G}Q)\qty{ \ket{\psi} + Q\ket{\phi} } \notag\\
    &= \qty{\bra{\psi'} + \bra{\phi'}Q}\mathcal{O}\qty{ \ket{\psi} + Q\ket{\phi} }\notag\\
    &= \mel{\psi'}{\mathcal{O}}{\psi}  + \mel{\psi'}{\mathcal{O}Q}{\phi} + \mel{\phi'}{Q\mathcal{O}}{\psi} + \mel{\phi'}{Q\mathcal{O}Q}{\phi} \notag\\
    &= \mel{\psi'}{\mathcal{O}}{\psi}  \pm\mel{\psi'}{Q\mathcal{O}}{\phi} \pm \mel{\phi'}{\mathcal{O}Q}{\psi}   \notag\\
    &= \mel{\psi'}{\mathcal{O}}{\psi}
    \label{eq:BRST_invariance_me}
.\end{align}
To go from the first line to the second, we used the fact that $Q$ annihilates BRST-exact and BRST-closed states, letting the second operator product act to the left and the third act to the right.
We then used the fact that $\mathcal{O}$ (anti-)commutes with the charge to move the charge so that it could act on the BRST-closed state to go from the third line to the fourth.

After the dust settles in \eqref{eq:BRST_invariance_me}, we  see that the BRST-exact addition to $\mathcal{O}$ had its effects cancel in the expectation value, likewise for the BRST-exact components of the state.
The idea then is to perform the type of symmetry breaking that typically arises in a mean-field calculation, but break the symmetry in the Hamiltonian (or action) with a BRST-exact term.
As a result, the effects of the symmetry breaking cancel and we directly obtain intrinsic physics when we calculate observables.
This automatic cancellation removes the need for a separate projection and alternate variational procedure.

Because the BRST charge is fermionic (Grassmann odd) and we want a BRST extension to our operator that is even in Grassmann parity, the operator that is transformed by $Q$ to yield a BRST-exact addition to a given operator should also be fermionic.
As such, we define an operator $\rho$ called a \textit{gauge fixing fermion}. This is a fermionic operator that contains the gauge fixing condition to give us a broken symmetry, so that we can extend the Hamiltonian%
\footnote{In principle, this can be done for any operator. The Hamiltonian is of immediate interest for the goal of improved mean-field approximations.} 
as 
\begin{align}
    \HBRST = H + \comm{Q}{\rho}_+
    \label{eq:HBRST_plain}
.\end{align}
Working with such an extension of the Hamiltonian, we will find that as long as our states obey BRST symmetry, the extra degrees of freedom will automatically cancel the gauge-dependent parts of a calculation in the calculation of observables, and this cancellation is precisely driven by the nilpotence of the BRST transformation.

To summarize, adding a dynamical collective coordinate to a theory makes local the redundancy of the collective motion of the system, so the symmetry can be treated as a gauge symmetry.
By enlarging the phase space of the theory to include fermionic (Grassmann) variables, we can consider nilpotent gauge transformations called BRST transformations.
A normal gauge transformation uses a (set of) c-number parameter(s), which we treat as dynamical variables; a BRST transformation uses Grassmann parameters that we also treat as variables.
This enlargement of the phase space that we are quantizing allows for the addition of BRST-exact terms to the Hamiltonian: terms that are the output of a BRST transformation.
We can then diagonalize the extended Hamiltonian $\HBRST$, which yields a set of BRST-closed eigenstates that reflects the extra terms and degrees of freedom, but \textit{only the gauge-invariant parts of the calculation survive}, with the rest of the calculation automatically canceling by nilpotency.
That is, only the part of the calculation that is independent of the collective symmetry survives, we ``automatically project'' the result without a distinct projection step, at the cost of more variables in the Hamiltonian to be diagonalized.

\subsection{Glossary of terms}\label{subsec:glossary}
Many terms and ideas have been introduced here.
For the sake of the new reader and those interested in exploring the technical literature, we include a short glossary of terms that are essential to understanding how this formalism works.
When relevant, we include both the notation and the definition for each quantity.
    \begin{itemize}
        \item \textbf{First-class constraint}~($F$):~A definite relationship between a momentum and other canonical variables, resulting in the momentum not being unique.
        For the treatment of collective symmetries, the constraint can be thought of the equation that defines the collective conserved quantity in terms of the original DoFs.
        
        \item \textbf{Gauge-fixing function}~($G$):~A function of the original variables that should \textit{not} commute with the constraint, but \textit{should} commute with the collective momentum.
        This generates a gauge slice (see Figs.~\ref{fig:gauge_orbits_fixing} and \ref{fig:phase_space_translational}).
        
        \item \textbf{BRST charge}~($Q$):~The generator of a BRST transformation, constructed (for the abelian case) with constraints and ghosts as
        \begin{align}
            Q = -\eta_a F_a + \pibar_a B_a .\notag
        \end{align}
        
        \item \textbf{BRST-exact}:~A state ($\ket{\chi}$) or operator ($\mathcal{A}$) that results from a BRST transformation on another state or operator:
        \begin{align}
            \ket{\chi} = Q\ket{\phi}\qc \mathcal{A} = \comm{Q}{\mathcal{G}}_{\pm}
        \end{align}
        for some general $\ket{\phi}$ or general operator $\mathcal{G}$.
        
        \item \textbf{BRST-closed}: %
        A state ($\ket{\psi}$) or operator ($\mathcal{O}$) that is annihilated upon BRST transformation. Thus, BRST-closed states (operators) contain gauge-invariant states (operators).
        \begin{align}
            Q\ket{\psi} = 0\qc \comm{Q}{\mathcal{O}}_{\pm} = 0
        .\end{align}
        \item \textbf{Gauge-fixing fermion}~($\rho$):~A Grassmann-valued operator that contains the gauge-fixing function. It must be Grassmann-valued so that the BRST-exact term formed from $\rho$ is bosonic, a necessary condition on operators that correspond to observables.

    \end{itemize}
This brief glossary is intended for non-experts and thus does not address all possible complications in general applications of the BRST framework. For a more complete exposition we refer the reader to Ref.~\cite{Henneaux:1992ig}.
    
\section{Hamiltonian Implementation}\label{sec:Hamiltonian_osc}
    In this section, we implement the BRST formalism for our simple system using the Hamiltonian framework of quantum mechanics~\cite{VanHolten:2001nj}, with the goal of diagonalizing the extended BRST Hamiltonian in \eqref{eq:HBRST_plain}. 
    This goal informs our choice of gauge fixing.
    The effect of gauge fixing in the BRST framework is to add multiple sectors of the Hamiltonian, which can be generalized and used for perturbative calculations, which we later demonstrate in Sec.~\ref{sec:zero_modes}.
    In Sec.~\ref{sec:pi_osc} we show how the corresponding gauge fixing and treatment of zero modes emerges in a path integral framework.
    
    The symmetry we are trying to break is the invariance of the system to translations of the center of mass, so the gauge fixing term should involve the center of mass $\frac{1}{2}(x_1 + x_2)$. 
    In general, gauge fixing picks a single representative out of any gauge-equivalent configurations of the system.
    For this system, that reduces to picking a single trajectory of the center of mass (e.g., the $G=0$ line in Fig.~\ref{fig:gauge_orbits_fixing} or the options in Fig.~\ref{fig:phase_space_translational}).
    At first we make a simple choice of fixing the CoM at zero, but other choices are permissible and can be useful to test for independence from the details of the gauge fixing.

    The Hamiltonian was presented at the end of Sec.~\ref{sec:setup}.
    To this, we want to add a BRST-exact term.
    We have to introduce two pairs of ghosts: one for the constraint $F = p_R - p_1 - p_2 = 0$ and one for the constraint $B = 0$.
    The charge is thus 
    \begin{align}
        Q = -\eta F + \pibar B,
        \label{eq:Qnoa}
    \end{align}
    as discussed in the motivation for Eq.~\eqref{eq:Q_BRST}.
    There are four possible terms that make up our gauge-fixing fermion $\rho$, which we have parametrized here by four distinct bosonic operators $a_i$ that are not fixed by any particular requirements to this point:
    \begin{align}
        \rho = a_1\eta + a_2\etabar + a_3\pi + a_4\pibar \label{eq:rho_param_1stquant}
    .\end{align}
    The ghost variables are defined so that 
    \begin{align}
        \acomm{\eta}{\pi} = \acomm{\etabar}{\pibar} = 1
    ,\end{align} 
    with all other anticommutators zero. 
    Performing the anticommutator,\footnote{The operator identity $\comm{Aa}{Bb}_+ = \comm{A}{B}ab + BA\comm{a}{b}_+$ for bosonic operators $A, B$ and fermionic operators $a, b$ is used repeatedly.} we get
    \begin{align}
        \comm{Q}{\rho}_+ = \comm{B}{a_1}\pibar\eta + \comm{B}{a_2}\pibar\etabar + a_2B + \comm{B}{a_3}\pibar\pi - \comm{F}{a_2}\eta\etabar - \comm{F}{a_3}\eta\pi - a_3F - \comm{F}{a_4}\eta\pibar
    .\end{align}
    There is a \textit{tremendous} amount of freedom in the choice of the gauge-fixing fermion.
    Because the symmetry is abelian, commutators with the constraint will not generate any structure factors, so it should be possible to pick a gauge-fixing function that decouples the ghosts from the other DoFs.
    
    We  now list considerations that lead to a particular choice of gauge-fixing fermion $\rho$.
    These considerations do not preclude other options, they simply lead to a convenient (e.g., quadratic) form of the BRST-extended Hamiltonian.
    \begin{enumerate}
        \item We would like to have $a_1, a_2, a_3$ contain powers of $\lambda$ no higher than one, so that the first three commutators are c-numbers and do not couple the ghosts to any of the other variables.
        \item We want $a_2, a_3, a_4$ to contain powers of $(x_1 + x_2)/2$ no higher than one so that the last three commutators are c-numbers.
        \item We want a quadratic Hamiltonian so we should have a linear $B$ term in $a_2$.
        \item For the sake of seeing the difference between this prescription and just using a Lagrange multiplier as a parameter rather than a dynamical variable, we should have $a_3$ solely be $\lambda$ so that it adds a constraint term to the Hamiltonian that is of the form that we would expect in that approach.
        \item The prior point means that there is no term of the form $\eta\pi$ in the Hamiltonian.
        We have the freedom to choose things to work out simply, so we choose to similarly make the $\etabar\pibar$ term vanish by choosing $a_2$ to have no $\lambda$ dependence.
        \item Hoping to get the ghosts in a form reminiscent of an oscillator Hamiltonian, we also choose to eliminate terms that mix the ``momentum'' of one ghost mode with the ``coordinate'' of the other, so $a_1$ has no $\lambda$ and $a_4$ has no CoM.
    \end{enumerate}
    Based on these considerations, $a_1$ and $a_4$ have no role they are \textit{required} to play and can safely be chosen to be zero.
    This satisfies points 1, 5, 6.
    We fixed our form of $a_3$ above, it should just be $\lambda$, so a candidate form of $\rho$ is
    \begin{align}
        \rho = \qty(\underbrace{\alpha\qty(\frac{x_1 + x_2}{2})}_{\text{Point 2}} + \underbrace{\beta}_{\text{Points 2 and 3}} - \underbrace{\frac{B}{2\xi}}_{\text{Point 3}})\etabar + \underbrace{\lambda}_{\text{Point 4}}\pi\label{eq:osc_gf_fermion},
    \end{align}
    where we have chosen our proportionality constant for $B$ so that it will look like a kinetic term in the ``ghost-less'' part of the BRST-exact term.
    In principle, we could have also had terms that couple $B$ to the CoM in $\rho$, but we elect to use our freedom to keep them decoupled for simplicity.
    The constant $\beta$ will just contribute a constant shift to the Hamiltonian when the anti-commutator is performed, so it can be chosen to be zero, the constant $\alpha$ will be maintained to keep the units consistent, but in a different form as we will soon arrive at.
    This form of the gauge-fixing fermion then makes a BRST-exact term:
    \begin{align}
        \comm{Q}{\rho}_+ = \alpha B\qty(\frac{x_1 + x_2}{2}) - \frac{B^2}{2\xi} + i\pi\pibar - i\alpha\eta\etabar - \lambda(p_R - p_1 - p_2)
        \label{eq:sectIVanticommutator}
    .\end{align}
    If we add this to our Hamiltonian, we get
    \begin{align}
        \HBRST &\equiv H + \comm{Q}{\rho}_+ \notag\\
         & = \frac{p_1^2 + p_2^2}{2m} + \potential{x_2}{x_1} + \alpha B\qty(\frac{x_1 + x_2}{2}) - \frac{B^2}{2\xi} + i\pi\pibar - i\alpha\eta\etabar - \lambda(p_R - p_1 - p_2)\label{eq:H_BRST_2body}
    .\end{align}
    To have the units stay consistent, $\lambda$ must dimensionally be a velocity, meaning $B$ must have dimensions of [mass]$\times$[length] in order to preserve the commutation relation with $\lambda$.
    This means $\alpha$ must have dimensions of a squared frequency.
    As such, we define an arbitrary frequency $\alpha \equiv -\Omega^2$ with the minus sign chosen only for algebraic convenience in what follows.
    The BRST-extended Hamiltonian is then
    \begin{align}
        \HBRST = \frac{p_1^2 + p_2^2}{2m} + \potential{x_2}{x_1} - \Omega^2 B\qty(\frac{x_1 + x_2}{2}) - \frac{B^2}{2\xi} + i\pi\pibar + i\Omega^2\eta\etabar - \lambda(p_R - p_1 - p_2)
    .\end{align}
    
    We now must diagonalize this Hamiltonian.
    We introduce abbreviated notation that can just as well denote $N$ particles:
    \begin{align}
        P \equiv \sum_{j}^{N}p_j \overset{N=2}{\longrightarrow} p_1 + p_2\qc 
        X \equiv \frac{1}{N}\sum_{j}^N x_j \overset{N=2}{\longrightarrow} \frac{1}{2}(x_1 + x_2)\qc 
        M \equiv Nm \overset{N=2}{\longrightarrow} 2m
          \label{eq:abbreviated-notation}.
    \end{align}
    This notation emphasizes that the procedure for diagonalizing this BRST-extended Hamiltonian is insensitive to the number of particles in the system.
    Equation~\eqref{eq:abbreviated-notation} \textit{does not} imply a change of coordinates; everything is still written and solved in terms of the original degrees of freedom.
    Rather it is simply to abbreviate terms in the Hamiltonian for the ease of algebraic manipulation and to manifest the generalization to $N$ particles.
    The Hamiltonian is then 
    \begin{align}
        \HBRST = \frac{P^2}{2M} + \frac{(p_1 - p_2)^2}{4m} + \potential{x_2}{x_1} - \Omega^2 BX - \frac{B^2}{2\xi} + i\pi\pibar + i\Omega^2\eta\etabar - \lambda(p_R - P)\label{eq:HBRST_osc}
    .\end{align}
    A noteworthy feature of this Hamiltonian is that it is manifestly quadratic, up to the choice of potential $\potential{x_2}{x_1}$, so we can expect oscillator solutions in some of the variables.
    A second important feature of this Hamiltonian is that the collective momentum $p_R$ commutes with it.
    On a practical level, this means we can treat $p_R$ as a constant rather than an operator, treating the operator as always acting on an eigenstate of $p_R$ since we know what momentum eigenstates look like.
    On a conceptual level, this is a reflection of the translational symmetry we are trying to break for our original DoFs: it shows itself with the collective coordinate rather than with the original DoFs.

        \subsection{Diagonalizing the BRST Hamiltonian}

            We discuss here how to diagonalize the BRST Hamiltonian for a quadratic form that we can solve exactly.
            This illustrates the BRST formalism and sets us up for perturbative expansions about a quadratic Hamiltonian.
            We must solve an extended Schrodinger equation that includes for each symmetry broken: a collective coordinate, a set of ghosts, and (optionally) a Lagrange multiplier as a dynamical variable.
            The contributions to the Hamiltonian from the BRST-exact term can be assessed and organized according to an approximation scheme of choice in exactly the same manner as the other terms in the Hamiltonian, as demonstrated in Chapter 7 of \cite{Bes:1990}.
            Although there will not be any approximation schemes needed in solving the present model, the \textit{mechanism} of cancellation is the same as what happens with an approximate Hamiltonian.
            This mechanism distinguishes the BRST program from the existing commonly used schemes: the restoration of the symmetry occurs as a result of working in the extended space, \textbf{not} as a separate modification of states.
            As a reminder, we use hats over operators only when there is a chance of confusion.
        \subsubsection{Spurious sector}
        To diagonalize the Hamiltonian, we begin by completing squares.
        We have two sets of coupled modes: one for the Lagrange multiplier and one for the CoM of the original DoFs, independent of how many particles are present.
        Let us examine the spectra for each.
        The Lagrange multiplier momentum terms are
        \begin{align}
            -\frac{1}{2\xi}\qty(B^2 + 2\xi\Omega^2BX + \xi^2\Omega^4X^2) + \frac{\xi\Omega^4}{2}X^2 = -\frac{(B + \xi\Omega^2X)^2}{2\xi} + \frac{\xi\Omega^4}{2}X^2
        .\end{align}
        The total momentum terms are 
        \begin{align}
            \frac{1}{2M}\qty(P^2 + 2MP\lambda + M^2\lambda^2) - \frac{M}{2}\lambda^2 = \frac{(P + M\lambda)^2}{2M} - \frac{M}{2}\lambda^2
        .\end{align}
        Each of these terms produces an apparent harmonic potential for the other.
        We can thus treat this sector of the Hamiltonian as oscillator modes that have unusual looking kinetic terms.
        Because all of the spurious center of mass contributions to the energy arise from the DoFs in this part of the Hamiltonian, we denote this the \textit{spurious sector} of the Hamiltonian.
        The Lagrange multiplier mode is strictly negative, and that is a feature that helps accomplish the cancellations we want.
        The kinetic terms seem to couple the two modes, but we still have a free parameter $\xi$.
        Let us see if we can choose $\xi$ to decouple the modes.
        The ``momenta'' for the modes contain all of the operator structure that would lead to coupling, so if we can choose a $\xi$ that makes $P + M\lambda$ commute with $B + \xi\Omega^2 X$, then the modes are decoupled.
        Computing the commutator and setting it to zero,
        \begin{align}
            0 &= \comm{P + M\lambda}{B + \xi\Omega^2X} \notag\\
            &= \xi\Omega^2\comm{P}{X} + M\comm{\lambda}{B}\notag\\
            &= \xi\Omega^2(-i) + M(i)\notag\\
            \implies \xi &= \frac{M}{\Omega^2}
        .\end{align}
        Plugging this choice of $\xi$ into the oscillator modes, we get for this sector of the Hamiltonian
        \begin{align}
             H_{\text{sp}} &= \frac{(P + M\lambda)^2}{2M} + \frac{M\Omega^2}{2}X^2 - \frac{\Omega^2(B + MX)^2}{2M} - \frac{M}{2}\lambda^2 - \lambda p_R \notag\\
              &= \frac{(P + M\lambda)^2}{2M} + \frac{M\Omega^2}{2}X^2 - \frac{\Omega^2(B + MX)^2}{2M} - \frac{M}{2}\qty(\lambda + \frac{p_R}{M})^2 + \frac{p_R^2}{2M}
        .\end{align}
        This is essentially a pair of standard oscillators\footnote{Note that we have obtained an oscillator in the CoM as one would get from a Lawson term~\cite{GLOECKNER1974313}. Unlike a Lawson term, however, this potential's effects will automatically cancel in the calculation of observables due to the other variables it was introduced with!} --- they just have unusual looking kinetic terms.
        We define creation and annihilation operators for the decoupled modes as 
        \begin{align}
            \Gammaop_P \equiv \sqrt{\frac{M\Omega}{2}}\qty(X + \frac{i}{M\Omega}(P + M\lambda))\qc \gamma_B = \sqrt{\frac{\Omega}{2M}}\qty(B + MX - i\frac{M}{\Omega}(\lambda+p_R/M))
        ,\end{align}
        with standard commutation relations 
        \begin{align}
            \comm{\Gammaop_P}{\Gammaop_P^\dagger} = 1\qc \comm{\gamma_B}{\gamma_B^\dagger} = 1 .
        \end{align}
        The spurious sector of the Hamiltonian is then 
        \begin{align}
            H_{\text{sp}} = \Omega\qty(\Gammaop_P^\dagger\Gammaop_P + \frac{1}{2}) - \Omega\qty(\gamma_B^\dagger\gamma_B + \frac{1}{2}) + \frac{p_R^2}{2M}
        .\end{align}
        
        The negative frequency of the Lagrange multiplier mode now needs to be addressed.
        Because of the negative frequency, the annihilation operator $\gamma_B$ actually \textit{raises} the energy by a factor $\Omega$ while its conjugate creation operator lowers the energy.
        As such, we define a different pair of operators $\Gammaop_B \equiv \gamma_B^\dagger$, so that the raising operator raises the energy and the lowering operator lowers the energy.
        This reinterpretation of the operators flips the commutation relation from what one expects it to be for an oscillator:
        \begin{align}
            \comm{\Gammaop_B^\dagger}{\Gammaop_B} = 1
        ,\end{align}
        but it results in a ground state (defined by $\Gammaop_B\ket{0} = 0$) that has lower energy than any of its excitations by this mode.
        The flipped commutation relation results in excited states having negative norms:
        \begin{align}
            \ev{\Gammaop_B\Gammaop_B^\dagger}{0} &= \ev{-1 + \Gammaop_B^\dagger\Gammaop_B}{0}\notag\\
            &= -1
        .\end{align}
        This is a feature, not a bug: the negative norms assist in the cancellation of the effects of symmetry breaking when calculating observables.%
        \footnote{States of negative norm can generally appear as artefacts of gauge fixing --- see Sec.~5.2 of Ref.~\cite{Mandl:2010bg}.
        In a first-quantized system such as this one, the Lagrange multiplier cleanly gives us the negative norm states as manifestly separated from excitations in our original degrees of freedom.\label{fn:lagrange_multiplier}}
        The final form of our spurious sector Hamiltonian is thus 
        \begin{align}
            H_{\text{sp}} &= \Omega\qty(\Gammaop_P^\dagger\Gammaop_P + \frac{1}{2}) - \Omega\qty(\Gammaop_B^\dagger\Gammaop_B - \frac{1}{2}) +\frac{p_{R}^2}{2M} \notag\\
            &= \Omega\qty(\Gammaop_P^\dagger\Gammaop_P - \Gammaop_B^\dagger\Gammaop_B + 1) + \frac{p_R^2}{2M}
        \end{align}
        Note that, aside from the general idea of translational invariance which was used to establish a definition for the charge that generated these terms, no information about the inter-particle interaction was used to arrive at this form of the spurious sector.
        Also worth reiterating is the fact that this procedure does not change with $N$ particles: the gauge-fixing fermion $\rho$ in Eq.~\eqref{eq:osc_gf_fermion} would be adjusted to the appropriate $N$-particle sum, but our notation for the total mass and the total CoM of the original degrees of freedom means that the diagonalization of this sector all follows exactly the same way.

        \subsubsection{Ghost sector}
            The ghosts decouple from the rest of the Hamiltonian here, though they are still necessary for the symmetry restoration in the calculation of observables.
            The ghost sector Hamiltonian is 
            \begin{align}
                H_{\text{gh}} = i(\pi\pibar + \Omega^2\eta\etabar) ,
            \end{align}
            which couples the two different pairs of ghosts with each other.
            We thus reparametrize the operators to pairs that decouple.
            This is conveniently done with~\cite{Bes:2002ajp}
            \begin{align}
                a = \frac{1}{\sqrt{2\Omega}}\pibar - i\sqrt{\frac{\Omega}{2}}\eta&\qc b = \frac{1}{\sqrt{2\Omega}}\pi + i\sqrt{\frac{\Omega}{2}}\etabar ,\notag\\
                \abar = \frac{i}{\sqrt{2\Omega}}\pi + \sqrt{\frac{\Omega}{2}}\etabar&\qc \bbar = \sqrt{\frac{\Omega}{2}}\eta - \frac{i}{\sqrt{2\Omega}}\pibar
            ,\end{align}
            which satisfy anticommutation relations 
            \begin{align}
                \comm{a}{\abar}_+ = 1\qc \comm{b}{\bbar}_+ = 1
            .\end{align}
            When substituted into the Hamiltonian, we find that the transformation yields
            \begin{align}
                H_{\text{gh}} = \Omega(\abar a + \bbar b - 1)
            \end{align}
            Treating these as fermionic oscillator variables, we see that $\abar$ and $\bbar$ correspond to raising operators, while the unbarred versions correspond to lowering operators.
            The zero point energy is exactly opposite to that of the spurious sector.
            Once again, no information about the inter-particle interaction was used in the diagonalization of this part of the Hamiltonian.
            Only the details of the gauge-fixing choice contribute to the ghost sector's dynamics, and this is entirely general as the gauge-fixing choice is based on the idea of breaking translational symmetry; no other feature of the interaction came into consideration.
            Despite their decoupling, the ghosts do contribute to the ground-state energy since they have the negative zero-point energy. 
            This contribution is, thanks to the BRST machinery, necessarily going to help cancel the contributions from the spurious sector.

        \subsubsection{Intrinsic sector}
            What remains of the BRST-extended Hamiltonian is the intrinsic part, i.e., the part that is translationally invariant:
            \begin{align}
                H_{\text{int}} = \frac{(p_2 - p_1)^2}{4m} + \potential{x_2}{x_1} .
                \label{eq:Hint}
            \end{align}
                This may need to be approximated depending on the form of the inter-particle potential.
                For the familiar case of a harmonic interaction between particles, 
                we can write the exact solution
            \begin{align}
                H_{\text{int}} = \sqrt{\frac{2k}{m}}\qty(\Gammaop^\dagger\Gammaop + \frac{1}{2})\qc \Gammaop = \qty(\frac{km}{8})^{1/4}\qty(x_2 - x_1 + \frac{i}{\sqrt{2mk}}(p_2 - p_1))
            .\end{align}
            For intrinsic potentials that do not have analytic solutions, we must use approximation methods, which are discussed below after the ground-state wavefunction is determined.

        \subsection{Ground state}
            The ground-state wavefunction of the system should be annihilated by all of the annihilation operators.
            Our BRST-extended Hamiltonian with an intrinsic harmonic interaction can now be written 
            \begin{align}
                \HBRST &= H_{\text{int}} + H_{\text{sp}} + H_{\text{gh}} \notag\\
                & \overset{V= kx^2/2}{\longrightarrow} \sqrt{\frac{2k}{m}}\qty(\Gammaop^\dagger \Gammaop + \frac{1}{2}) + \frac{p_R^2}{2M} + \Omega\qty(\Gammaop_P^\dagger\Gammaop_P - \Gammaop_B^\dagger\Gammaop_B + \abar a + \bbar b) 
            .\end{align}
            Before we solve for the ground-state wavefunction, it is worth making a stronger statement about the spectrum here.
            The BRST charge can be written in terms of the oscillator operators as
            \begin{align}
                Q &= \pibar B - \eta F\notag\\
                  &= \sqrt{\frac{\Omega}{2}}\qty(a + i\bbar)\sqrt{\frac{M}{2\Omega}}\qty(\Gammaop_B^\dagger + \Gammaop_B - \Gammaop_P^\dagger - \Gammaop_P) + \frac{1}{\sqrt{2\Omega}}\qty(\bbar + ia)i\sqrt{\frac{M\Omega}{2}}\qty(\Gammaop_P^\dagger - \Gammaop_P + \Gammaop_B - \Gammaop_B^\dagger)\notag\\
                  &= \sqrt{M}\qty(i\bbar(\Gammaop_B - \Gammaop_P) + a(\Gammaop_B^\dagger - \Gammaop_P^\dagger))\label{eq:brst_q_osc}
            .\end{align}
            The BRST-closed states are the ones that are annihilated by this charge.
            The ground state is the state that is annihilated by all 5 lowering operators here. 
            This state is BRST-closed since the BRST charge contains a lowering operator in each term.
            The excitations in the $P, B, a, b$ modes yield states that are BRST-exact, not BRST-closed.
            
            To begin demonstrating this, we look at the states with excitations in the ghost modes.
            There are four fermionic states: $n_a, n_b = 0, 1$.
            There are many states with $n_b = 1$, the simplest of which is the one that is in the ground state in the $a, B, P$ modes.
            Although this state is annihilated by $Q$, it is BRST-exact since, up to normalization,
            \begin{align}
                \ket{n_a = 0, n_b = 1, n_B = 0, n_P = 0} \propto Q\ket{n_a = 0, n_b = 0, n_B = 1, n_P = 0}
            .\end{align}
            For the sake of compactness, we keep state labels in the above order for this demonstration, so that we can simply write, for example, $\ket{0, 0, 1, 0}$ for the RHS of the above equation.
            It is readily checked that the same state also arises from the state with $n_P = 1, n_B = 0$.
            States with $n_b = 1$, nonzero $n_B$ or $n_P$, and $n_a = 0$ are similarly seen to be BRST-exact.
            States with $n_b = 1$ and \textbf{both} $n_B\neq 0$, $n_P\neq 0$ are less obvious. 
            As a sample calculation to show that such a state is BRST-exact, let us examine the state $\ket{0, 1, 1, 1}$.
            It can be quickly verified that, up to normalization
            \begin{align}
                Q\ket{0, 0, 1, 2} \propto (\ket{0, 1, 0, 1} - \ket{0, 1, 1, 1})
            .\end{align}
            Now, as already outlined, the first state on the RHS is BRST exact since $\ket{0, 1, 0, 1}\propto Q\ket{0, 0, 0, 2}$.
            So, we have 
            \begin{align}
                Q\ket{0, 0, 1, 2} \propto Q\ket{0, 0, 0, 2} - \ket{0, 1, 1, 1} \implies \ket{0, 1, 1, 1} \propto Q(\ket{0, 0, 1, 2} - \ket{0, 0, 0, 2})
            .\end{align}
            The same kind of argument can now be extended to higher values of $n_B, n_P$.
            We have focused on $n_b = 1, n_a = 0$ because a state with $n_a = 1$ sees $n_a$ lowered by the charge while $n_P$ and $n_B$ are raised, so such a state is manifestly not BRST-closed.
            This means that only $n_b = 0, n_a = 0$ is BRST-closed and not BRST-exact.
            Keeping these fixed while letting $n_B, n_P\neq 0$ is manifestly not BRST-closed.
            As such, the only BRST-closed state which is not BRST-exact in these four modes is the ground state of all four modes.
            The only part of the spectrum of this Hamiltonian that survives will thus be the intrinsic part.

            By solving the equations 
            \begin{align}
                \Gammaop\ket{0} = 0\qc \Gammaop_P\ket{0} = 0\qc \Gammaop_B\ket{0} = 0  ,
            \end{align}
            we can get the ground-state wavefunction as a function of $x_1, x_2, \lambda, R$.
            Because the ghosts decouple, their wavefunction will just be multiplied with that of the bosonic DoFs.
            Solving the above three equations in coordinate space (including $\lambda$ and $R$ as coordinates) identifies the components of the wavefunction,
            \begin{align}
                \Gammaop\ket{0} = 0 &\implies \Psi_0 \sim \exp{-\frac{1}{2}\sqrt{\frac{mk}{2}}(x_2 - x_1)^2} ,\\
                \Gammaop_P\ket{0} = 0 &\implies \Psi_0 \sim \exp{-\frac{M\Omega}{2}\qty(\frac{x_1 + x_2}{2})^2 - iM\lambda \qty(\frac{x_1 + x_2}{2})} ,\\
                \Gammaop_B\ket{0} = 0 &\implies \Psi_0 \sim \exp{-iM\lambda\qty(\frac{x_1 + x_2}{2}) + \frac{p_R}{\Omega}\lambda + \frac{M}{2\Omega}  \lambda^2} .
            \end{align}
            Combining these components, the overall ground-state wavefunction of the bosonic part is, up to normalization
            \begin{align}
                \Psi_0 \propto \exp{-\frac{1}{2}\sqrt{\frac{mk}{2}}(x_2 - x_1)^2 -\frac{M\Omega}{2}\qty(\frac{x_1 + x_2}{2})^2 - iM\lambda \qty(\frac{x_1 + x_2}{2}) + \frac{p_R\lambda}{\Omega} + \frac{M\lambda^2}{2\Omega} + ip_R R}\label{eq:psi0_brst_osc} .
            \end{align}
            The $R$ piece of the wavefunction is given by the fact that the Hamiltonian commutes with $p_R$, so the $R$ dependence of the wavefunction must be a momentum eigenstate, i.e., a plane wave.
            Applying the Hamiltonian in coordinate representation to this state yields the eigenvalue 
            \begin{align}
                E_0 = \frac{p_R^2}{2M} + \frac{1}{2}\sqrt{\frac{2k}{m}}
            .\end{align}
            The only contribution from the original DoFs is the intrinsic ground state frequency; the collective contribution is given purely in terms of the collective coordinate that we added.
            The CoM oscillator potential was added via a BRST-exact term, so its effects were canceled in the BRST-closed ground state.

            In order to discuss the calculation of observables in this state, we must discuss normalizability.
            As is manifest from the form of the ground-state wavefunction \eqref{eq:psi0_brst_osc}, the traditional norm on wavefunctions cannot give a sensible result with the $\lambda$ integration.
            We must therefore use a modified norm.
            One option for an alternative norm is given in Ch.~3 of \cite{Bes:1990} as
            \begin{align}
                \braket{g}{h} = \int_{-\infty}^\infty\dd{\lambda} \qty[g(-i\lambda)]^*h(i\lambda) .
                \label{eq:osc_modified_norm}
            \end{align}
            This preserves hermiticity of the $\lambda$ and $B$ operators, and otherwise satisfies all the requisite features of an inner product while giving us a normalizable wavefunction from the $\lambda$ degree of freedom, which contains all of the unusual norm properties associated with gauge-fixing.
            If any non-trivial functions of $\lambda$ are to be calculated with an expectation value, then this norm induces the coordinate representation of the operators 
            \begin{align}
                \lambdaop\psi = i\lambda\psi\qc \Bop\psi = -\pdv{\lambda}\psi .
            \end{align}

In summary, we added a gauge-fixing term to our Hamiltonian through a BRST-exact operator. 
The extra degrees of freedom associated with the BRST extension to the Hamiltonian were mixed with the original degrees of freedom, but two sectors of the Hamiltonian were ultimately written in a simply solved form that did not depend on the number of particles or choice of inter-particle potential. 
BRST-closed states were manifestly separable and straightforward to solve, despite the generality of the prescription, and ultimately we found an eigenvalue that reflected the intrinsic spectrum and a collective contribution from a part of the wavefunction that manifestly factorized.
Of course, a real system requires approximation schemes; much of the rest of this paper is dedicated to addressing them.

\section{Variational functionals}
    \label{sec:PAV_and_VAP}
    In a many-body system, a reference state is usually constructed by performing a variational optimization on a product ansatz of single-particle states.
    In order to restore symmetry, the projection occurs as a separate step, either before or after the variation~\cite{ringschuck, greiner:1972}.
    Projection After Variation (PAV) projects the optimized ground state, and thus yields a state which is not necessarily a minimum of the variational energy landscape.
    Variation After Projection (VAP) performs the projection first, so the variational minimum that is found is the true minimum of a variational energy functional, though it is not the same energy functional as what would be obtained without the projection.
    Following Sec.~\ref{sec:Hamiltonian_osc}, it is natural to imagine that the BRST version of the variational functional would be based on the expectation value of $\Hint$ from \eqref{eq:Hint}.
    In the next subsection we will contrast these three approaches
    and then return in Sec.~\ref{subsec:BRST_variational_functional}
    to show how a 
    BRST energy functional, if treated exactly and with proper treatment of gauge fixing, ultimately yields the same functional as VAP. 
    But the BRST path is different and along the way offers more opportunities for approximation and EFT power counting.

\subsection{General energy functionals}
  \label{subsec:general_functionals}
  
    In anticipation of generalizing to $N$ particles, we consider a parametrized single-particle product ansatz for the PAV and VAP calculations using the original Hamiltonian. 
    We begin with the PAV calculation.
    For a symmetric product ansatz with variational parameters $(\alpha_1, \alpha_2, \ldots \alpha_n) \equiv \alphavec$
    \begin{align}
        \Phi(\xvec; \alphavec) = \phi(x_1; \alphavec)\phi(x_2; \alphavec) ,
        \label{eq:Phi_simple}
    \end{align}
    we first compute the energy expectation value 
    \begin{align}
        \widetilde{E}_H(\alphavec) = \frac{\ev{H}{\Phi}}{\braket{\Phi}} ,
        \label{eq:E_H}
    \end{align}
    and then minimize it with respect to the parameters $\alphavec$
    \begin{align}
        0 = \nabla_{\alphavec}\widetilde{E}_H(\alphavec)\eval_{\alphavec = \alphavec_{\PAV}^*} .
    \end{align}
    (Note that we will get different results for $\alphavec_{\PAV}^*$ using $H$ and $\Hint$ in general, see below.)
    Next we apply a projection operator to the optimized wavefunction.
    The projection operator to project onto a state of momentum $q$ is (for $N$ particles)
    \begin{align}
        \Phat(q) = \int\dd{a}\exp\Bigl\{ 
        ia\Bigl(\sum_{j=1}^N \widehat{p}_j - q\Bigr)
        \Bigr\} ,
        \label{eq:Pofq}
    \end{align}
    where we suppress for now the (infinite) integration limits on $a$.
    To project a wavefunction in coordinate space onto zero momentum, we simply choose $q=0$ and write the momentum operators in their coordinate representation, as derivatives:
    \begin{align}
        \Phat(0)\equiv \Phat_0 = \int\dd{a}\exp\Bigl\{a\sum_{j=1}^N \nabla_j\Bigl\} .
    \end{align}
    The effect of the exponential is to shift the wavefunction from $\phi(x_1, x_2,\ldots)$ to $\phi(x_1 + a, x_2 + a, \ldots)$, and the integration is then summed over all shifts.
    This constructs a momentum eigenstate with total momentum equal to zero.
    The final result for the $N=2$ wavefunction is
    \begin{align}
        \Phi_{\PAV}(\xvec;\alphavec_{\PAV}^*) = \int\dd{a}\phi(x_1 + a; {\alphavec^*_{\PAV}})\phi(x_2 + a; {\alphavec^*_{\PAV}}) . 
    \end{align}
    To evaluate $E_{\PAV}$, we first need to define a projected functional to evaluate at $\alphavec_{\PAV}^*$.

    So now we consider the VAP calculation. 
    We start with the same product ansatz, but we perform the projection first. 
    This means that we must first calculate 
    \begin{align}
        \Phi_{\VAP}(\xvec; \alphavec) = \int\dd{a}\phi(x_1 + a; \alphavec)\phi(x_2 + a; \alphavec)
    ,\end{align}
    and then use this projected wavefunction in the variational energy functional.
    In this case, for every value of $\alphavec$ our variational ansatz is in the intrinsic space (it only depends on $x_1 - x_2$ by construction).
    In contrast, the PAV ansatz for general values of $\alphavec$ will have dependence on $X$, i.e., it will be outside the intrinsic space.
    
If we identify the norm and Hamiltonian kernels associated with a translation by $a$ (with $\alphavec$ dependence implicit) as:
\begin{align}
    N(a) &\equiv \ev{e^{ia\sum_{j=1}^N \widehat{p}_j}}{\Phi} , \\
    H(a) &\equiv \ev{H e^{ia\sum_{j=1}^N \widehat{p}_j}}{\Phi} ,
\end{align}
then we can write the VAP energy functional as
\begin{equation}
    E_{\VAP}[\Phi] = \frac{\int\dd{a}H(a)}{\int\dd{a}N(a)} .
    \label{eq:VAP}
\end{equation}
The result for $E_{\PAV}$ comes from \eqref{eq:VAP} evaluated at the fixed $\alphavec_{\PAV}^*$ value.
Expanding numerator and denominator in \eqref{eq:VAP} about $a=0$ to quadratic order defines the Gaussian Overlap Approximation (GOA). 

As noted above, 
we might naively expect based on the results in Sec.~\ref{sec:Hamiltonian_osc} that we can take the wave function ansatz to have the spurious and ghost sectors in their ground state and therefore decoupled from the $x_1, x_2$ dependence.
Then we would simply evaluate the energy functional using the same product ansatz in \eqref{eq:Phi_simple} together with the intrinsic Hamiltonian $\Hint$ from \eqref{eq:Hint}.
Indeed, this is the prescription sometimes used in many-body calculations to (partially) remove potential center-of-mass contamination.
In the following section we use explicit choices for the product ansatz and two-body potential to explore the behavior of the PAV, VAP, and $\Hint$-only functionals.
    
\subsection{Example potentials and ans\"atze } \label{subsec:variational_examples}

    Consider first a simple one-parameter Gaussian ansatz:
        \begin{align}
            \Phi(\xvec, \alphavec) = e^{-\frac{1}{2}\alpha x_1^2} e^{-\frac{1}{2}\alpha x_2^2} 
            = e^{-\frac{1}{4}\alpha(x_1-x_2)^2}e^{-\alpha X^2},
            \label{eq:gaussian_product_ansatz}
        \end{align}
        with $\alpha$ positive for normalizability.
        We have highlighted in \eqref{eq:gaussian_product_ansatz} that the dependence on $x_1-x_2$ factors from the dependence on $X = \frac12(x_1 + x_2)$.
Let us see the implications for the procedures from Sec.~\ref{subsec:general_functionals}.

For illustration we first use a harmonic interaction, for which the integrations are analytic:
    \begin{align}
        H = \frac{p_1^2 + p_2^2}{2m} + V(x_2 - x_1)
       \ \longrightarrow\  \Hho   = -\frac{\partial_1^2 + \partial_2^2}{2m} + \frac{k}{2}(x_2 - x_1)^2 .
    \end{align}
    Other potentials are of course admissible and demonstrated below, the only restriction on $H$ is that it be translationally invariant (and thus commute with $P$).
From \eqref{eq:E_H}
    \begin{align}
        \widetilde E_H(\alpha) = \frac{\alpha(km + \alpha^2)}{2m\alpha^2} ,
    \end{align}
    which is minimized for
    \begin{align}
        \alpha^* = \sqrt{km} 
        \quad \Longrightarrow \quad
        \widetilde E_H(\alpha^*) = \sqrt{\frac{k}{m}}.
        \end{align}
    If we now project with this \emph{fixed} value of $\alpha$ onto a state of zero momentum, we compute 
        \begin{align}
            \psi^{\proj}(\xvec, {\alpha}^*) &= \int_{-\infty}^{+\infty}\dd{a}\exp{-\frac{1}{2}\sqrt{km}(x_1 + a)^2}\exp{-\frac{1}{2}\sqrt{km}(x_2 + a)^2} \notag\\
             & = \sqrt{\frac{\pi}{\sqrt{km}}}\exp{-\frac{1}{4}\sqrt{km}(x_2 - x_1)^2} .
        \end{align}
        This projected wavefunction is a function of $x_2 - x_1$ only, as expected, and yields
        \begin{align}
            E_{\PAV}(\alpha^*) = \frac{\ev{P_0HP_0}{\Psi}}{\ev{P_0P_0}{\Psi}} &= \frac{\mel{\psi^{\proj}}{\Hho}{\Psi}}{\braket{\psi^{\proj}}{\Psi}}
            = \frac{3}{4}\sqrt{\frac{k}{m}}.
        \end{align}
        This is lower than the energy from the plain variational principle because the wavefunction used in the final calculation of the energy does not include the contamination from the $X$ dependence.
        However, it is not as low as it could be, since the variation was done with the $X$ dependence still present.

        For the VAP calculation, we use the projected form of the wavefunction for the variation, so we are only varying within the projected (intrinsic) subspace.
        We begin by projecting the trial wavefunction 
        \begin{align}
            \psi^{\proj}(\xvec, {\alpha}) &= \int\dd{a} e^{-\frac{1}{2}\alpha (x_1 + a)^2}e^{-\frac{1}{2}\alpha (x_2 + a)^2} \notag\\
            &=\sqrt{\frac{\pi}{\alpha}}\exp{-\frac{\alpha}{4}(x_2 - x_1)^2} .
        \end{align}
        We now use this to perform a variational calculation of the energy
        \begin{align}
            E_{\VAP}({\alpha}) = \frac{\ev{P_0HP_0}{\Psi}}{\ev{P_0P_0}{\Psi}} = \frac{\mel{\psi^{\proj}}{H}{\Psi}}{\braket{\psi^{\proj}}{\Psi}} ,
            \label{eq:E_VAP_alpha}
        \end{align}
        where now the energy is a function of ${\alpha}$ whereas in the PAV prescription, ${\alpha^*}$ was already determined without any symmetry restoration, and therefore minimizes a different functional from the one that yields the best minimum.
        The original Hamiltonian $\Hho$ is still used here, and we get
        \begin{align}
            E_{\VAP}({\alpha}) = \frac{\alpha^2 + 2km}{4m\alpha} .
            \label{eq:E_VAP}
        \end{align}
        Minimizing with respect to ${\alpha}$ parameters for positive values, we find 
        \begin{align}
            \alpha^* = \sqrt{2km}  \implies E_\VAP({\alpha}^*) 
            = \frac{1}{\sqrt{2}}\sqrt{\frac{k}{m}} .
        \end{align}
        With the correct projected functional form of the trial wavefunction, we obtain the exact ground state energy; this is not surprising because our ansatz includes the exact ground-state wavefunction.
        
        For the $\Hint$ calculation, we use the intrinsic sector Hamiltonian as this is the only one that is not diagonalized in general by our choice of BRST extension in Sec.~\ref{sec:Hamiltonian_osc}:
        \begin{align}
            \Hint &= \frac{p_1^2 + p_2^2}{2m} - \frac{(p_1 + p_2)^2}{2(2m)} + V(x_2 - x_1)\notag\\
              &= -\frac{\partial_1^2 + \partial_2^2 - 2\partial_1\partial_2}{4m} + \frac{k}{2}(x_2 - x_1)^2
            \label{eq:H_intinsic_variational}
        .\end{align}
        The Hamiltonian and normalization integrals are straightforward;
        the resulting energy function to minimize is the same as \eqref{eq:E_VAP}, i.e.,  $E_{\Hint}(\alpha)= E_{\VAP}(\alpha)$. 
        In retrospect, it is easy to see that the equivalence of $E_{\Hint}(\bf{\alpha})$ and $E_{\VAP}(\bf{\alpha})$ for \emph{any} ansatz for which the $x_1-x_2$ dependence factors from the $X$ dependence and for \emph{any} potential depending only on $x_1-x_2$.
        In particular, for $E_{\Hint}$, the Hamiltonian acts only on the $x_1-x_2$ dependence while the $X$ dependence cancels between numerator and denominator.
        For $E_{\VAP}$, the $X$ dependent part of $H$ will give zero on $\psi^{\proj}$, the remaining $X$ dependence will cancel, and the $x_1-x_2$ parts of the functionals are the same.

\begin{figure}[htb]
    \centering
    \begin{subfigure}[t]{0.45\textwidth}
        \centering
        \includegraphics[width=0.99\textwidth]{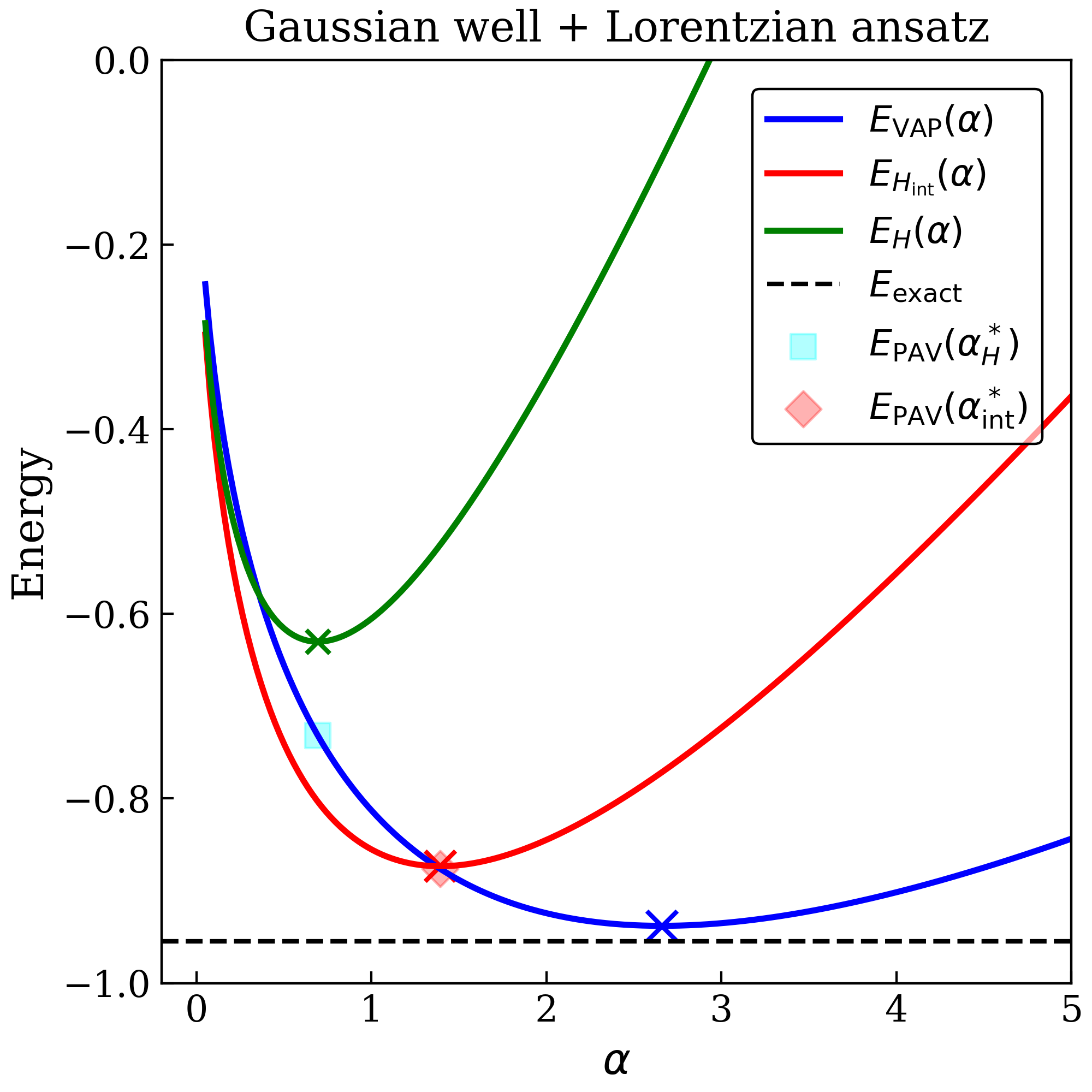}
    \end{subfigure}
        \hfill
    \begin{subfigure}[t]{0.45\textwidth}
        \centering
        \includegraphics[width=0.99\textwidth]{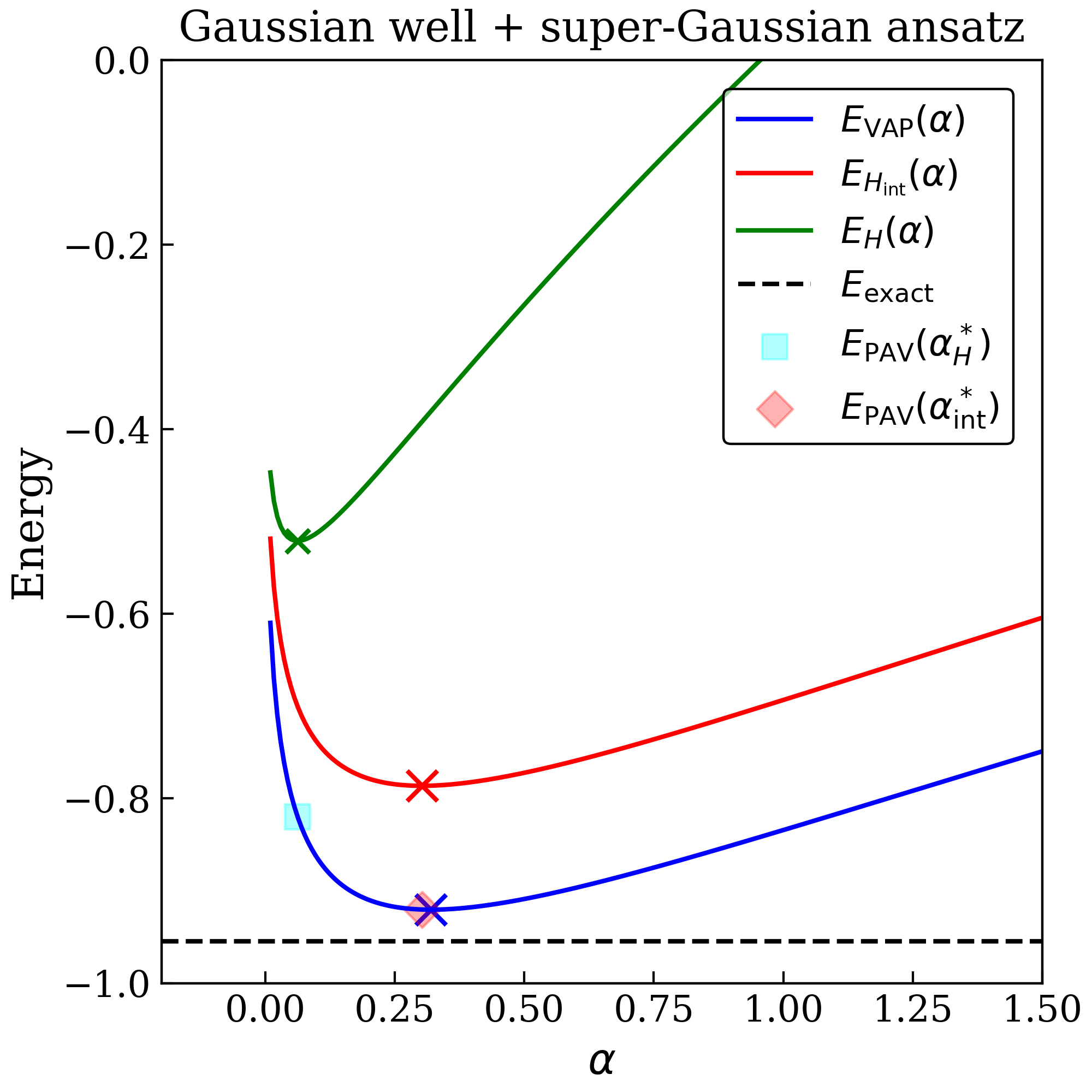}                
    \end{subfigure}%
    \caption{Energy as a function of variational parameter $\alpha$ for a Gaussian potential well with (a) a Lorentzian product ansatz and (b) a super-Gaussian product ansatz. Results for VAP, $\Hint$, and plain $H$ are compared to the exact result and PAV evaluated with $\alpha^*_H$ and $\alpha^*_{\Hint}$.}   
   \label{fig:Var_E_gaussian-well}        \end{figure}

\begin{figure}[htb]
    \centering
    \begin{subfigure}[t]{0.45\textwidth}
        \centering
        \includegraphics[width=0.99\textwidth]{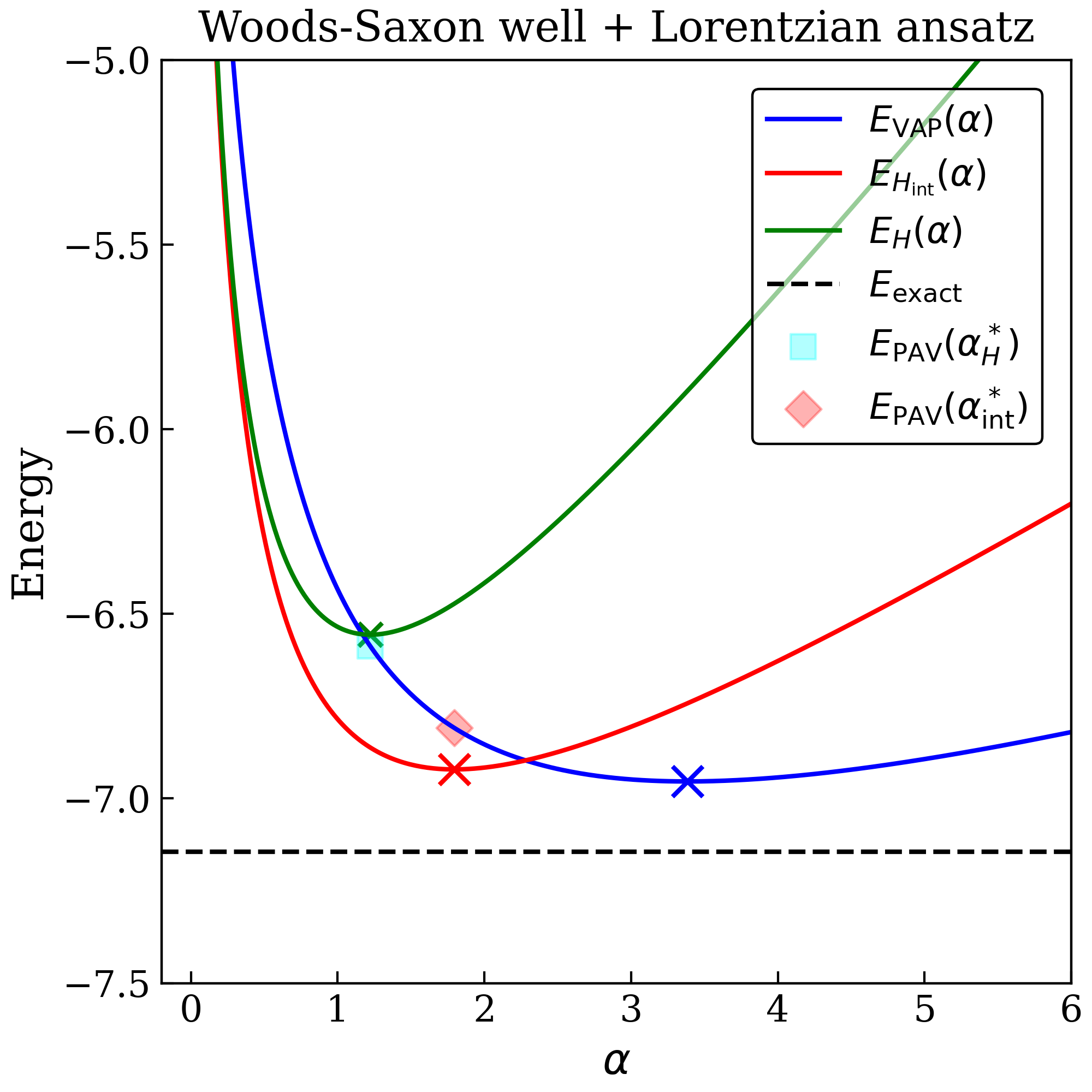}
    \end{subfigure}
    \hfill
    \begin{subfigure}[t]{0.45\textwidth}
        \centering
        \includegraphics[width=0.99\textwidth]{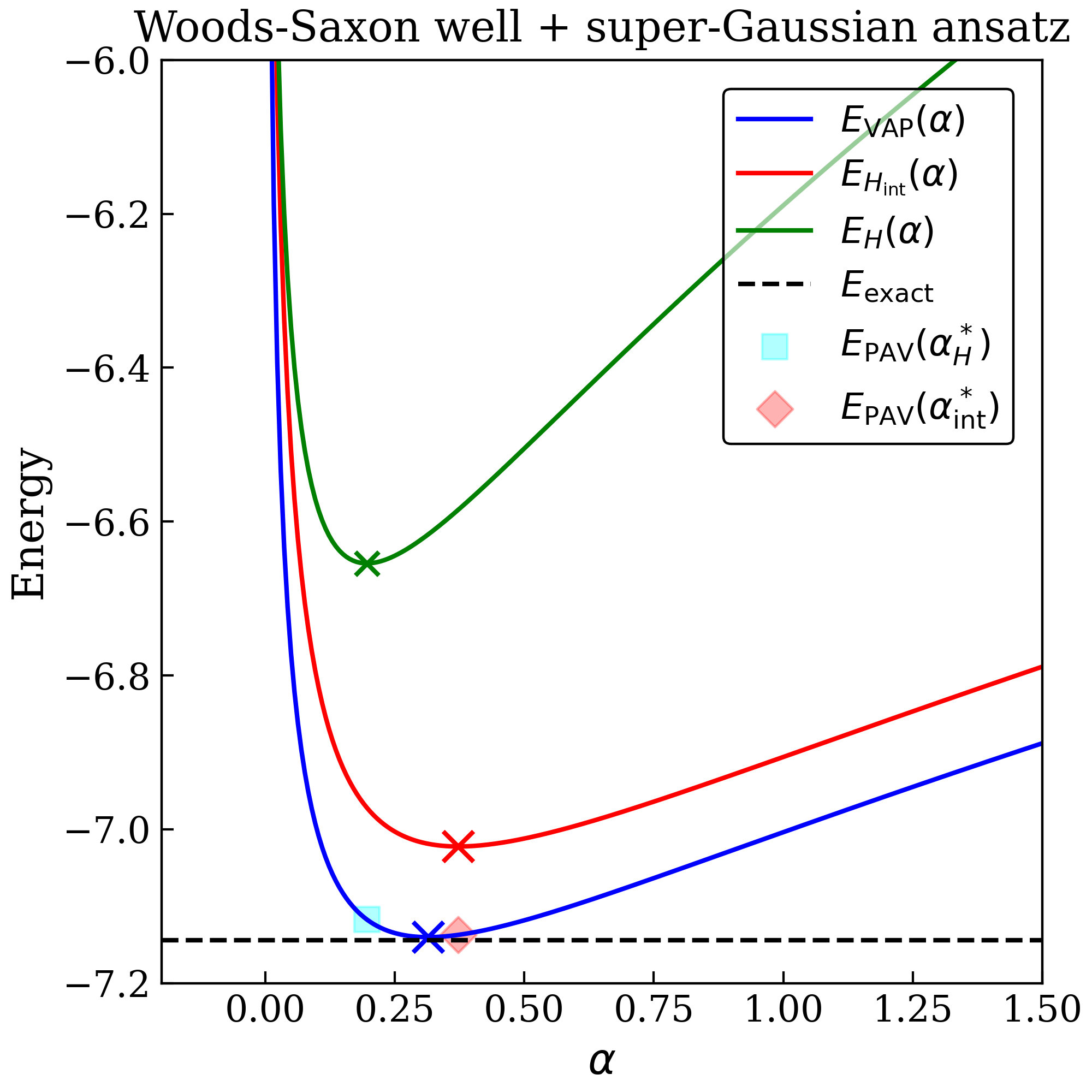}                
    \end{subfigure}%
    \caption{Energy as a function of variational parameter $\alpha$ for a Woods-Saxon potential well with (a) a Lorentzian product ansatz and (b) a super-Gaussian product ansatz. Results for VAP, $\Hint$, and plain $H$ are compared to the exact result and PAV evaluated with $H$ and $\Hint$.}
    \label{fig:Var_E_Woods-Saxon}
\end{figure}

Therefore, to find differences we need to look at a wider range of ans\"atze and potentials.
We show results for some representative choices in Figs.~\ref{fig:Var_E_gaussian-well} and \ref{fig:Var_E_Woods-Saxon}.
The wavefunction ans\"atze are Lorentzian and
super-Gaussian (quartic, specifically):
    \begin{align}
        \Psi_{\text{Lorentzian}}(\xvec, {\alpha}) = \frac{1}{\alpha x_1^2 + 1}\frac{1}{\alpha x_2^2 + 1} 
        \quad\text{and}\quad
        \Psi_{\text{super-Gaussian}}
        = \exp{-\alpha x_1^4}\exp{-\alpha x_2^4}.
            \label{eq:lorentzian_ansatz}
    \end{align}
and the potentials are Gaussian and Woods-Saxon wells:
\begin{align}
  V_{\text{Gauss}}(x_1 - x_2) &= -V_0 e^{-\mu(x_1-x_2)^2}
  \quad\text{with}\quad V_0=2,\ \mu=1 ,\\
  V_{\WS}(x_1 - x_2) &= \frac{-V_0}{1 + e^{(|x_1-x_2| - R_0)/a_{\WS}}} 
    \quad\text{with}\quad V_0=8,\ R_0=2.5,\ a_{\WS}=0.6 .
\end{align}
All of the examples in the figures exhibit the same features:
\begin{itemize}
    \item $E_{H}$ lies strictly above $E_{\Hint}$ for all $\alpha$ values, but only the minimum is guaranteed to be above $E_{\VAP}$.
    \item The minimized value of $E_{\PAV}$ relative to the minimized $E_{\Hint}$ depends on whether $E_H$ or $E_{\Hint}$ is used to evaluate the projected energy and on particular details. 
    This value lies on the projected functional $E_{\VAP}(\alpha)$, but not at its minimum because the $\alpha$ was chosen to minimize $E_{H}(\alpha)$.
    \item The $E_{\Hint}$ minimum is always above the $E_{\VAP}$ minimum. Thus, the equivalence found for factorizable ans\"atze is not general.
\end{itemize}
This last point means that we need to turn to a full BRST treatment to recover the VAP results.

\subsection{BRST symmetry and VAP}
  \label{subsec:BRST_variational_functional}

Here we show how an energy functional in the extended BRST space deals with a product ansatz.
In Secs.~\ref{subsec:general_functionals} and \ref{subsec:variational_examples}, we were able to use a product of single-particle wave functions $\Phi(\xvec;\alphavec)$ in  \eqref{eq:Phi_simple} as our ``seed'' trial function  for the VAP and PAV functionals, even though it was not translationally invariant, because we included a separate projection step.
In the BRST framework, the analogous projection to a BRST-closed state can arise within the functional through an integration in the extended space.
In the context of gauge theory, using a projection operator corresponds to the Dirac prescription for the calculation of observables.
That is, the use of a projection operator is equivalent to picking gauge-invariant states and operators from the outset, as the calculation is restricted to states that solve the constraint $F=0$.
Still to be explored is the potential for gauge fixing, which can be done consistently by extending the Hamiltonian via a BRST-exact term. 
Gauge-fixing is non-trivial for the variational method that yields mean-field solutions, and a tractable prescription for doing so is, at this time, unclear. 
But we will keep the full BRST machinery to make the connection to the path integral formulation where it is important to make consistent approximations without spurious contamination. 

\subsubsection{Alternative parametrization of a BRST extension}

To implement constraint(s) and gauge-fixing choice(s) in deriving an energy functional or transition matrix elements that preserves the key result of Eq.~\eqref{eq:BRST_invariance_me}, we turn to an alternate parametrization of a BRST extension to gauge-invariant operators.
That is, we perform a BRST extension of an operator that is equivalent to what is exemplified in Eq.~\eqref{eq:H_BRST_2body}, but written in a more convenient manner for the calculations of matrix elements. 
In particular,
we multiply  a general (gauge-invariant) operator $\mathcal{G}_0$ whose expectation value we wish to calculate by an exponentiated BRST-exact operator:%
\footnote{The overall phase in the exponential is arbitrary. 
Choosing it as $i$ corresponds to particular analytic continuations that are convenient for our discussion, but does not affect physical observables.} 
\begin{align}
    \mathcal{G}_0 \to \mathcal{G}_0%
   \exp{i\comm{\rhotilde}{Q}_{+}} .
   \label{eq:Mrhotilde}
\end{align}
This exponential has a term that is just the identity operator, and then powers of the BRST-exact operator $\comm{\rhotilde}{Q}_{+}$.
To demonstrate that this is equivalent to the picture of BRST extensions that were shown in Sec.~\ref{sec:brst_ped_formalism}, we must show that the exponential can be written as a sum of a term that does not modify\footnote{Up to an overall c-number scaling, which can be reabsorbed into a normalization factor.} $\mathcal{G}_0$, and a BRST-exact term.

The series expansion of the exponential yields 
\begin{align}
    e^{i\comm{\rhotilde}{Q}_{+}} = 1 + i\comm{\rhotilde}{Q}_+ + \frac{1}{2}i^2\comm{\rhotilde}{Q}_+^2 + \frac{1}{3!}i^3\comm{\rhotilde}{Q}_+^3 \ldots \,.
\end{align}
The identity operator will leave $\mathcal{G}_0$ unaffected, which is one of our criteria above. 
The remaining part is
\begin{align}
    e^{i\comm{\rhotilde}{Q}_{+}} - 1 &= \qty(i + \frac{1}{2}i^2\comm{\rhotilde}{Q}_+ + \frac{1}{3!}i^3\comm{\rhotilde}{Q}_+^2 + \ldots)\comm{\rhotilde}{Q}_+\notag\\
    &= \qty(\sum_{n=0}^{\infty}\frac{1}{(n+1)!}i^{n+1}\comm{\rhotilde}{Q}_{+}^n)\comm{\rhotilde}{Q}_+ .
\end{align}
Now, 
we note that everything on the right side is 
BRST-exact,
and as such has zero anticommutator with the charge $Q$.
As such, we can write the result as the anticommutator 
\begin{align}
    e^{i\comm{\rhotilde}{Q}_{+}} - 1 &= \comm{\qty(\sum_{n}\frac{1}{(n+1)!}i^{n+1}\comm{\rhotilde}{Q}_{+}^n)\rhotilde}{Q}_+  \notag \\
   & \equiv \comm{\mathcal{S}}{Q}_{+}
.\end{align}
This is an operator being transformed by the BRST charge, so ultimately we have that our exponential is equivalent to
\begin{align}
    e^{i\comm{\rhotilde}{Q}_{+}}  = 1 + \comm{\mathcal{S}}{Q}_{+} .
\end{align}
As such, the multiplication of the exponentiated BRST-exact operator with the original gauge-invariant operator $\mathcal{G}_0$ is 
\begin{align}
    \mathcal{G}_0 e^{i\comm{\rhotilde}{Q}_{+}} &= \mathcal{G}_0 + \mathcal{G}_0\comm{\mathcal{S}}{Q}_+\notag\\
    &= \mathcal{G}_0 + \comm{\mathcal{G}_0\mathcal{S}}{Q}_+ ,\end{align}
where the gauge invariance of $\mathcal{G}_0$ was used to move it into the anticommutator. 
Thus, $\mathcal{G}_0 \to \mathcal{G}_0\exp{i\comm{\rhotilde}{Q}_{+}}$ is nothing more than a different parametrization of the type of extension already discussed in Sec.~\ref{sec:brst_ped_formalism} and explicitly used for diagonalization in Sec.~\ref{sec:Hamiltonian_osc}.
Stated symbolically, we are still doing $\mathcal{G}_0 + \comm{Q}{\rho}_+$ as indicated in Eq.~\eqref{eq:BRST_invariance_me}, but now with 
\begin{align}
    \rho = \mathcal{G}_{0}\sum_{n}\frac{1}{(n+1)!}i^{n+1}\comm{Q}{\rhotilde}_+^n\rhotilde
\end{align}
This form is helpful for the calculation of matrix elements%
\footnote{Rather than choosing terms to add directly to the operator, this parametrization allows for the engineering of terms in the exponent as an equivalent manner of gauge fixing when calculating matrix elements of the original operator.}
and will be shown to yield a direct path to reproducing a VAP calculation, as described above.

\subsubsection{Additional freedom in the variational ansatz}
\label{subsubsect:variational-ansatz}

A physical wave function must be BRST closed, which means that it is annihilated by the charge $Q$. 
As before, we have $Q = -\eta F + \pibar B$, with $F = p_R - P \rightarrow -i(\partial_R - \partial_{x_1} - \partial_{x_2})$, $B = -i\partial_\lambda$, and $\pibar = \partial_{\etabar}$.
The form of $Q$ suggests two paths to achieve $Q\Psi = 0$ for physical $\Psi$.
One way is to work with an ansatz without ghosts and solve the constraint $F = 0$, 
which, as discussed at the start of Sec~\ref{subsec:BRST_variational_functional}, is equivalent to standard projection.
However, the ghosts give us additional freedom, which may lead to a useful alternative.
Such alternatives will still need to be explored in future work, however there is a result that is known in the literature~\cite{Henneaux:1985kr, SHVEDOV20022} that reproduces the Dirac prescription (that is, VAP) using the extra DoFs introduced by the BRST quantization machinery.
For the sake of pedagogy and potential inspiration for extensions, we will now illustrate this procedure.

The constraint term in the charge can be eliminated if the ansatz already has a ghost and we exploit the nilpotence of Grassmann numbers rather than the solving of the constraint in order to achieve annihilation by the charge. 
If we want the wavefunction to have zero ghost number (and be Grassmann even), then we must include not just $\eta$ but $\etabar$.
To have this kind of ansatz still be annihilated by the charge then, we must have a function that is annihilated by $B$, which means it is independent of $\lambda$.
So, we must have something that is of the form\footnote{This form can also be motivated from a BRST cohomological argument, as done in Ch.~14 of \cite{Henneaux:1992ig}.} 
\begin{align}
    \Psi = \Phi(x_1, x_2, R)\etabar\eta \label{eq:ghost_variational_ansatz}
,\end{align}
with $\Phi$ allowed to be a product ansatz if desired, though we do not make this choice in the derivation.
The $\lambda$ integration for such an ansatz is determined entirely by the $\lambda$ dependence of the gauge-fixing form.
The choice $\rhotilde = \lambda\pi$ yields for the BRST extension of the original Hamiltonian $H_{0}$, 
\begin{align}
    H_{0} \to H_{0}e^{-i\lambda F - \pi\pibar}.
\end{align}
Using the ansatz Eq.~\eqref{eq:ghost_variational_ansatz} with some set of parameters $\alpha$, we can compute the expectation value in coordinate space by calculating
\begin{align}
    E(\alpha) = \frac{\ev{H_{0}\exp{-i\lambda F - \pi\pibar}}{\Psi;\alpha}}{\ev{1 \exp{-i\lambda F - \pi\pibar}}{\Psi;\alpha}}
\end{align}
with integrations over $x_j$, $\lambda$, $\eta$, $\etabar$, $R$ in the coordinate space evaulation of the expectation values. 
We will demonstrate the recovering of the VAP integration on the numerator, the denominator follows exactly the same steps.

Because the choice of $\rhotilde = \lambda\pi$ does not fix a gauge, the operators in the exponential commute with each other,\footnote{A gauge-fixing function must not commute with the constraint in order to pick out a gauge slice, which would amount to fixing some function of the coordinates $x_j$, $R$. Not doing so means that our Hamiltonian still commutes with $P$ and $p_R$, so we have not picked a gauge slice, c.f.\ Fig~\ref{fig:gauge_orbits_fixing}.} so we are free to separate the exponential into a product:
\begin{align}
    e^{-i\lambda F - \pi\pibar} = e^{-i\lambda F} e^{-\pi\pibar}.
\end{align}
The ghost exponential is written as a first-order Taylor expansion by Grassmann properties, and we find for the numerator that 
\begin{align}
    \ev{H_{0}e^{-i\lambda F}e^{-\pi\pibar}}{\Psi;\alpha} = \int\dd{x_1}\dd{x_2}\dd{R}\dd{\lambda}\int\dd{\etabar}\dd{\eta}\Phi^*\eta\etabar\qty(\frac{p_1^2 + p_2^2}{2m} + \potential{x_2}{x_1})e^{-i\lambda F}(1 - \pi\pibar)\etabar\eta\Phi.
\end{align}
The $\pi$ operators are, in coordinate space, derivatives with respect to their conjugate ghost coordinate.
Counting the number of ghosts, we can see that only the $\pibar\pi$ term in the expanded exponential yields a nonzero Grassmann integration. 
The identity operator term leaves two powers of $\eta$ and $\etabar$ in the integrand, which is zero by Grassmann properties. 
As such, we take the derivatives with respect to $\eta, \etabar$ and perform the Grassmann integration, leaving (up to a negative sign which can be omitted as it will occur in the denominator and cancel)
\begin{align}
    \ev{H_{0}e^{-i\lambda F}e^{-\pi\pibar}}{\Psi;\alpha} = \int\dd{x_1}\dd{x_2}\dd{R}\dd{\lambda} \Phi^*\qty(\frac{p_1^2 + p_2^2}{2m} + \potential{x_2}{x_1})e^{-i\lambda F}\Phi.
\end{align}
We have so far not made any assumptions about properties of $\Phi$. 
The Hamiltonian $H_{0}$ manifestly commutes with $p_R$, so we now postulate that the wavefunction factorizes into a plane wave in $R$ and some other function that only depends on the original degrees of freedom, $\Phi = \Phi_0(x_1, x_2)\exp{ik_{R}R}$.
Inserting this into our expectation value integral, we find that
\begin{align}
    \ev{H_{0}e^{-i\lambda F}e^{-\pi\pibar}}{\Psi;\alpha} = \int\dd{x_1}\dd{x_2}\dd{R}\dd{\lambda} \Phi_{0}^*e^{-ik_{R}R} \qty(\frac{p_1^2 + p_2^2}{2m} + \potential{x_2}{x_1})e^{-i\lambda F}e^{ik_{R}R}\Phi_{0}.
\end{align}
Now, the plane wave in $R$ is an eigenstate of the constraint operator, since it is a momentum eigenstate and the constraint operator is $p_R - \sum_{j}p_j$.
Because there is no other operator that responds to the collective coordinate $R$ in the expectation value, we can replace the $p_R$ with its eigenvalue $k_R$, and then move the plane wave to cancel with its complex conjugate:
\begin{align}
    \ev{H_{0}e^{-i\lambda F}e^{-\pi\pibar}}{\Psi;\alpha} = \int\dd{x_1}\dd{x_2}\dd{R}\dd{\lambda} \Phi_{0}^* \qty\bigg(\frac{p_1^2 + p_2^2}{2m} + \potential{x_2}{x_1})\exp\qty\Big{-i\lambda \qty\Big(k_R - \sum_{j}p_j)}\Phi_{0}.
\end{align}
Regulating the $R$ integration by putting the system in a finite volume $\mathcal{V}$, we are left with 
\begin{align}
    \ev{H_{0}e^{-i\lambda F}e^{-\pi\pibar}}{\Psi;\alpha} = \mathcal{V}\int\dd{x_1}\dd{x_2}\dd{\lambda} \Phi_{0}^* \qty\bigg(\frac{p_1^2 + p_2^2}{2m} + \potential{x_2}{x_1})\exp\qty\Big{-i\lambda \qty\Big(k_R - \sum_{j}p_j)}\Phi_{0},
    \label{eq:Dirac_projection}
\end{align}
which only differs cosmetically from the action of the projection operator Eq.~\eqref{eq:Pofq} on the wavefunction used in the calculation of the energy functional. 
We were able to restrict ourselves to only gauge-invariant states via the extended space integration.
This was achieved with the Lagrange multiplier becoming the integration variable for a projection integral, which is equivalent to the Dirac prescription of restricting states to those that are gauge-invariant in the language of gauge theory.
We can thus think of the standard projection procedures as a subset of possible BRST extensions of an operator.

\subsubsection{Variational calculations and gauge fixing}

The choice $\rhotilde = \lambda\pi$ reduces to the VAP result by using the integration in the extended space to construct a projection operator for the gauged symmetry. 
Extremely minimal assumptions were made on the state in order to achieve this result.
We have yet to find a choice of $\rhotilde$ that fixes a gauge rather than producing a gauge-invariant state that is suitable for single-particle basis-expansion methods.
The main difficulty is that a gauge-fixing operator must not commute with the constraint, meaning that the factorization of the exponentials is not guaranteed, making the evaluation of the expectation value extremely non-trivial. 
The use of a gauge-fixing extension with the ansatz Eq.~\eqref{eq:ghost_variational_ansatz} is appealing, as alternative choices of gauge-fixing fermion should in principle allow for a complete circumventing of a projection integral while still yielding an intrinsic result, in the spirit of the diagonalization in Sec.~\ref{sec:Hamiltonian_osc}.

We have two lines of thought for possibly performing such a calculation.
The first is to ``Trotterize'' a gauge-fixing exponential so that the operator ordering of the exponential can be assessed via insertions of complete sets of states, and a suitable calculation with such an ordering can be performed.
Such a procedure would be mechanically similar to time slicing arguments used to derive path integrals, but with no time stepping involved between the intermediate states (see Appendix~\ref{sec:recentered} for an example of how this plays out).
The second is to pick a gauge-fixing operator that enables the order-by-order neglect of Baker-Campbell-Hausdorff terms in some sort of approximation scheme.
An incomplete set of considerations for picking a $\rhotilde$ that facilitates such an expansion is 
\begin{itemize}
    \item The (Grassmann-even) ansatz that allows for annihilation by the charge without solving the constraint requires $\eta$ and $\etabar$, which will yield zero in expectation values unless the gauge-fixing occurs with a term that has both $\pi$ and $\pibar$.
    As such, to use the parametrization of Eq.~\eqref{eq:rho_param_1stquant} on $\rhotilde$ in the exponential, the gauge-fixing \textit{must} occur in $a_3$.
    \item The other $a$ terms do not directly contribute, so they can be set to zero for a minimal implementation of gauge fixing.
    This does not necessarily mean that nonzero $a_1$, $a_2$, $a_4$ cannot be useful, as they may provide a way to obtain operator algebra identities in the exponential.
    At minimum, the operator structure inside the exponential is then
    \begin{align}
        -\comm{a_{3}}{F}\pi\eta - F a_{3} + \comm{a_{3}}{B}\pi\pibar &= i\sum_{j}\pdv{a_{3}}{x_j}\pi\eta - F a_{3} + i\pdv{a_{3}}{\lambda}\pi\pibar\notag\\
        &= i\sum_{j}\pdv{a_{3}}{x_j}\pi\eta - a_{3}F - i\sum_{j}\pdv{a_{3}}{x_j} + i\pdv{a_{3}}{\lambda}\pi\pibar
    ,\end{align}
    up to a global phase.
    \item If we want to avoid a projection integral in favor of a cancellation of spurious contributions to the observable in the spirit of Sec.~\ref{sec:Hamiltonian_osc}, we should avoid the $\lambda$ dependence of $a_{3}$ being a simple linear relationship.
    Since the derivative with respect to $\lambda$ occurs multiplying the Grassmann derivatives $\pi\pibar$ that yield a nonzero result, we can engineer the $\lambda$ dependence of $a_3$ to have a derivative that yields some desired form.
    For example, we can obtain a Gaussian in $\lambda$ if we have $a_3$ depend on $\lambda$ as $\erf(\varphi\lambda)$ for some dimensionful constant $\varphi$ that is arbitrary in the way $\alpha$ was in Sec.~\ref{sec:Hamiltonian_osc}.
    \item The series of BCH terms that are generated by attempting to fix a gauge is driven by commutators of $a_3$ with the constraint $F$, so the exponential will schematically be 
    \begin{align}
        \exp{i\sum_{j}\pdv{a_{3}}{x_j}\pi\eta - a_{3}F - i\sum_{j}\pdv{a_{3}}{x_j}}\exp{i\pdv{a_{3}}{\lambda}\pi\pibar}\exp{\sum_{j}\pdv{a_{3}}{x_j} + \pdv[2]{a_{3}}{x_j} + \ldots}
    .\end{align}
    The derivatives in $x$ will keep getting higher with higher BCH orders, since higher BCH terms will induce increasingly nested commutators with the constraint. 
    If each differentiation with respect to $x$ brings an additional power of an expansion parameter, such as $1/N$, then the BCH series could potentially be truncated in a systematic manner, resulting in the gauge being fixed at a given order of the parameter.
    Such a scheme could be realized with an exponential $x$ dependence of $a_3$.
\end{itemize}
Other such lines of thought can likely be discovered, the hope is to demonstrate ways of thinking about the implementation of a gauge-fixed variational calculation that could prove feasible in future work.

    \section{Path Integral Implementation}\label{sec:pi_osc}
        The path integral treatment enables the direct calculation
        of effective actions, which can serve as nuclear density functionals~\cite{Furnstahl:2019lue, sharma2025}.
        The construction of the path integral is itself somewhat subtle here, particularly because of the altered inner product that must be used for the $\lambda$ mode as discussed at the end of Sec.~\ref{sec:Hamiltonian_osc}.
        In the path integral formalism, the issue of the zero mode due to the symmetry presents itself in a direct manner: we get a trace log of the spectrum of the theory, which includes the zero mode manifestly.
        It is instructive to see this in the naive treatment of our toy model, both to serve as an indicator of how the calculation of the free energy reflects the symmetry for a real system, and to contrast against the BRST treatment of the problem.
        To that end, we seek to compute the ground state energy with a Euclidean path integral.
        This is done via the path integral representation of the partition function (c.f.\ Sec 3.2 of~\cite{Altland:2006}, Sec 2.1 of~\cite{wen_manybody}, Sec 2.2 of~\cite{Negele:1988vy}), taken to zero temperature.
        
        We start with a partition function, constructed in the standard manner of inserting complete sets of states at imaginary time slices as described in any of~\cite{Altland:2006, wen_manybody, Negele:1988vy, Kleinert_piqm:2004ev}:
        \begin{align}
            Z = \int\mathcal{D}x_1\mathcal{D}x_2\exp{-\int\dd{\tau}\qty(\frac{m}{2}\qty(\dot{x}_1^2 + \dot{x}_2^2) + \potential{x_2}{x_1})}
        .\end{align}
        For the familiar example of the oscillator system, the action $A$ in this partition function can be written in matrix form, integrating the kinetic terms by parts to get the derivatives as operators acting to the right on a vector of $x$ coordinates as 
        \begin{align}
            A = \int\dd{\tau}\frac{m}{2}\mqty[x_1 & x_2]\mqty[-\dv[2]{\tau} & 0\\ 0 & -\dv[2]{\tau}]\mqty[x_1\\x_2] + \frac{k}{2}\mqty[x_1 & x_2]\mqty[1 & -1\\ -1 & 1]\mqty[x_1\\x_2]\label{eq:unfixed_action_oscillator} .
        \end{align}
        This has manifest time translational symmetry, so we write it in Fourier space with $\omega_0^2\equiv k/m$:\footnote{The actual intrinsic frequency to the problem should use the reduced mass $\mu = m/2$ rather than the single-particle mass but such a concept is not necessarily generalizable to a many-body system, and the eventual intent is to use this for a single-particle picture of a many-body system, so we do not use it. As a result, our intrinsic mode will be proportional to $\omega_0$, not equal to it.}
        \begin{align}
            A = \frac{m}{2}\int\frac{\dd{\omega}}{2\pi}\mqty[x_1 & x_2]_{-\omega}\mqty[\omega^2 + \omega_0^2 & -\omega_0^2\\ -\omega_0^2 & \omega^2 + \omega_0^2]\mqty[x_1\\x_2]_\omega
        .\end{align}
        The partition function is thus a functional Gaussian integral with a 2x2 matrix as the kernel, which can be performed using textbook techniques (c.f.\ discussions surrounding ``Gaussian Integral'' listings in indices of \cite{Altland:2006, Kleinert_piqm:2004ev, Stone:2000}).
        In particular, the result for a multivariate Gaussian integral (summation convention implied on repeated indices for the rest of the section),
        \begin{align}
            \frac{1}{\pi^{n/2}}\int\dd[n]{x}\exp{-x_{i} M_{ij} x_{j}} = \det{M}^{-1/2}
        ,\end{align}
        is taken to the continuum limit here, with the index on $x$ above corresponding to a frequency value in the action. 
        We can exponentiate the determinant according to 
        \begin{align}
            \det{A}^{-1/2} = \exp{-\frac{1}{2}\ln\det{M}}
        ,\end{align}
        and use the standard matrix identity $\ln\det{M} = \Tr\ln{M}$ in order to get
        \begin{align}
            \det{A}^{-1/2} = \exp{-\frac{1}{2}\Tr\ln{M}}
        \end{align}
        Applying this result to our partition function, we see that we have to trace over the continuum matrix that represents the dynamics of our system, and the matrix corresponding to each particle index:
        \begin{align}
            Z = \qty(\det\mqty[\omega^2 + \omega_0^2 & -\omega_0^2\\ -\omega_0^2 & \omega^2 + \omega_0^2])^{-1/2} = \exp{-\frac{1}{2}\Tr\ln\mqty[\omega^2 + \omega_0^2 & -\omega_0^2\\ -\omega_0^2 & \omega^2 + \omega_0^2]}
        .\end{align}
        As such, our trace includes a time integration and a sum of the eigenvalues of the 2x2 matrix. 
        The matrix is readily diagonalized; through the above discussion on the continuum gaussian integral properties, we find\footnote{The partition function has a UV divergence associated with the construction of the path integral measure. Regulating such a divergence is necessary if the result is not divided by a reference partition function. The regulation does not affect the frequency dependence that is the main point we are after, so we report the un-regulated results}
        \begin{align}
            Z = \exp{-\frac{1}{2}\int\dd{\omega}\ln(\omega^2(\omega^2 + 2\omega_0^2))}
        .\end{align}
        We see the intrinsic mode at $\omega = \sqrt{2k/m} = \sqrt{2}\omega_0$ and a zero frequency mode.\footnote{Note that we are working in Euclidean spacetime, so the frequencies are on the imaginary axis, hence the poles appearing as $\omega^2 + \omega_0^2$ rather than $\omega^2 - \omega_0^2$.}
        
        The zero mode is precisely because of the translational symmetry of the CoM, and it is behind the difficulties of performing perturbative calculations in the many-body case as well.
        This can be readily seen by adding a CoM potential to the above action and following the same procedure of performing the functional Gaussian integrals.
        Adding an artificial potential of this form is one possibility, but we show instead that we can start from a Euclidean phase-space path integral for the original system and ``insert 1'' in different ways to recover the same extension and cancellation as our BRST-extended Hamiltonian in Sec.~\ref{sec:Hamiltonian_osc}.
        The role of BRST symmetry is less overt here.
        BRST symmetry was originally discovered~\cite{Becchi1975, Tyutin1975} as a symmetry of the Faddeev-Popov path integral, which is what we will arrive at here.
        The BRST-closed quantities are gauge-invariant quantities by construction, and anything that is not BRST-invariant must cancel by the symmetry in the path integral.\footnote{A familiar analogy: the $\ell\neq 0$ multipole moments of a spherically symmetric charge distribution must vanish by symmetry in the volume integral.}

        We begin with a ``phase space'' path integral for the original problem. 
        The phase space path integral integrates over positions and momenta, rather than just positions. 
        We should start here for two reasons
        \begin{itemize}
            \item The constraint machinery is built on momenta, not velocities.
            \item This is the closer analogy to the many-body path integral in second quantization: there we integrate over coherent states of the creation and annihilation operators, and do not proceed to integrate out any variables. For this simple system, the analogous starting point is the phase space path integral without the momenta integrated out.
        \end{itemize}
        We thus start with 
        \begin{align}
            Z = \int\prod_{i}^{N(=2)}\DD{x_i}\DD{p_i}e^{-A_0}\qc A_0 \equiv \int_{\tau}-ip_j\dot{x}_j + H(x, p)
        .\end{align}
        We introduce a collective momentum $p_R$ and a gauge-fixing function $G$ into our partition function by integrating them into the functional integral via delta functions:
        \begin{align}
            Z = \int\DD{x_i}\DD{p_i}\DD{p_R}\DD{G}\delta(p_R - p_1 - p_2)\delta(G)e^{-A_0}\qc p_R - p_1 - p_2 \equiv F
        .\end{align}
        The gauge-fixing function $G$ should in general depend on the DoFs we wish to fix, and it should depend explicitly on the parameters that describe the gauge transformation~\cite{Alessandrini:1978qd, Alessandrini:1978ki}.
        Here, that means that $G$ should be a function of the $x_j$ that breaks the translational invariance of the system.
        Because the gauge transformation is a translation, we can define a parameter $R$ that describes the distance by which we translate the system.
        We would like to convert the functional integration over $G$ to a functional integration over a coordinate, since it is not feasible to do any calculations with an integration over an arbitrary function $G$.
        We can do this by writing $G$ as not just a function of the original degrees of freedom $x_j$, but also of the gauge transformation parameter as it transforms the original DoFs.
        Here, we are gauging translation so we denote the gauge parameter $R$, so that we have $G = G(x_1 - R, x_2 - R)$.\footnote{It can be readily verified by following the argument that shifting in the positive or negative direction does not change anything about the final results, only some signs that must be consistent with each other throughout the calculation.}
        If other variables are introduced to the path integral (as will be the case here), then it will need to be assessed whether those variables must be fixed by $G$ as well.
        Regardless of the variables involved, we can convert the integration over $G$ to an integration over the gauge variable $R$ using the standard formula for a change in variables, just with a functional integral:
        \begin{align}
            Z = \int\DD{x_i}\DD{p_i}\DD{p_R}\DD{R}\delta(F)\delta(G(x_1 - R, x_2 - R))\qty|\fdv{G_\tau}{R_{\tau'}}|e^{-A_0}
        ,\end{align}
        where the temporal dependence of $G(x_j(\tau) - R(\tau))$ and $R(\tau)$ has been denoted with a subscript, both for compactness and to indicate the ``indices'' of the functional jacobian for the sake of familiarity.
        
        The goal now is to exponentiate these three new additions to the functional integral.
        This is helpful in principle because it allows us to see the constraint and gauge fixing as modifications to the effective Lagrangian or Hamiltonian, giving us an avenue to understand how the constraint and gauge function affect the theory.
        It is also helpful in practice because for a system that we cannot solve exactly, having a clearly written action that takes all of the gauge-fixing machinery into account allows us to write Feynman rules and perform standard field theory techniques to compute observables.
        We use different methods for exponentiating the different terms.

        We begin with the functional Jacobian.
        A general property of Grassmann numbers~\cite{Negele:1988vy} is that 
        \begin{align}
            \int\prod_i\dd{\bar{\theta}_i}\dd{\theta_i}\exp{\bar{\theta}_j A_{jk}\theta_k} = \det{A}
        .\end{align}
        We employ this identity in reverse to write the functional determinant as an exponential:
        \begin{align}
            \qty|\fdv{G_\tau}{R_{\tau'}}| = \int\DD{\etabar}\DD{\eta}\exp{\int_{\tau,\tau'}\etabar_\tau (\fdv*{G_\tau}{R_{\tau'}})\eta_{\tau'}}
        .\end{align}
        The gauge-fixing term is less obvious.
        We use a trick initially introduced by 't Hooft~\cite{THOOFT1971173}: the introductions of these terms into the partition function is not changed if, rather than fixing $G = 0$, we fix $G = c$ for some function $c(\tau)$ that does not depend on any of the other variables or parameters in the problem.\footnote{This can be viewed as a realization of the arbitrary frames of reference that the system can be viewed in: we are averaging over the CoM being an arbitrary function of time, so the system as a whole moves arbitrarily with time.}
        We can then integrate $Z$ over $c$ with a normalized Gaussian weight $\exp{-\int_\tau \xi c(\tau)^2/2}$, where $\xi$ is a constant and the normalization is absorbed into the integration measure.
        This leaves the LHS unaffected while on the RHS, we get a Gaussian in the constraint:
        \begin{align}
            \int\DD{R}\DD{c}\delta(G(x_j - R(\tau)) - c)e^{-\int_\tau \xi c^2/2} = \int\DD{R}\exp{-\int_{\tau}\frac{\xi}{2} G^2}\label{eq:thooft_osc}
        \end{align}
        This Gaussian may be (functionally) Fourier transformed, which is not necessary for solving the system but it is helpful for seeing connections to the Hamiltonian picture, as we will see after some further development.
        
        The constraint delta functional does not benefit from such a trick -- it defines a coordinate and it is not helpful to define the coordinate with an arbitrary function $c$ in the way we did for the gauge-fixing functional.
        As such, we must express the delta functional as an exponential using a standard representation.
        Two such options are with a (functional) Fourier representation, or a zero-width normalized Gaussian.
        Both are valid, though the Gaussian width parameter then must be taken to zero in the calculation of any observable~\cite{Alessandrini:1978qd}.
        In order to correspond with the Hamiltonian calculation in Sec.~\ref{sec:Hamiltonian_osc}, we choose the Fourier representation (factors of $2\pi$ absorbed into the functional integration measure):
        \begin{align}
            \delta(F) = \int\DD{\lambda}\exp{-i\int_\tau \lambda_\tau F}
        .\end{align}

        With all of these in mind, our partition function is now
        \begin{align}
            Z &= \int\DD{x_i}\DD{p_i}\DD{p_R}\DD{\lambda}\DD{R}\DD{\etabar}\DD{\eta}
            \notag \\
            & \quad\null\times\exp{\int_\tau ip_j\dot{x}_j - \frac{p_1^2 + p_2^2}{2m} - \potential{x_2}{x_1} - i\int_\tau \lambda(p_R - p_1 - p_2) - \int_\tau \frac{\xi G^2}{2} + \int_{\tau, \tau'}\etabar_{\tau}\qty(\fdv{G}{R})\eta_{\tau'}}
        .\end{align}
        We have added variables to the partition function, we should assess how they should transform under gauge transformations. 
        Aside from the gauge-fixing terms at the end, the action should be invariant under collective translations.
        We know that the coordinates $x_j$ transform additively: $x_j \to x_j + \epsilon$, and the momenta do not change. 
        The gauge parameter $R$ must transform additively as well, by construction.
        That leaves the $\lambda$ transformation law.
        If we perform the transformation on the $x_j$, then the $p^2/2m$ and potential energy terms are unaffected, but the ``Legendre transform'' terms $p\dot{x}$ pick up a shift of the form $(p_1 + p_2)\dot{\epsilon}$.
        This would be canceled if we had $\lambda \to \lambda - \dot{\epsilon}$ and there was a term of the form $ip_R\dot{R}$ in the action.
        The overall transformation laws then are 
        \begin{align}
            x_j \to x_j + \epsilon\qc R \to R - \epsilon\qc \lambda \to \lambda - \dot{\epsilon} ,
        \end{align}
        which corresponds to the transformation laws obtained from the canonical approach in Sec.~\ref{sec:brst_ped_formalism}.

        We now have to pick a gauge-fixing function $G$. 
        There is a tremendous amount of freedom here, we can pick anything that involves the CoM.
        For this discussion, we let correspondence with the Hamiltonian picture be our guide, and choose 
        \begin{align}
            G(x_1 - R, x_2 - R) &= \alpha\qty(\frac{x_1 - R + x_2 - R}{2}) - (\dot{\lambda} + \ddot{R})\notag\\
            &= \alpha\qty(\frac{x_1 + x_2}{2} - R) - \dot{\lambda} - \ddot{R}
        .\end{align}
        The functional derivative is then 
        \begin{align}
            \fdv{G_\tau}{R_{\tau'}} = -\alpha\delta(\tau - \tau') - \delta''(\tau - \tau')
        .\end{align}
        There are no other degrees of freedom in this functional derivative, so the ghosts decouple\footnote{This is generally possible for abelian symmetries, but not for non-abelian! The choice of the gauge-fixing function $G$ \textit{can} also have the ghosts couple to the other DoFs for abelian symmetries if desired, it is just \textit{possible} to decouple the ghosts for an abelian symmetry and this is generally simpler.} and we can immediately integrate them out, yielding a functional determinant that only depends on $\alpha$. 
        As such, we omit the ghost terms till the end of the calculation.

        To address the rest of the action, we first change to variables for the intrinsic frame: 
        \begin{align}
            x_j \to x_j + R
        \end{align}
        so that our action (without the ghost term) becomes 
        \begin{align}
            A &= \int_\tau -ip_1\dot{x}_1 - ip_2\dot{x}_2 - i(p_1 + p_2)\dot{R} + \frac{p_1^2 + p_2^2}{2m} + \potential{x_2}{x_1} + i\lambda(p_R - p_1 - p_2) + \frac{\xi}{2}(\alpha X - \dot{\lambda})^2\notag\\
              &\to \int_\tau -ip_1\dot{x}_1 - ip_2\dot{x}_2 - ip_R\dot{R} + \frac{p_1^2 + p_2^2}{2m} + \potential{x_2}{x_1} + i\lambda(p_R - p_1 - p_2) + \frac{\xi}{2}(\alpha X - \dot{\lambda})^2
        ,\end{align}
        where we used the fact that there is a delta functional fixing $p_R = p_1 + p_2$ to switch the coefficient of $\dot{R}$.
        We also note that the Fourier transform of the gauge fixing term at $R = 0$ corresponds with the $\HBRST$ from Eq.~\eqref{eq:HBRST_osc}:
        \begin{align}
            e^{-\int_{\tau}\xi G^2/2} &= \int\DD{B} \exp{\int_\tau iBG - \frac{B^2}{2\xi}}\notag\\
                           &= \int\DD{B} \exp{\int_\tau iB\alpha X - iB\dot{\lambda} - \frac{B^2}{2\xi}}
        .\end{align}
        Then, the action reads 
        \begin{align}
            \int_\tau -ip_1\dot{x}_1 - ip_2\dot{x}_2 - ip_R\dot{R} - iB\dot{\lambda} + \frac{p_1^2 + p_2^2}{2m} + \potential{x_2}{x_1} - i\lambda(p_R - p_1 - p_2) - i\alpha BX + \frac{B^2}{2\xi}
        .\end{align}
        For a suitable choice of $\alpha$, we have a Euclidean ``phase space'' action (an action written in terms of the canonical variables rather than only coordinates and velocities) for something very close to $\HBRST$. 
        The difference in the $B$ and $\lambda$ terms comes from the fact that the Lagrange multiplier modes have a modified norm, as discussed and demonstrated in Eq.~\eqref{eq:osc_modified_norm}.
        As such, to construct a path integral directly from $\HBRST$ would have these different norms showing up in the calculation of matrix elements of the Hamiltonian in the $\lambda, B$ pair of coordinates.
        Rather than going through that construction, which is feasible for this problem but becomes quite difficult to manage cleanly in more general systems, we have arrived at the correct result through the insertion of delta functionals.
        This manner of arriving at the correct gauge-fixed path integral is far more tractable to generalize to a many-body (second quantized) description of a system.

        To actually compute the zero-temperature partition function, we do not need $B$ and thus integrate it back out. 
        As such, we must compute 
        \begin{align}
            Z &= \int\DD{x_i}\DD{p_i}\DD{p_R}\DD{\lambda}\DD{R} \times\notag\\
            &\quad\exp{\int_\tau ip_j\dot{x}_j +ip_R\dot{R} - \frac{p_1^2 + p_2^2}{2m} - \potential{x_2}{x_1} + \int_\tau i\lambda(p_R - p_1 - p_2) - \int_\tau \frac{\xi}{2}\qty(\alpha X - \dot{\lambda})^2}
        .\end{align}
        We begin by integrating out the original momenta $p_1, p_2$. 
        This is the same general process as a textbook phase space path integral, but the terms we actually obtain are different from textbook treatments because of the coupling to extra variables. 
        Collecting all the $p_1$ terms together ($p_2$ follows exactly the same steps, with a cosmetic relabeling $1\longleftrightarrow 2$),
        \begin{align}
            \frac{p_1^2}{2m} -ip_1(\dot{x}_1 + \lambda) &= \frac{1}{2m}\qty(p_1^2 - 2mip_1(\dot{x}_1 + \lambda) - m^2(\dot{x}_1 + \lambda)^2) + \frac{m}{2}(\dot{x}_1 + \lambda)^2\notag\\
            &= \frac{1}{2m}(p_1 - im(\dot{x}_1 + \lambda))^2 + \frac{m}{2}(\dot{x}_1 + \lambda)^2
        .\end{align}
        All of the $p_1$ dependence has been collected into a single quadratic term, and can thus be integrated out as a Gaussian integral.
        Doing this for $p_2$ as well, our partition function is now
        \begin{align}
            Z = \int\DD{x_i}\DD{p_R}\DD{\lambda}\DD{R} \exp{\int_\tau ip_R\dot{R} - \frac{m}{2}(\dot{x}_1 + \lambda)^2 - \frac{m}{2}(\dot{x}_2 + \lambda)^2 - \potential{x_2}{x_1} + i\lambda p_R - \frac{\xi}{2}\qty(\alpha X - \dot{\lambda})^2}
        .\end{align}
        Now, to do the $R$ integration, we recall how the path integral is constructed. 
        The integral is a product of time sliced matrix elements of the evolution operator for time steps $\varepsilon$.
        In the Lagrangian, we only get $R$ dependence in the ``Legendre transformation'' term $ip_R \dot{R}$, which is the continuum limit of the time-sliced expression 
        \begin{align}
            \varepsilon \qty(\ldots + i{p_R}_j\qty(\frac{R_j - R_{j-1}}{\varepsilon}) + i{p_R}_{j-1}\qty(\frac{R_{j-1} - R_{j-2}}{\varepsilon}) + \ldots) = \ldots + -i({p_R}_j - {p_R}_{j-1})R_{j-1} + \ldots
        \end{align}
        in the exponent.
        For the integrations over each $R_j$, this is just the Fourier representation of a delta function.
        These delta functions fix the value of $p_R$ to be the same at each time slice as what was obtained in the time slice before: $p_R$ is conserved.
        This is a general feature that can be expected for any momentum that commutes with the Hamiltonian.
        As such, we can perform the $R$ integration, which demotes the $p_R$ functional integration to a one-dimensional integral over the conserved value of $p_R$ throughout the trajectories being integrated over (notationally, $\DD{p_R}\to\dd{p_R}$):
        \begin{align}
            Z = \int\DD{x_i}\dd{p_R}\DD{\lambda} \exp{-\int_\tau \frac{m}{2}(\dot{x}_1 + \lambda)^2 + \frac{m}{2}(\dot{x}_2 + \lambda)^2 + \potential{x_2}{x_1} - i\lambda p_R + \frac{\xi}{2}\qty(\alpha X - 
            \dot{\lambda})^2}
        .\end{align}
        Before doing the other two integrations, it is worth picking a value of $\xi$.\footnote{Formally, the choice of $\xi$ should not matter as it was introduced with the gauge-fixing function, and our results should be independent of the choice of gauge fixing. In practice, picking a convenient value makes the calculations much simpler.}
        As in the Hamiltonian analysis in Sec.~\ref{sec:Hamiltonian_osc}, we note that units force $\alpha$ to be a squared frequency, which we denote as $\alpha \equiv -\Omega^2$.
        There are no commutators to do here because these are not operators.
        However, there is still coupling in the action between the $x_j$ and $\lambda$ degrees of freedom.
        Just as in the Hamiltonian case, we choose $\xi$ with the intent of decoupling the DoFs, but here this is reduced to an inspection of the terms in the action when we expand the quadratic pieces.
        Expanding everything that couples an $x$ to $\lambda$, we get
        \begin{align}
            \int_\tau \frac{m}{2}\qty(\dot{x}_1^2 + \dot{x}_2^2) + m(\dot{x}_1 + \dot{x}_2)\lambda + m\lambda^2 + \frac{\xi}{2}\Omega^4 X^2 + \frac{\xi}{2}\dot{\lambda}^2 + \xi\Omega^2 X\dot{\lambda}
        .\end{align}
        The cross terms need to cancel.
        Recognizing $m(\dot{x}_1 + \dot{x}_2)\lambda = 2m\dot{X}\lambda$ and then integrating either this term or the final term by parts, we get that we need to choose $\xi$ so that 
        \begin{align}
            (2m - \xi\Omega^2) = 0 \implies \xi = \frac{2m}{\Omega^2} = \frac{M}{\Omega^2}
        ,\end{align}
        exactly the same result as in the Hamiltonian treatment.
        With this choice of $\xi$, the modes are now decoupled so the partition function factorizes.
        For the $x$ modes, we have 
        \begin{align}
            \int\DD{x_i}\exp{-\int_\tau \frac{m}{2}\qty(\dot{x}_1^2 + \dot{x}_2^2) + \potential{x_2}{x_1} + \frac{M\Omega^2}{2}X^2}\label{eq:x_action_fixed_osc}
        .\end{align}
        For the $\lambda$ mode, we have 
        \begin{align}
            \int\dd{p_R}\DD{\lambda}\exp{-\int_\tau m\lambda^2 + \frac{M}{2\Omega^2}\dot{\lambda}^2 - i\lambda p_R} &= \int\dd{p_R}\DD{\lambda}\exp{-\int_\tau \frac{M}{2\Omega^2}\dot{\lambda}^2 + \frac{M}{2}\qty(\lambda - i\frac{p_R}{M})^2 + \frac{p_R^2}{2M}}\notag\\
            &= \int\dd{p_R}e^{-\beta \frac{p_R}{2M}}\int\DD{\lambda}\exp{-\frac{1}{\Omega^2}\int_\tau \frac{M}{2}\dot{\lambda^2} + \frac{M\Omega^2}{2}\lambda^2}
        ,\end{align}
        where the simple constant shift $\lambda \to \lambda + ip_R/M$ was employed to go to the second line.
        The overall factor of $M/\Omega^2$ is the normalization of the Gaussian integral introduced by the 't Hooft trick Eq.~\eqref{eq:thooft_osc}, so it will cancel in the final result.
        The collective momentum integration yields a familiar contribution to the partition function: it is the kinetic energy for a free particle of mass $M$. 
        The $\lambda$ integral can be done exactly and, if we specialize to an oscillator potential between the particles, the $x$ integrals can be done exactly as well.
        It is now plainly visible that the center of mass in $X$ has an apparent oscillator potential with the same frequency as that of $\lambda$.
        These two parts will contribute a factor of 
        \begin{align}
            \exp{-\frac{1}{2}\int_\omega\ln(\qty(\omega^2 + \frac{2k}{m})\qty(\omega^2 + \Omega^2))}\exp{-\frac{1}{2}\int_\omega\ln\qty(\omega^2 + \Omega^2)}
        .\end{align}

        Now to address the ghosts.
        The ghost part of the action is 
        \begin{align}
            \int_{\tau\tau'}\etabar_{\tau}\qty(\Omega^2 - \partial_\tau^2)\delta(\tau - \tau')\eta_{\tau'}
        .\end{align}
        Performing the $\tau'$ integration and Fourier transforming, we immediately get
        \begin{align}
            \int_{\omega}\etabar_{\tau}(\omega^2 + \Omega^2)\eta_{\tau}
        .\end{align}
        The ghost contribution to the partition function is thus
        \begin{align}
            \exp{\int_\omega \ln(\omega^2 + \Omega^2)}
        ,\end{align}
        so the total partition function is
        \begin{align}
            Z &= \underbrace{\exp{-\frac{1}{2}\int_{\omega}\ln(\omega^2 + \frac{2k}{m})}}_{(x_1, x_2)}\underbrace{\exp{-\int_{\omega}\ln(\omega^2 + \Omega^2)}}_{\lambda, (x_1, x_2)}\underbrace{\exp{\int_{\omega}\ln(\omega^2 + \Omega^2)}}_{(\etabar,\eta)}\int\dd{p_R}e^{-\beta p_R^2/2M}\notag\\
            &= \exp{-\frac{1}{2}\int_{\omega}\ln(\omega^2 + \frac{2k}{m})}\int\dd{p_R}e^{-\beta p_R^2/2M}
        .\end{align}
        We see that we get the same intrinsic mode that we did when we naively performed the path integral.
        What is new is that the CoM motion from the original DoFs (which are labeled $(x_1, x_2)$ in the underbraces above) is no longer a zero mode as it has a finite frequency $\Omega$, whose contribution to the free energy was then canceled out with the help of the Lagrange multiplier mode and the ghosts.
        The collective motion shows up as a factorized contribution to the partition function in a manner that is familiar from textbook treatments of statistical mechanics.

    \subsection{Semi-classical approximation}
        Approximating the path integral by treating it as integrating out fluctuations about the classical solution yields the semi-classical approximation to the partition function.
        At next-to-leading order, the saddlepoint evaluation of the path integral yields what would be the RPA energy in the field theory context~\cite{Furnstahl:2002gt}.
        Let us examine the effect of gauge fixing on the saddle-point of the partition function.
        The initial partition function, before integrating the momenta out, is 
        \begin{align}
            Z_{0} = \int\DD{p_1}\DD{p_2}\DD{x_1}\DD{x_2}\exp{-\int_{\tau}-ip_1\dot{x}_1 - ip_2\dot{x}_2 + \frac{p_1^2 + p_2^2}{2m} + V(x_2 - x_1)}
        .\end{align}
        The action is 
        \begin{align}
            A_{0} = \int_{\tau}-ip_1\dot{x}_1 - ip_2\dot{x}_2 + \frac{p_1^2 + p_2^2}{2m} + V(x_2 - x_1)
        .\end{align}
        Minimizing the action yields the classical EoM in imaginary time.
        These are the trajectories that contribute the greatest to the functional integration in the partition function.\footnote{For a simpler example of this same idea (c.f.\ Eq.~(2.197) in~\cite{Negele:1988vy}), suppose we have the integral $$\int\dd{x}\exp{-\ell f(x)},$$ where $f\to+\infty$ when $x\to\pm\infty$. Assuming the integrand has a single peak at $x = x_0$, defined by $$\pdv{f}{x}\eval_{x_{0}} = 0,$$ so when $\ell$ gets large, the integral can be approximated by expanding the function around $x_0$, yielding $e^{-f(x_0)}$ to lowest order. The semi-classical expansion follows from this idea, taking $\ell = 1/\hbar$.}
        Let us find these trajectories for our original action.
        The saddlepoint equations for this action are 
        \begin{align}
            \fdv{A_0}{p_j(\tau)} = -i\dot{x}_j + \frac{p_j(\tau)}{m} \qc \fdv{A_{0}}{x_j(\tau)} = i\dot{p}_j + \pdv{V}{x_j}
        .\end{align}
        The first equation tells us that $p_j = im\dot{x}_j$, which is what we would expect classically with imaginary time. 
        Differentiating this equation with respect to time and using the second equation tells us that $m\ddot{x}_j = \pdv{V}{x_j}$, which reflects the inversion of the potential that Wick rotations induce.
        Note that we can add and subtract the EoMs for the two particles, which yields 
        \begin{align}
            m(\ddot{x}_1 + \ddot{x_2}) = 0\qc m(\ddot{x_2} - \ddot{x}_1) = \pdv{V}{x_2} - \pdv{V}{x_1}\label{eq:osc_eom_stock}
        .\end{align}
        We have free motion of the CoM while the potential only affects the relative separation between the particles, as expected. 
        The two derivatives of the potential work out to be the same term due to the negative sign in the separation of the particles, and the resultant factor of two can be moved to the LHS to convert the particle mass to the reduced mass if desired.

        Now, we can find the corresponding equations for the gauge-fixed action that has the extra DoFs.
        Our action is 
        \begin{align}
            A = \int_\tau -ip_1\dot{x}_1 - ip_2\dot{x}_2 - ip_R\dot{R} + \frac{p_1^2 + p_2^2}{2m} + \potential{x_2}{x_1} + i\lambda(p_R - p_1 - p_2) + \frac{\xi}{2}(\alpha X - \dot{\lambda})^2 + \etabar_{\tau}\qty(\Omega^2 - \partial_\tau^2)\eta_{\tau}
        .\end{align}
        The saddlepoint equations are, with $\alpha = -\Omega^2$ and $\xi = 2m/\Omega^2$, 
        \begin{align}
            \fdv{A}{p_j} = -i\dot{x}_j + \frac{p_j}{m} - i\lambda &\qc \fdv{A}{x_j} = i\dot{p}_j + \pdv{V}{x_j} + \frac{m\Omega^2}{2}(x_1 + x_2) + {m}\dot{\lambda}\\
            \fdv{A}{p_R} = -i\dot{R} + i\lambda &\qc \fdv{A}{R} = i\dot{p}_R\\
            \fdv{A}{\lambda} = i(p_R - p_1 - p_2) - \frac{2m}{\Omega^2}\ddot{\lambda} + {m}(\dot{x}_1 + \dot{x}_2) &\qc \fdv{A}{\etabar} = (\Omega^2 - \partial_\tau^2)\eta
        .\end{align}
        Setting these to zero tells us a number of things about the classical behavior of the system.
        The equations from differentiating the action with respect to the collective coordinate and momentum tell us that the collective momentum is conserved, and that the Lagrange multiplier coincides with the collective velocity.
        Differentiating the $\fdv{A}{p}$ equations with respect to time and using the $\fdv{A}{x}$ equations as we did with the original description of the system, we find 
        \begin{align}
            m(\ddot{x}_1 + \ddot{x}_2) = m\Omega^2(x_1 + x_2) \qc m(\ddot{x}_2 - \ddot{x}_1) = \pdv{V}{x_2} - \pdv{V}{x_1}
        .\end{align}
        The ghost equation and the Lagrange multiplier equation remain. 
        Plugging the $\fdv{A}{p}$ equations into the Lagrange multiplier equation gets it in a simple form, if we make the replacement $\tilde{\lambda} = \lambda + ip_R/2m$.
        The ghost equation is already in a simple form. 
        The two equations are
        \begin{align}
            \ddot{\tilde{\lambda}} = \Omega^2\tilde\lambda \qc \ddot{\eta} = \Omega^2\eta
        .\end{align}
        Everything is a Euclidean oscillator, except for the separation between the $x_j$.
        In particular, we have a symmetry-breaking force with a finite (but arbitrary) frequency in the CoM.
        The cancellation between the terms upon plugging in solutions to these known ODEs proceeds similarly to what was shown in prior sections; we can already see the $\Omega$ parameter occurring in the ghost, Lagrange multiplier, and CoM modes. 
        
        The saddlepoint equation for a second-quantized system is the Hartree equation, which is the leading term of the semi-classical approximation to the quantum theory. 
        We already obtain symmetry breaking and cancellation of the symmetry breaking terms at the Hartree level with the gauge-fixed action. 
        The gauging of the symmetry is thus very promising for systematic many-body generalizations~\cite{bes_bcs_brst, Bes:1990}.

\section{Zero modes and perturbation theory}\label{sec:zero_modes}

In this section we consider how the BRST framework offers an alternative path to address zero frequency modes that can cause infrared divergences when trying to apply conventional perturbation theory.

In general, a zero mode manifests itself as a divergence while calculating observables in the unperturbed state.
We can see this at the level of wavefunctions, and (for this paper, Euclidean) Green's functions. 
Although these are formally equivalent approaches to the analysis of the system, surveying the problem from both perspectives to demonstrate the differences in the manifestation of the zero mode is helpful.
In the two-body example, the unperturbed wavefunction is a function of $x_2 - x_1$, and a computation of any amplitude or expectation value requires an integration over $x_1$ and $x_2$.
However, this is equivalent to an integration over a relative coordinate and the center of mass, and the wavefunction only depends on the relative coordinate, so the CoM integration diverges.
Symbolically, for some operator $\mathcal{O}$,
\begin{align}
    \mel{\phi}{\mathcal{O}}{\psi} &= \int\dd{x_1}\dd{x_2}\phi^*(x_2 - x_1)\mathcal{O}\psi(x_2 - x_1) \notag\\
    &= \int\dd{(x_2 - x_1)}\phi^*(x_2 - x_1)\mathcal{O}\psi(x_2 - x_1)\int\dd{X}1
.\end{align}
In a system with few particles, the freedom corresponding to the zero mode can be manifestly factorized exactly like this via Jacobi coordinates. 
However, in a many-body system with anti-symmetrized states where 
matrix elements are computed in terms of individual particle coordinates or momenta, a regulator of some sort must be present to avoid divergences associated with this zero mode.
Such a regulated divergence cancels in the final calculation of any observable, but its presence in the intermediate stages of the calculation can cause problems.

For example,
the divergence can be regulated by putting the system in an artificial trap, as the CoM then has a finite frequency associated with it. 
This manifests in the above matrix element being modified, in CoM coordinates, to
\begin{align}
    \int\dd{(x_2 - x_1)}\phi^*(x_2 - x_1)\mathcal{O}\psi(x_2 - x_1)\int\dd{X}\exp{-M\Omega_T X^2/2}
,\end{align}
up to choices of how the trap is implemented.
The trap can then be removed as a limiting value of a trap parameter at the end of the calculation.
This practice works well for the case of translation, because the symmetry operation does not mix coordinates.
For rotations (in space or in phase angle), the symmetry operation mixes the independent degrees of freedom for the system.
As a result, the limit to removing the trap can be complicated or even incorrect.
This is demonstrated in the case of a 2D system with a rotational symmetry in appendix 7.B of~\cite{Bes:1990}.

From the Green's function perspective, in the Fourier representation, we can see the zero mode manifestly.
It is useful to briefly review how Green's functions are obtained in the path integral framework.
A Green's function is a time-ordered expectation value of two fields at different points in time.\footnote{In a second-quantized picture, this should be spacetime. However, for the first-quantized description, our ``fields'' are particle coordinates which only depend on time, so the Green's function is a function of (Euclidean or real) time.}
In the path integral, the (Euclidean) time ordering arises by construction, and we have 
\begin{align}
    \mathcal{G}(\tau, \tau') = \ev{x(\tau)x(\tau')} = \int\DD{x} x(\tau) x(\tau')e^{-A(x)}
.\end{align}
For simplicity, it can be helpful to think of a ``discretized'' example of how to compute such a function.
In a multivariable Gaussian integral, if we want to compute the expectation value 
\begin{align}
    \ev{x_i x_j} = \int\dd[n]{x} x_i x_j e^{-\frac{1}{2} x_k M_{kl} x_l}
,\end{align}
we can do so by introducing a ``source term'' which we can take derivatives with respect to before setting it to zero at the end of the calculation:
\begin{align}
    \ev{x_i x_j} = \pdv{J_i}\pdv{J_j}\int\dd[n]{x} \exp{-\frac{1}{2}x_k M_{kl} x_l - J_k x_k}\eval_{\vb{J} = 0}.
\end{align}
We can complete the square in the argument of the exponential:
\begin{align}
    \frac{1}{2} x_k M_{kl} x_l + J_k x_k = \frac{1}{2}\qty(x_k + M^{-1}_{mk}J_m) M_{kl} (x_l + M^{-1}_{nl}J_{n}) - \frac{1}{2}J_m M^{-1}_{mn} J_n
,\end{align}
and change integration variables $\vb{x} \to \vb{x} + M^{-1}\cdot J$ so that the Gaussian $x$ integration can be performed, yielding a determinant of $M$ that will cancel in normalization, leaving 
\begin{align}
    \ev{x_i x_j} \propto \pdv{J_i}\pdv{J_j} \exp{\frac{1}{2}J_m M^{-1}_{mn} J_n}\eval_{\vb{J} = 0} = M^{-1}_{ij}.
\end{align}
The Green's function calculation follows exactly like this, one must simply take the indices to the continuum as a (Euclidean or real) time.

Treating the unperturbed Green's function as arising from evolution due to a harmonic (quadratic) part of a Hamiltonian/action, we find that the correlator between particle coordinates\footnote{The first-quantized nature of the two-body system as analyzed here means this correlator is the analog of the standard many-body Green's function in a more realistic system.} is (c.f.\ the action in Eq.~\eqref{eq:unfixed_action_oscillator})
\begin{align}
    \ev{\mathrm{T}x_i(\tau) x_j(\tau')} = \int\frac{\dd{\omega}}{2\pi} \mathcal{G}_0(\omega) e^{-i\omega(\tau - \tau')}\qc \mathcal{G}_0(\omega) = \frac{1}{\omega^2(\omega^2 + 2k/m)}
    \mqty[\omega^2 + k/m & k/m\\
            k/m & \omega^2 + k/m]\label{eq:osc_correlator}
,\end{align}
with time ordering symbol $\mathrm{T}$.
The zero mode is manifest as a pole of the Green's function: the sole $\omega^2$ term in the denominator yields an infrared divergence.
For example, an intrinsic interaction of the form $(x_2 - x_1)^k$ will involve evaluations of the Green's function at the same time, so $\tau = \tau'$. 
This immediately gives an IR divergence from doing the frequency integral in Eq.~\eqref{eq:osc_correlator}.
Putting each of the particles in an artificial trap of the form $V_{\text{Trap}} = m\Omega_T^2 x^2/2$ modifies the Green's function so that 
\begin{align}
    \mathcal{G}_0(\omega) = \frac{1}{(\omega^2 + \Omega_T^2)(\omega^2 + \Omega_T^2 + 2k/m)}
        \mqty[\omega^2 + \Omega_T^2 + k/m & k/m\\
            k/m & \omega^2 + \Omega_T^2 + k/m]\label{eq:osc_gf_trap}
,\end{align}
which has poles at strictly nonzero (Euclidean) frequencies.
The trap frequency $\Omega_T$ can be taken to zero after the calculation has been performed.
This works for translational freedom, but can encounter difficulties with symmetries that mix DoFs, as noted above. 

Performing these calculations in the extended space associated with BRST symmetry gives us 
an alternative
way of avoiding the divergences associated with zero modes. 
In the wavefunction picture, we now use an operator that has been gauge-fixed via BRST exact term. 
The unperturbed wavefunction depends on a greater set of variables, and is solved for in a manner quite like the procedure used to find the oscillator ground state wavefunction, Eq.~\eqref{eq:psi0_brst_osc}. 
Solving a gauge-fixed unperturbed problem yields a gauge-fixed unperturbed wavefunction.
Fixing a gauge with a condition that fails to commute with $P$ but commutes with $p_R$ means that integrations over the original coordinates do not reflect zero modes anymore.
Instead, the mode corresponding to the CoM in the original DoFs has a finite and arbitrary frequency, whose contributions are automatically eliminated by the ghosts and Lagrange multiplier DoFs without taking limits. 
This persists in the calculations of matrix elements in a perturbative scheme. 

The fact that a gauge-fixed basis ensures that matrix elements of gauge-invariant operators are gauge-fixed bears strong analogy to the idea that a given angular momentum state $\ket{jm}$ is preserved under the action of a rotationally invariant operator. 
In both cases, one has a continuous symmetry whose generator commutes with the operator whose matrix elements are sought out, and a state that picks a particular ``orientation'' in the set of symmetrically equivalent states. 
The simplest case for translational symmetry in the system we have already discussed is one where an oscillator solution is part of the unperturbed intrinsic problem.
This can be the case for a system in an external harmonic trap, and can generally be the case for any problem if Variational Perturbation Theory~\cite{sharma2025,Kleinert_piqm:2004ev} is employed.
Let us suppose that instead of a simple harmonic interaction, we have a quartic anharmonicity:
    \begin{align}
          \potential{x_2}{x_1} = \frac{k}{2}(x_2 - x_1)^2 + g(x_2 - x_1)^4
    ,\end{align}
so that $H_{\text{int}}$ is not exactly solvable anymore. 
In this case, the unperturbed wavefunction is still exactly what was obtained in Eq.~\eqref{eq:psi0_brst_osc}.
We can perturb around this harmonic solution, so that we have oscillator eigenstates for our unperturbed eigenstates:
    \begin{align}
       H_{\text{int}} = \underbrace{\frac{p_1^2 + p_2^2}{2m} - \frac{P^2}{2M} + \frac{k}{2}(x_2 - x_1)^2}_{H_0} + \underbrace{g(x_2 - x_1)^4}_{H_1}
    \end{align}
Calculating matrix elements with $\HBRST$ proceeds following the textbook prescription for perturbation theory with the ground-state wavefunction $\Psi_0$ of Eq.~\eqref{eq:psi0_brst_osc}
    \begin{align}
          \ev{H_1}{\Psi_0} = \int\dd{\lambda}\int\dd{x_1}\dd{x_2}\Psi_{0}^*(x_1, x_2, \lambda, R)\, g(x_2 - x_1)^4\, \Psi_{0}(x_1, x_2, \lambda, R)
    .\end{align}
Note that the integral is performed over the original coordinates and $\lambda$, since the $R$ degree of freedom and the ghosts factorize in the wavefunction and thus do not need to be involved in the calculation.
Also note that the $\lambda$ integration must be carried out with the altered norm Eq.~\eqref{eq:osc_modified_norm}.
The result for the first-order perturbative correction to the ground-state energy is 
    \begin{align}
          E_0^{(1)} = \frac{3g}{m^2\omega_0^2}
    ,\end{align}
where $\omega_0 = \sqrt{2k/m}$ is the intrinsic ground state frequency, read off from the pure harmonic result above.

In the Green's function perspective, we have a gauge-fixed propagator. 
The gauge-fixed propagator has the zero mode made finite as in the case of the trap, but there is no need to take the trap frequency to zero at the end of the calculation. 
Instead, there are additional propagators and contributions to the energy from the additional DoFs, which have the same frequency as the CoM, but with opposite statistics in the form of ghosts. 
The result is a negation of the contributions from the CoM to the observables of the system in the unperturbed propagator.
Any perturbative calculation uses Wick's theorem to compute corrections to an observable in terms of the propagator of the theory, so the zero mode's contributions continue to not cause issues with a gauge-fixed propagator.
Once again looking at our gauge-fixed translational harmonic system with an intrinsic quartic anharmonicity, we find for our unperturbed Green's function (c.f.\ the action in Eq.~\eqref{eq:x_action_fixed_osc} written using only $m$ and not $M$)
\begin{align}
    \mathcal{G}(\omega) = \frac{1}{m(\omega^2 + \frac{2k}{m})(\omega^2 + 4\Omega^2)}
    \mqty[\omega^2 + \frac{k}{m} + 2\Omega^2 & -\frac{k}{m} + 2\Omega^2\\
            -\frac{k}{m} + 2\Omega^2 & \omega^2 + \frac{k}{m} + 2\Omega^2]
.\end{align}
The frequency $\Omega$ from the gauge-fixing function is explicitly present to make what was the zero mode have a finite frequency.
To perturbatively correct the ground state energy, then, we must compute the expectation value of the perturbation via Wick's theorem:
\begin{align}
    \ev{(x_2 - x_1)^4} &= \ev{x_2^4} + \ev{x_1^4} + 6\ev{x_2^2 x_1^2} - 4\ev{x_2^3 x_1} - 4\ev{x_2 x_1^3} \\
    &= 3 \mathcal{G}_{22}(0)^2 + 3 \mathcal{G}_{11}(0)^2 + 6\qty[\mathcal{G}_{22}(0)\mathcal{G}_{11}(0) + 2\mathcal{G}_{12}(0)\mathcal{G}_{21}(0)] - 4[3\mathcal{G}_{22}(0)\mathcal{G}_{21} + 3\mathcal{G}_{21}(0)\mathcal{G}_{11}(0)]
,\end{align}
and the Green's functions at zero time are evaluated safely
\begin{align}
    \mathcal{G}_{22}(0) = \mathcal{G}_{11}(0) &= \int\frac{\dd{\omega}}{2\pi} \frac{\omega^2 + k/m + 2\Omega^2}{(\omega^2 + 2k/m)(\omega^2 + 4\Omega^2)}\notag\\
    &= \frac{1}{4m}\sqrt{\frac{m}{2k}} + \frac{1}{8m\Omega}\\
    \mathcal{G}_{12}(0) = \mathcal{G}_{21}(0) &= \int\frac{\dd{\omega}}{2\pi}\frac{k/m - 2\Omega^2}{(\omega^2 + 2k/m)(\omega^2 + 4\Omega^2)}\notag\\
    &= -\frac{1}{4m}\sqrt{\frac{m}{2k}} + \frac{1}{8m\Omega}
,\end{align}
with the CoM mode regulated by the gauge fixing.
After some algebra, the total first order correction is thus found to be
\begin{align}
    E_{0}^{(1)} = \frac{3g}{2km} = \frac{3g}{m^2\omega_0^2}
,\end{align}
with $\omega_0$ once again equal to $\sqrt{2k/m}$.
Because the gauge fixing is accomplished in a BRST-symmetric manner, the ghosts have already eliminated the spurious zeroth-order contributions to the energy (see the end of Sec \ref{sec:Hamiltonian_osc} or \ref{sec:pi_osc} for wavefunctions and Green's functions, respectively).

In any perturbative calculation, the information about the unperturbed state is encoded in the Green's function or the set of basis states of the system. 
In both cases, fixing a gauge 
is sufficient to remove the effect of zero modes entirely from the perturbation series for the observable of choice. 
The demonstration for this simple system is correspondingly simple. 
However, BRST symmetry offers a systematic way of addressing zero modes in a consistent way for arbitrary symmetries.
Exactly the same ideas and the same machinery can be used for more complicated systems with more complicated (and, with some additional terms, even non-Abelian) symmetries. 

In the wavefunction picture, it is often the case that the basis being used gets truncated. 
The question of the compatibility of truncations with BRST symmetry is a natural one. 
The form of the charge is general, but its representation in terms of excitation operators for eigenstates of the system is not.
With the form of gauge-fixing function that was chosen at the start of this section, we found that we could write the BRST charge $Q$ in terms of ladder operators for the various degrees of freedom in the gauge-fixing scheme, Eq.~\eqref{eq:brst_q_osc}.
The operators involved tell us that the charge mixes states with excitations in the spurious CoM mode, the Lagrange multiplier mode, and the ghost modes.
As such, basis truncations which affect these modes will see breaking of BRST symmetry as an artifact of the approximation.
Truncations in the ghost space are not necessary as the ghost space is so small: we have only two different modes, each of which can only be excited or not. 
This will generally be true: we will only have one or two sets of ghosts (depending on whether a Lagrange multiplier is used for a given symmetry) for each symmetry that is gauged.
Each set of ghosts can occupy a very small set of states because of their fermionic character.
For an abelian symmetry (translation, particle number, cylindrical), the ghosts can decouple from the other DoFs, resulting in a set of (at most) two equations to solve per symmetry, which means that working with the exact solution in the ghost sector is quite feasible in the abelian case.

Although the particular representation of the BRST charge in terms of the ladder operators above is unique to the choice of gauge-fixing function, it is feasible that the charge can be represented in terms of excitation modes of operators in broader cases. 
The possibility of such operators forming closed sets of states or factorizing the states into irreducible representations can be a point of interest in future investigations.

\section{Summary and generalizations}\label{sec:summary}
    The addition of a collective coordinate to a translationally invariant system enables the promotion of a global translational symmetry to a local translational symmetry, allowing the collective symmetry to be treated as a gauge symmetry.
    As such, the ideas of gauge invariance and BRST invariance can be used to formulate new ways to address the problem of symmetry restoration. 
    The redundancy induced by the collective coordinate must be reduced in order to have well-defined observables and time evolution. 
    Eliminating the redundancy amounts to either restricting the states and operators considered to those which are gauge invariant, or breaking the gauge symmetry, which is equivalent to the addition of symmetry-breaking potentials to the Hamiltonian/action.
    If we break the gauge symmetry in a manner that is BRST-invariant, then the nilpotent properties of the BRST variables allow for the automatic cancellations of gauge-dependent features of a calculation, without necessarily requiring a projection integral (or its approximations).
    This was demonstrated in both the wavefunction and path integral frameworks. 
    
    The form of BRST-exact term that was added to the Hamiltonian in Sec.~\ref{sec:Hamiltonian_osc} yielded a simple harmonic oscillator in the CoM of the original degrees of freedom.
    An abbreviated notation was chosen to highlight the fact that the procedure to diagonalize this part of the Hamiltonian does not depend on the number of particles: $P$ could be the sum of momenta of $N$ particles, $X$ the CoM of $N$ particles, and so on.
    The diagonalization procedure in every sector except the intrinsic sector proceeds exactly the same for $N$ particles.
    Also worth noting is that the diagonalization did \textit{not} depend at all on the interaction between the particles.
    This form of BRST extension is thus applicable beyond the scope of the simple toy model that was used to illustrate it in this pedagogical guide. 
        
    The BRST-symmetric fixing of the gauge in the path integral was chosen to correspond to the Hamiltonian picture, and we indeed got the same type of oscillator prescription using similar techniques.
    The possibility of a different BRST-exact term that yields a different intrinsic Hamiltonian is still being explored,
    with the parametrization of the gauge-fixing fermion presented at the start of \ref{sec:Hamiltonian_osc} being a natural starting point.
    Worth noting, however, is that BRST extensions work based on the symmetry broken: the prescriptions discussed in this paper will work for any first-quantized description of a system with collective translational symmetry.
    For translational symmetry with the choice of $\rho$ in Eq.~\eqref{eq:osc_gf_fermion}, we obtained the familiar result that we must subtract $T_{\com}$ from the Hamiltonian, however, we have the extra degrees of freedom to eliminate spurious contributions to the energy.
    This is needed because subtracting $T_{\com}$ from the Hamiltonian makes the operator intrinsic, but not the wavefunction.
    The BRST framework tells us how to do both together consistently.
    Finding a prescription for rotational and particle number symmetries would mean a similarly useful or convenient modified Hamiltonian that can be written down before numerics are involved.
    
        In a second-quantized system, we have fermionic degrees of freedom rather than bosonic DoFs, and the constraint must then be local in time but not in space in order for a collective coordinate to represent the total momentum of a second-quantized many-body system.
    This changes the form of a candidate BRST extension, but not the applicability of the formalism.
    BRST quantization for the nuclear many-body problem offers, at the broadest level, a framework for the consistent handling of collective symmetries in the computation of observables. 
    Central to this framework is the BRST charge $Q$, whose behavior when acting on a wavefunction or commuting with an operator tells us about the BRST symmetry properties of either. 
    By respecting BRST symmetry in the formulation of approximations and general calculations of observables, we can get automatic symmetry restoration with a factorized collective contribution to wavefunctions.
    Consistency of the symmetry restoration with approximation schemes is manifest because the machinery to restore the symmetry is coming in as an effective modification to the Hamiltonian/action for the theory: through the addition of a BRST-exact term in the Hamiltonian perspective or through insertion of delta functionals in the path integral perspective.
    Different choices of BRST extensions to the Hamiltonian/action may yield calculations that differ in the details of the symmetry restoration; this is a new frontier to explore.
    The tools to do so have been presented in this paper.
Work is in progress to apply these tools to several controlled problems,
including 
the seniority model~\cite{ringschuck, greiner:1972} 
and
a system of $N$ particles with degeneracy $N$ as an exactly solvable benchmark~\cite{Engel:2006qu}.

\begin{acknowledgments}
We thank S. Jaiswal, B. Lem, H. Merritt, S.~Sundberg, A. Vaidya for engaging discussions.  This research was supported by the National Science Foundation Award Number PHY-2514765 and by the NUCLEI SciDAC program under award DE-FG02-96ER40963. 
\end{acknowledgments}
        
    \bibliography{BRST_refs}

\appendix
    
\section{Constraints and Classical Electromagnetism}\label{app:constraints_em}
    In order to more firmly root the ideas discussed in this paper with the foundation expected from a standard graduate-level physics education, we show how the formalism of constraints and gauge transformations manifests in the familiar example of classical electromagnetism.
    In particular, we will show how constraints naturally arise from the Lagrangian for classical electrodynamics, and that those constraints generate precisely the familiar gauge transformations of the vector potential.
    Because we see gauge freedom in the fields, we work in free field theory for simplicity - no matter fields.
    The (relativistic) vector potential is defined~\cite{Landau:1975pou} as\footnote{We use the metric signature $\eta = \text{diag}(+,-,-,-)$}
    \begin{align}
        A^\mu = (\phi, \vb{A})
    ,\end{align}
    and it is used to construct the field tensor
    \begin{align}
        F^{\mu\nu} = \partial^{\mu}A^{\nu} - \partial^{\nu}A^{\mu}
    .\end{align}
    The Lagrangian and Lagrangian density are then
    \begin{align}
        L = -\frac{1}{4}\int\dd[3]{x}\mathcal{L} = -\frac{1}{4}\int\dd[3]{x} F_{\mu\nu}F^{\mu\nu}
    .\end{align}
    Trying to pass to the Hamiltonian picture~\cite{Jose_Saletan:1998}, we must find the canonical momenta conjugate to the vector potential components:
    \begin{align}
        \Pi^{\sigma} &= \pdv{\mathcal{L}}{(\partial_0 A_\sigma)} \notag\\
                  &= -\frac{1}{4}\eta^{\alpha\mu}\eta^{\beta\nu} 
                  \pdv{(\partial_0 A_\sigma)} 
                  \qty{\qty(\partial_{\mu}A_{\nu} - \partial_{\nu}A_{\mu})\qty(\partial_{\alpha}A_{\beta} - \partial_{\beta}A_{\alpha})}\notag\\
                  &= F^{\sigma 0}
    .\end{align}
    There is a difficulty here: $\Pi^0 = 0$, so it is not an independent canonical variable.
    This is a constraint.
    We get another constraint by ensuring that this one is obeyed as time evolves.\footnote{This is a concern here, but not in the main discussion of this paper with collective coordinates, because the constraint arising from the introduction of a collective coordinate manifestly commutes with the Hamiltonian, so it is automatically obeyed through time evolution}
    We accomplish this via the Euler-Lagrange equation for $\mathcal{L}$, from which we see 
    \begin{align}
        \partial_{\nu}\pdv{\mathcal{L}}{(\partial_{\nu}A_{\mu})} &= 0\notag\\
        \implies \partial_{t}\Pi^{\mu} + \partial_{i} \pdv{\mathcal{L}}{(\partial_{i}A_{\mu})} &= 0
    .\end{align}
    Requiring that the constraint $\Pi^0 = 0$ does not evolve in time means 
    \begin{align}
        0 = \partial_{t}\Pi^{0} = - \partial_{i} \pdv{\mathcal{L}}{(\partial_{i}A_0)}
    .\end{align}
    Differentiating the Lagrangian density using the same decomposition as what we used to find the momentum $\Pi^\sigma$, we find this secondary constraint to be
    \begin{align}
        0 = -\partial_{i}\pdv{\mathcal{L}}{(\partial_i A_{0})} &= -\partial_i F^{0 i}\notag\\
        &= \partial_{i} \Pi^{i}\notag\\
        &= -\nabla\cdot\qty(\pdv{t}\vb{A} + \grad\phi)\label{eq:gauss_law}
    .\end{align}
    This is Gauss's law in free space.
    We know that Gauss's law holds for all time, so we do not need to check if there are more constraints to set to zero by examining the time evolution of this one.
    We thus have two constraints:
    \begin{align}
        F_1(\xvec) = \Pi^0(\xvec) \qc F_2(\xvec) = \partial_i\Pi^i(\xvec)
    \end{align}
    In the classical formalism, the constraints generate gauge transformations through the Poisson bracket.
    The Poisson bracket for a classical theory of a set of fields $\psi_I$ is defined (see Sec.~9.3 of ~\cite{Jose_Saletan:1998}) as
    \begin{align}
        \poisson{a}{b} = \int\dd[3]{x}\qty[\fdv{a}{\psi_I}\fdv{b}{\Pi^I} - \fdv{a}{\Pi^I}\fdv{b}{\psi_I}]
    ,\end{align}
    for arbitrary function(als) $f,g$ of the canonical fields.
    Here, the functional derivatives are to be taken at equal time.
    Let us examine the transformation generated by our constraints on the vector potential $A^\mu$.
    For finite degrees of freedom, we have one transformation parameter for every component of our generator,\footnote{For rotations, we infinitesimally rotate the coordinate vector by parameters $\epsilon_a$ with
    \begin{align}
        \epsilon_a\poisson{\ell^a}{x^b}
    \end{align}
    for angular momentum components $\ell$.
    } so for a field theory, we must have a function $\epsilon(\xvec)$ as our transformation parameters.
    A general infinitesimal transformation generated by our constraints on some function(al) $f$ of our canonical fields is then
    \begin{align}
        \delta_{\epsilon}f = \int\dd[3]{x} \Bigl[\epsilon_1(\xvec)\poisson{F_1(\xvec)}{f} + \epsilon_2(\xvec)\poisson{F_2(\xvec)}{f}\Bigr]
    .\end{align}
    Let us compute the Poisson brackets with the constraints.
    The first constraint is simply $\Pi^0$, so we can readily calculate the transformation law
    \begin{align}
        \poisson{\Pi^0(\xvec)}{A^{\mu}(\xvec')} &= \poisson{\Pi^0(\xvec)}{A^0(\xvec')}\notag\\
        &= -\eta^{0\mu}\delta(\xvec - \xvec')
    .\end{align}
    The second constraint is easier to think of in terms of the canonical variable $\Pi$ than it is to think of in the familiar form at the end of Eq~\eqref{eq:gauss_law}.
    The Poisson bracket is
    \begin{align}
        \poisson{\partial_{i}\Pi^i}{A^{\mu}} = -\eta^{\mu i}\partial_{i}\delta(\xvec - \xvec')
    .\end{align}
    We so far have, for our general transformation,
    \begin{align}
        \delta_{\epsilon}A^{\mu}(\xvec') &= \int\dd[3]{x}\qty[-\eta^{0\mu}\epsilon_{1}(\xvec)\delta(\xvec - \xvec') - \eta^{\mu i}\epsilon_{2}(\xvec)\partial_{i}\delta(\xvec - \xvec')]\notag\\
        &= -\int\dd[3]{x}\delta(\xvec - \xvec')[\eta^{0\mu}\epsilon_{1}(\xvec) - \eta^{\mu i}\partial_{i}\epsilon_{2}(\xvec)]\notag\\
        &= -\eta^{0\mu}\epsilon_1(\xvec') + \eta^{\mu i}\partial_{i}\epsilon_2(\xvec') 
    .\end{align}
    There are further restrictions on $\epsilon_1$ that are illuminated if we construct the Hamiltonian.
    The Lagrangian density written in terms of the momenta as 
    \begin{align}
        \mathcal{L} = -\frac{1}{4}\qty(F_{0\nu}F^{0\nu} + F_{i\nu}F^{i\nu}) = -\frac{1}{4}\qty(\Pi_{\nu}\Pi^{\nu} + \Pi_{i}\Pi^{i} + F_{ij}F^{ij})
    .\end{align}
    The Hamiltonian is thus obtained by inverting the momentum relationship for the time derivatives of $A$ in the usual way for the space components, leaving the time component untouched for a moment since we know it has complications:
    \begin{align}
        H &= \int\dd[3]{x}\Pi^{\nu}\dot{A}_{\nu} - \mathcal{L}\notag\\
        &= \int\dd[3]{x}\Pi_0 \dot{A}^{0} + \Pi_i(\partial^{i}A^{0} - \Pi^{i}) + \frac{1}{4}(\Pi_{0}\Pi^{0} + 2\Pi_{i}\Pi^{i} + F_{ij}F^{ij})\notag\\
        &= \int\dd[3]{x}\Pi_{0}\dot{A}^{0} - A^{0}\partial^{i}\Pi_{i} - \frac{1}{2}\Pi_{i}\Pi^{i} + \frac{1}{4}\Pi_{0}\Pi^{0} + \frac{1}{4}F_{ij}F^{ij}
    .\end{align}
    We see that $A^{0}$ appears as a Lagrange multiplier on one of the constraints, and its time derivative multiplies another.
    As argued in Sec.~\ref{sec:brst_ped_formalism}, in order to maintain consistency, a Lagrange multiplier must transform as the time derivative of the gauge parameter that adds the other degrees of freedom.
    This fixes $\epsilon_1 = \dot{\epsilon}_2$, which means that the transformation law generated by the constraints is ultimately 
    \begin{align}
        \delta_{\epsilon}A^{\mu}(\xvec', t) = -\eta^{0\mu}\dot{\epsilon}_2(\xvec', t) + \eta^{\mu i}\partial_{i}\epsilon_2(\xvec', t)
    .\end{align}
    Using the components of the metric, this becomes the familiar
    \begin{align}
        \delta_{\epsilon}A^{\mu}(\xvec', t) = -\partial^{\mu}\epsilon_{2}(\xvec', t)
    \end{align}

    \section{Alternate Path integral Implementation}
    Here, we provide an alternate prescription for implementing gauge fixing in the path integral formulation. 
    In particular, the implementation here does not use Lagrange multipliers, so this path integral calculation corresponds to the minimal enlargement of phase space required in order to utilize BRST symmetry.%
    \footnote{In the canonical language, this calculation is equivalent to using the minimal BRST charge of Eq.~\eqref{eq:Qminimal}.}
    We calculate the ground state energy of the same system as in Sec.~\ref{sec:pi_osc} in order to highlight the differences and similarities between the two.
    Having this comparison makes clear some general features of gauge fixing with BRST symmetry, which is discussed at the end of the calculation.
    
    The partition function for the original theory is
        \begin{align}
            Z = \int\DD{x_i}\DD{p_i}\exp{\int_\tau ip_j\dot{x}_j - \frac{p_1^2 + p_2^2}{2m} - \potential{x_2}{x_1}}
        ,\end{align}
        and a collective momentum can be introduced via delta function as in Sec.~\ref{sec:pi_osc}:
        \begin{align}
            Z = \int\DD{x_i}\DD{p_i}\DD{p_R}\delta(p_R - p_1 - p_2)\exp{\int_\tau ip_j\dot{x}_j - \frac{p_1^2 + p_2^2}{2m} - \potential{x_2}{x_1}}
        .\end{align}
        We also fix the gauge with the Faddeev-Popov procedure, as before:
        \begin{align}
            Z = \int\DD{x_i}\DD{p_i}\DD{p_R}\DD{R}\delta(p_R - p_1 - p_2)\delta(G(x_1-R,x_2-R))\qty|\fdv{G(R(\tau))}{R(\tau')}|\exp{\int_\tau ip_j\dot{x}_j - \frac{p_1^2 + p_2^2}{2m} - \potential{x_2}{x_1}}
        .\end{align}
        We use the 't Hooft trick to exponentiate the gauge-fixing function $G$ as before:
        \begin{align}
            Z = \frac{1}{\mathcal{N}}\int\DD{x_i}\DD{p_i}\DD{p_R}\DD{R}\delta(p_R - p_1 - p_2)\qty|\fdv{G_\tau}{R_{\tau'}}|\exp{\int_\tau ip_j\dot{x}_j - \frac{p_1^2 + p_2^2}{2m} - \potential{x_2}{x_1} - \frac{\xi}{2}G(x_1 - R, x_2-R)^2}
        ,\end{align}
        where $\mathcal{N}$ is the normalization from the Gaussian integral and will be omitted notationally for the remainder of the derivation.
        Unlike the prescription in Sec.~\ref{sec:pi_osc}, we do not use a Fourier representation of the constraint's delta function.
        The Fourier representation yields a Lagrange multiplier as a degree of freedom to enforce the constraint, which has a nice interpretation, but it is not the only way to exponentiate a delta function.
        We can also write the delta function as a Gaussian with width $\mu$\footnote{This should not be confused with the reduced mass of the system, as we are not using relative coordinates (and thus, not using the concept of reduced mass) anywhere in our description!} that is taken to zero.
        Making use of the abbreviations from above in order to simplify expressions and highlight applicability to higher particle numbers, 
        \begin{align}
            Z = \lim_{\mu\to 0}\int\DD{x_i}\DD{p_i}\DD{p_R}\DD{R}\qty|\fdv{G(R(\tau))}{R(\tau')}|\exp{\int_\tau ip_j\dot{x}_j - \frac{p_1^2 + p_2^2}{2m} - \potential{x_2}{x_1} - \frac{\xi}{2}G(X - R)^2 - \frac{(p_R - P)^2}{2\mu}}
        .\end{align}
        We now change variables to the ``rest frame'' as before, taking $x_j \to x_j + R$, which generates a ``Legendre transform'' term for $R$. 
        For brevity, we do not explicitly write the $\mu\to 0$ limit in every step, though we must always keep in mind that we only recover the partition function we set out to calculate in the $\mu\to 0$ limit.
        Our partition function is now
        \begin{align}
            Z = \int\DD{x_i}\DD{p_i}\DD{p_R}\DD{R}\qty|\fdv{G(R(\tau))}{R(\tau')}|\exp{\int_\tau ip_j\dot{x}_j +ip_R\dot{R} - \frac{p_1^2 + p_2^2}{2m} - \potential{x_2}{x_1} - \frac{\xi}{2}G(X)^2 - \frac{(p_R - P)^2}{2\mu}}
        ,\end{align}
        with action 
        \begin{align}
            A = \int_{\tau}-ip_j\dot{x}_j - ip_R\dot{R} + \frac{p_1^2 + p_2^2}{2m} + \potential{x_2}{x_1} + \frac{\xi}{2}G(X)^2 + \frac{(p_R - P)^2}{2\mu}
        \end{align}
        We begin by integrating out the momentum $p_R$.
        The relevant terms in the action are 
        \begin{align}
            \frac{1}{2\mu}\qty(p_R^2 - 2p_R P - 2\mu i p_R\dot{R}) = \frac{1}{2\mu}\qty(p_R - (P + i\mu\dot{R}))^2 - \frac{1}{2\mu}(P + i\mu\dot{R})^2
        .\end{align}
        The $p_R$ integral is now Gaussian and can be performed.
        What remains in the Lagrangian is 
        \begin{align}
            L \to -ip_j\dot{x}_j + \frac{p_1^2 + p_2^2}{2m} + \potential{x_2}{x_1} + \frac{\xi}{2}G(X)^2 + \frac{P^2}{2\mu} - \frac{P^2}{2\mu} - iP\dot{R} + \frac{\mu}{2}\dot{R}^2
        ,\end{align}
        where the last three terms come from the completion of the square in $p_R$.
        The $P^2/2\mu$ terms clearly cancel.
        We now turn our attention to the other momentum integrations.
        The process for $p_1$ and $p_2$ differs only by a cosmetic relabeling. 
        In the case of $p_1$, we get 
        \begin{align}
            \frac{1}{2m}(p_1^2 - 2mip_1(\dot{R} + \dot{x}_1)) = \frac{1}{2m}\qty(p_1 - im(\dot{R} + \dot{x}_1))^2 + \frac{m}{2}(\dot{x}_1 + \dot{R})^2
        .\end{align}
        Performing the Gaussian integrations in each momentum, we get for our Lagrangian
        \begin{align}
            L \to \frac{\mu}{2}\dot{R}^2 + \frac{m}{2}\qty{(\dot{x}_1 + \dot{R})^2 + (\dot{x}_2 + \dot{R})^2} + \frac{\xi}{2}G(x)^2 + \potential{x_2}{x_1}
        .\end{align}
        We now assess the coupling to $R$.
        Gathering the $R$ terms, which are only time derivatives, we find 
        \begin{align}
            \frac{\mu + M}{2}\qty(\dot{R}^2 + 2\dot{R}\frac{m}{\mu + M}(\dot{x}_1 + \dot{x}_2)) = \frac{\mu + M}{2}\qty(\dot{R} + \frac{m}{\mu + M}(\dot{x}_1 + \dot{x}_2))^2 - \frac{m^2}{2(\mu + M)}(\dot{x}_1^2 + \dot{x}_2^2)
        ,\end{align}
        where $M = 2m$ and can be seen from how this equation was derived to actually be $Nm$ for $N$ particles.
        The coupling to the velocities in $x_j$ can be removed by changing the $R$ variable as
        \begin{align}
            R \to R - \frac{m}{\mu + M}(x_1 + x_2)
        ,\end{align}
        which coincides in the $\mu\to 0$ limit with shifting by the CoM of the original coordinates.
        Our Lagrangian is now 
        \begin{align}
            L = \frac{\mu + M}{2}\dot{R}^2 - \frac{m}{\mu + M}\frac{m}{2}(\dot{x}_1 + \dot{x}_2)^2 + \frac{m}{2}(\dot{x}_1^2 + \dot{x}_2^2) + \frac{\xi}{2}G(X)^2 + \potential{x_2}{x_1}
        .\end{align}
        Now specializing to a gauge-fixing function $G(X) = \alpha X$ and a harmonic interaction between the particles $V(x) = \frac{k}{2}x^2$, we get a harmonic trap for our particles as we expected, which removes the zero mode as discussed at the beginning of Sec.~\ref{sec:pi_osc}.
        As in that section, the contributions of this trap to the energy will automatically cancel.
        Writing the $x_j$ part of the Lagrangian as a matrix and going to Fourier space, we get 
        \begin{align}
            L = \mqty[x_1 & x_2]\qty{\frac{m^2}{2(M + \mu)}\mqty[\omega^2 & -\omega^2\\ -\omega^2 & \omega^2] + \frac{m\mu}{2(M + \mu)}\mqty[\omega^2 & 0\\ 0 & \omega^2] + \frac{\xi\alpha^2}{8}\mqty[1 & 1\\ 1 & 1] + \frac{k}{2}\mqty[1 & -1\\ -1 & 1]}\mqty[x_1 \\ x_2]
        .\end{align}
        This is just two matrices: a term proportional to the identity, and a term proportional to $\sigma_x$, so the diagonalization is textbook.
        The diagonalized matrix in the Lagrangian is 
        \begin{align}
            \frac{m}{2(M + \mu)}\mqty[\mu\omega^2 + \frac{2}{m}(M+\mu) \frac{\xi\alpha^2}{4} & 0\\ 0 & (M+\mu)\omega^2 + \frac{2}{m}(M+\mu)k]
        .\end{align}
        In the path integral, this becomes a trace log in the eigenvalues:
        \begin{align}
            Z = \exp{-\frac{1}{2}\tr\ln(\qty(\frac{m\mu}{2(\mu + M)}\omega^2 + \frac{\xi\alpha^2}{4})\qty(\frac{m}{2}\omega^2 + k))}\int\DD{R}e^{-\int_\tau \frac{\mu + M}{2}\dot{R}^2}\qty|\fdv{G_{\tau}}{R_{\tau'}}|
        .\end{align}
        The intrinsic mode is visible in the trace log, along with a spurious mode corresponding to the CoM in the original DoFs with a frequency dependent on the gauge-fixing parameters.
        This is not a zero mode now, it has a finite (Euclidean) frequency.
        
        Now, we must address the ghosts.
        The choice of gauge fixing function here was $G(X) = \alpha X$ which transforms under a gauge transformation as 
        \begin{align}
            \alpha X \to \alpha X + \alpha R
        ,\end{align}
        so the derivative is simply $\alpha\delta(\tau - \tau')$.
        As such, the Jacobian is going to be 
        \begin{align}
            \qty|\fdv{G_{\tau}}{R_{\tau'}}| &= \det{\alpha\delta(\tau - \tau')}\notag\\
            &= \exp{\tr\ln(\alpha\delta(\tau - \tau'))}
        .\end{align}
        The total partition function then simplifies to 
        \begin{align}
            Z = \int\DD{R}e^{-\int_\tau \frac{\mu + M}{2}\dot{R}^2}\exp{-\frac{1}{2}\tr\ln(\qty(\frac{m\mu}{2(\mu + M)}\omega^2 + \frac{\xi\alpha^2}{4})\qty(\frac{m}{2}\omega^2 + k))}\exp{\tr\ln(\alpha\delta(\tau - \tau'))}
        .\end{align}

        Now, we recall that our result is only the true partition function in the limit $\mu\to 0$, as that is what we started with.
        Taking this limit now, we see that the spurious mode's dynamical piece (the time derivative part) disappears, and the collective partition function acquires the correct mass.
        We are now left with 
        \begin{align}
            Z = Z_{\text{coll}}\cdot\exp{-\frac{1}{2}\tr\ln(\qty(\frac{\xi\alpha^2}{4})\qty(\frac{m}{2}\omega^2 + k))}\exp{\tr\ln(\alpha\delta(\tau - \tau'))}
        .\end{align}
        Fourier transforming the ghost part and exploiting log properties, we get
        \begin{align}
            Z = \exp{-\frac{1}{2}\tr\ln(\qty(\frac{\xi\alpha^2}{4}))}\exp{-\frac{1}{2}\tr\ln(\qty(\frac{m}{2}\omega^2 + k))}\exp{\tr\ln(\alpha)}
        .\end{align}
        The normalization from 't Hooft's trick was $\exp{-\frac{1}{2}\tr\ln(\xi/2)}$.
        The ghost part can be rewritten with log properties and combined with the normalization as 
        \begin{align}
            \exp{\frac{1}{2}\tr\ln(\frac{\xi\alpha^2}{2})}
        .\end{align}
        The extra factor of two can be grouped into the $\tr\ln$ regulation that is required for all of these functional integral calculations (c.f.\ the discussion before Eq. 11 in \cite{sharma2025} and subsequent citations).

        It is worth commenting on the difference between this prescription and that of Sec.~\ref{sec:pi_osc}.
        Introducing a Lagrange multiplier adds another degree of freedom which is required to transform under a gauge transformation, which in turn requires a different gauge-fixing condition.
        Regardless of which degrees of freedom must be fixed, the derivative of the gauge-fixing condition is precisely the ghost action.
        A Lagrange multiplier that fixes a constraint on canonical momenta (as will always be the case when introducing a collective coordinate in this manner for any symmetry) will necessarily transform with the time derivative of the gauge parameter, as demonstrated in Sec.~\ref{sec:brst_ped_formalism}.
        This means that the presence of the Lagrange multiplier as an enforcer of the constraint is what gives the ghosts dynamics, since its gauge transformation rule brings time derivatives into the Jacobian that becomes the ghost action.
        In this scenario, the dynamical contribution of the ghosts to the ground state energy precisely cancels the contributions from the dynamics of the spurious part of the system - the CoM and the Lagrange multiplier.
        In the alternate prescription presented in this section, there was no Lagrange multiplier, so the ghosts could not have dynamics because there is nothing to transform with a derivative that can get picked up by the Jacobian of $G$. 
        The enforcement of the constraint was controlled by the width parameter $\mu$, and this is precisely what eliminated the frequency (time derivative) dependence of the contribution from the spurious dynamics.
        The ghosts once again worked with the enforcement of the constraint to cancel the remainder of the spurious dynamical contribution.

        The manner of constraint enforcement affects the ghost action through the potential presence of other degrees of freedom that must be fixed by $G$.
        The choice of gauge-fixing condition $G$ manifestly affects the ghost action because the ghost action is just the Jacobian of $G$ under gauge transformations, and $G$ also gives the finite frequency to the zero mode of the theory, so the ghosts \textit{must} cancel the frequency contributions from the gauge fixing.

\section{Ansatz in the extended coordinate space}\label{sec:recentered}

Here we consider an alternative to projection by using gauge fixing and working on a selected gauge slice with a BRST-closed variational ansatz~\cite{Batalin:1994rd,Marnelius:1993az,Marnelius:1993ba,Marnelius:1998pc,SHVEDOV20022,Held:2025mai}. 
Note the distinction from projecting a product state, which includes an integral over the translation orbit, in contrast to defining the variational state directly as a recentered reference-state ansatz (as opposed to a Jacobi-coordinate ansatz).
For $N$ particles we would start from our original reference state $\Phi_{\alphavec}$ and define a recentered (``rec'') wave function: 
\beq
 \Phi_{\alphavec}^{\text{rec}}(x_1,\ldots,x_N)
 = \Phi_{\alphavec}(x_1 - X, \ldots x_N-X) .
\eeq
If we start with a Slater determinant reference, this would no longer be an independent-particle product in the laboratory variables because $X$ couples all particles, but it is built directly from the same single-particle orbitals/reference state.
This is a different variational family and requires an alternative to the usual many-body machinery build for lab-frame product references  (e.g., such as Monte Carlo sampling).

To see how the gauge fixing comes about, we return to $e^{i\comm{\rhotilde}{Q}_{+}}$ but now with
\beq
  \rhotilde_X = X\etabar + \lambda\pi . 
\eeq
This is in the same class as that used in Sec.~\ref{sec:Hamiltonian_osc} but simpler than Eq.~\eqref{eq:osc_gf_fermion} as we restrict ourselves here to the simplest gauge fixing and structure in the extended space.
(We emphasize again the great freedom we have in making these choices.)
With this choice,
\beq
  \comm{\rhotilde}{Q}_{+} = BX - \lambda F -i\pibar\pi - i\eta\etabar .
\eeq
For bookkeeping, we introduce a ``time'' parameter $t > 0$ into the exponential gauge-fixing operator:
\beq
  e^{i t \comm{\rhotilde}{Q}_{+}} =
  e^{it(BX - \lambda F)}e^{t(\pibar\pi + \eta\etabar)} ,
\eeq
and take advantage of the commuting of bosonic and ghost operators to factorize the exponents.
We can now treat the two sectors independently, but noting that $[BX,\lambda F] \neq 0$ and analogously with the ghosts.

Since $B = -i\partial_\lambda$ and $F = +i\partial_X$, 
the bosonic exponent is
\beq
  it(BX - \lambda F) = t\bigl(X\partial_\lambda + \lambda\partial_X \bigr) .
\eeq
When acting on a ``coordinate space'' function of $X$ and $\lambda$, this operator mixes those coordinates like a boost (or a rotation with imaginary angle). 
When exponentiated, the action on a function $f(X,\lambda)$ is then just to shift the arguments:
\beq
  e^{t(X\partial_\lambda + \lambda\partial_X)}
  f(X,\lambda) =
  f\bigl(\cosh(t)X + \sinh(t)\lambda, \cosh(t)\lambda + \sinh (t)X)  ,
\eeq
which means
\beq
 \mel{X,\lambda}{e^{it(BX - \lambda F)}}{X',\lambda'} 
 = \delta\bigl(X' - \cosh(t)X - \sinh(t)\lambda\bigr)
  \, \delta\bigl(\lambda' - \cosh(t)\lambda - \sinh(t) X\bigr) .
  \label{eq:X-lambda-mel}
\eeq 

Now we should consider the structure of our ansatz in the extended space.
If we choose the $\lambda$ dependent wave function to be localized at $\lambda=0$, i.e., a delta function (or a regularized version for which we take the limit at the end), then the matrix element \eqref{eq:X-lambda-mel} will be applied with $\lambda = 0$:
\begin{align}
  \mel{X,\lambda=0}{e^{it(BX - \lambda F)}}{X',\lambda'=0} &= \delta\bigl(X' - \cosh(t) X\bigr)
    \, \delta\bigl(-\sinh(t)X\bigr)  \notag \\
   &= \frac{1}{\sinh(t)}\delta(X)\,\delta(X')
   = \frac{1}{\sinh(t)}\delta(X)\,\delta(X - X') .
\end{align}
Thus we explicitly see the gauge condition $X=0$ emerging, but only after we have chosen a wave function for the $\lambda$ dependence.

The residual dependence on $\sinh(t)$ will be canceled by analogous manipulations in the ghost sector.
Here we take the zero-ghost-number wave function to be one rather than $\etabar\eta$ as in Sec.~\ref{subsubsect:variational-ansatz}.
With $\pi = \partial_\eta$ and $\pibar = \partial_{\etabar}$, the ghost exponent is
\beq
  t(\pibar\pi + \eta\etabar)
  = t\bigl(\partial_{\etabar}\partial_\eta + \eta\etabar \bigr) .
\eeq
Acting once on $1$ gives $\eta\etabar$, while acting twice gives
\beq
 t(\partial_{\etabar}\partial_\eta + \eta\etabar)t(\partial_{\etabar}\partial_\eta + \eta\etabar) 1
 = t^2(\partial_{\etabar}\partial_\eta + \eta\etabar)\eta\etabar
 = t^2 ,
\eeq
where we have used $(\eta\etabar)^2 = 0$.
We can use these results to exponentiate and act on the wave function:
\beq
 e^{t(\partial_{\etabar}\partial_\eta + \eta\etabar)} 1 
 = \cosh(t) + \sinh(t)\, \eta\etabar .
\eeq
The ghost sector integral, with $\int d\etabar\,d\eta\, \eta\etabar = 1$, yields
\beq
  \int d\etabar\,d\eta\, [\cosh(t) + \sinh(t)\, \eta\etabar] = \sinh(t) ,
\eeq
which cancels the $1/\sinh(t)$ from the bosonic sector.
Note that the ansatz with the ghost state $1$ and sharp $\lambda$ dependence will be BRST closed if a recentering ansatz is used, because $Q\Psi_\alpha = (-\eta F + B\pibar)\Psi_\alpha = 0$ from $F$ giving zero acting on the $X$ dependence and from $\pibar 1 = 0$.

We can derive the same result from Trotterization.
We return to the bosonic part and apply the Lie-Trotter formula~\cite{reed1980methods}:
\beq
  e^{it(BX - \lambda F)} = \lim_{\Nt\to\infty}\bigl[e^{i\epsilon BX}\, e^{-i\epsilon\lambda F}\bigr]^{\Nt} ,
  \qquad \epsilon=\frac{t}{\Nt} .
\eeq
We can imagine inserting complete sets of phase-space states $(\lambda, B)$ and $(X,-F)$ between every time slice and using
\beq
   \braket{\lambda}{B} = \frac{1}{\sqrt{2\pi}}e^{iB\lambda}
\eeq
to find (focusing here only on the $\lambda$ and $B$ bras and kets and labeling a particular slice with $j$):
\begin{align}
  \mel{\lambda_{j+1}}{e^{i\epsilon B X_j}}{\lambda_j}
  &= \int \frac{dB_j}{2\pi} e^{iB_j(\lambda_{j+1} - \lambda_j - \epsilon X_j)}   \notag \\
  &= \delta(\lambda_{j+1}-\lambda_j - \epsilon X_j) .
\end{align}
So we see, as before, that the action here at finite $\Nt$ is to shift $\lambda$, here by $\epsilon X$ in each time step:
\beq
  \lambda_{j+1} - \lambda_j = \epsilon X_j .
\eeq 
Similarly, from the $\lambda F$ exponential we get (again, at finite $\Nt$):
\beq
  X_{j+1} - X_j = \epsilon \lambda_{j+1} .
\eeq
If we now take the continuum limit, $\epsilon \to 0$, these become coupled differential equations for $X$ and $\lambda$,
\begin{align}
  \dot X = \lambda , \\
  \dot \lambda = X ,
\end{align}
with boundary conditions given by the choice of ansatz.
The solutions are again $\cosh$ and $\sinh$ functions.
The ghost sector proceeds analogously and we again get canceling $\sinh$ factors.

\begin{figure}[tbh!]
    \centering
    \begin{subfigure}[t]{0.45\textwidth}
        \centering
        \includegraphics[width=0.99\textwidth]{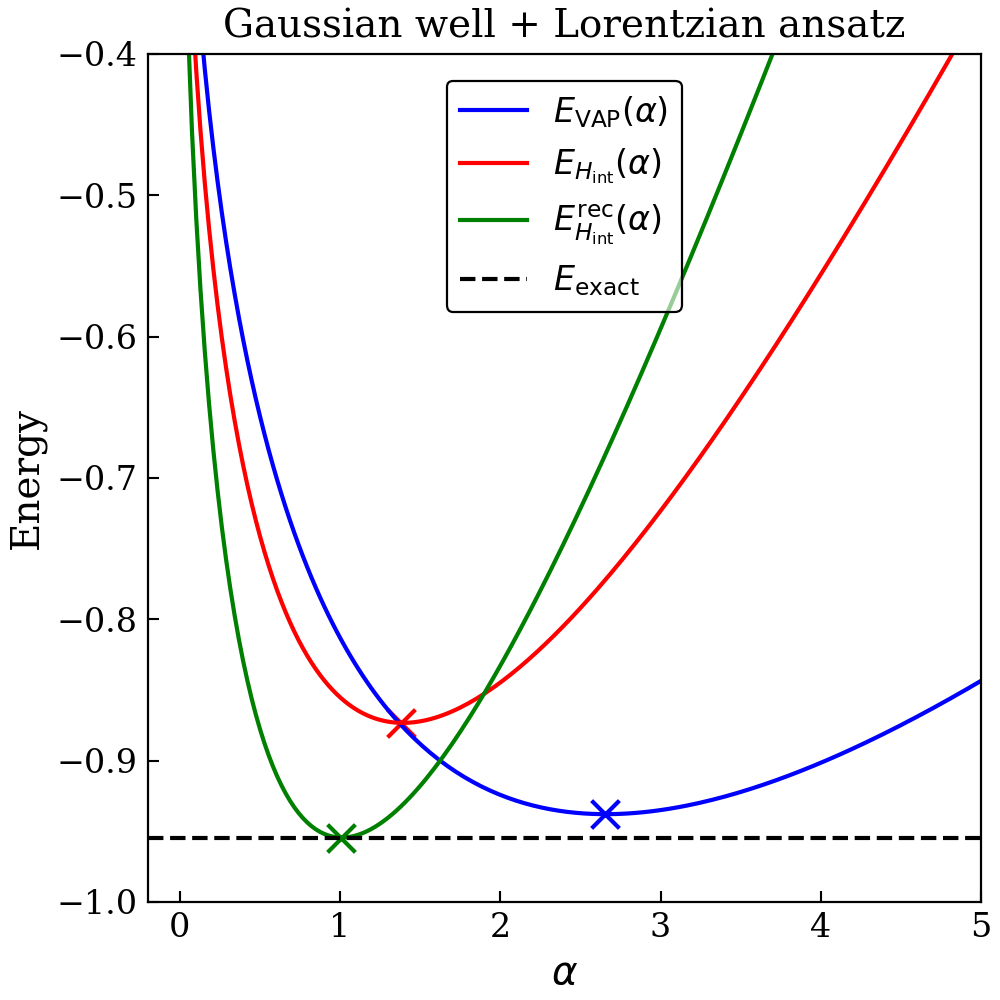}
    \end{subfigure}
        \hfill
    \begin{subfigure}[t]{0.45\textwidth}
        \centering
        \includegraphics[width=0.99\textwidth]{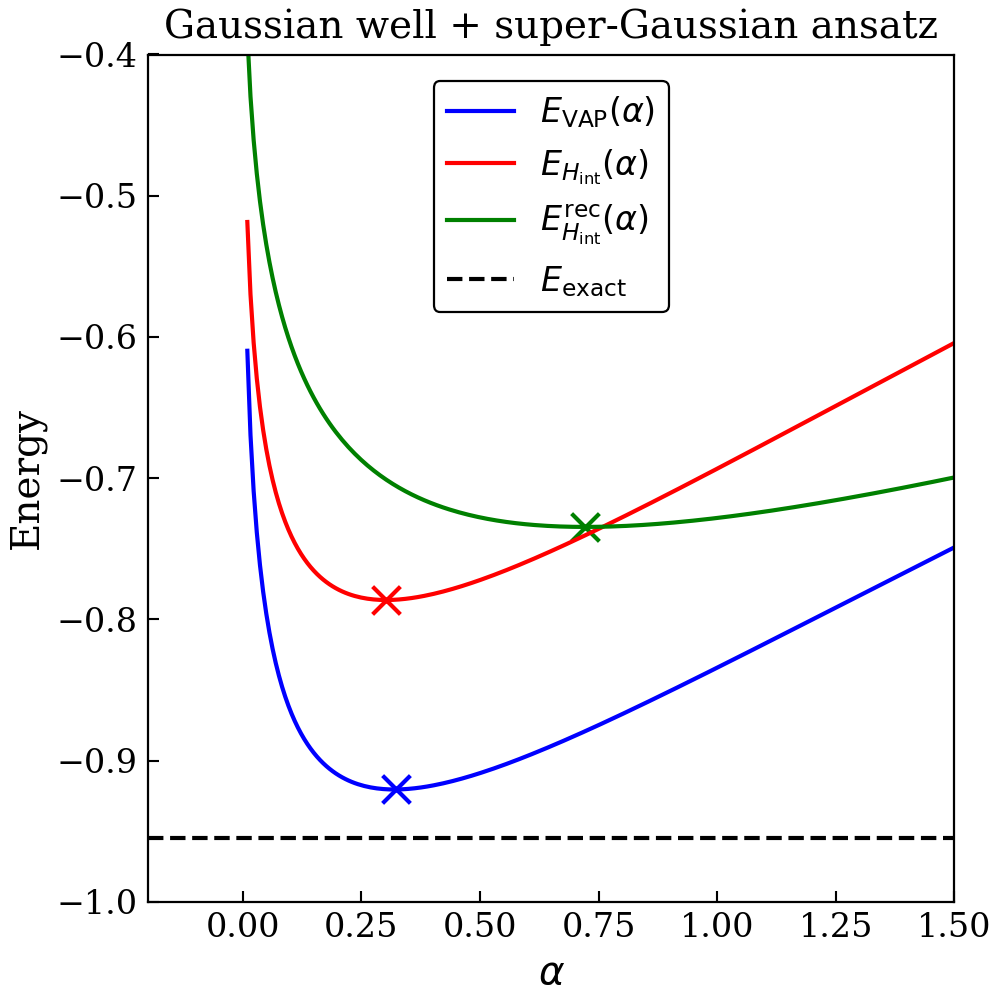}                
    \end{subfigure}%
    \caption{Energy as a function of variational parameter $\alpha$ for a Gaussian potential well with (a) a Lorentzian product ansatz and (b) a super-Gaussian product ansatz. Results for VAP, $\Hint$, and the ansatz evaluated on a recentered gauge slice are compared to the exact result.}   
   \label{fig:Var_E_gaussian-well_slice_exact}        \end{figure}

\begin{figure}[tbh!]
    \centering
    \begin{subfigure}[t]{0.45\textwidth}
        \centering
        \includegraphics[width=0.99\textwidth]{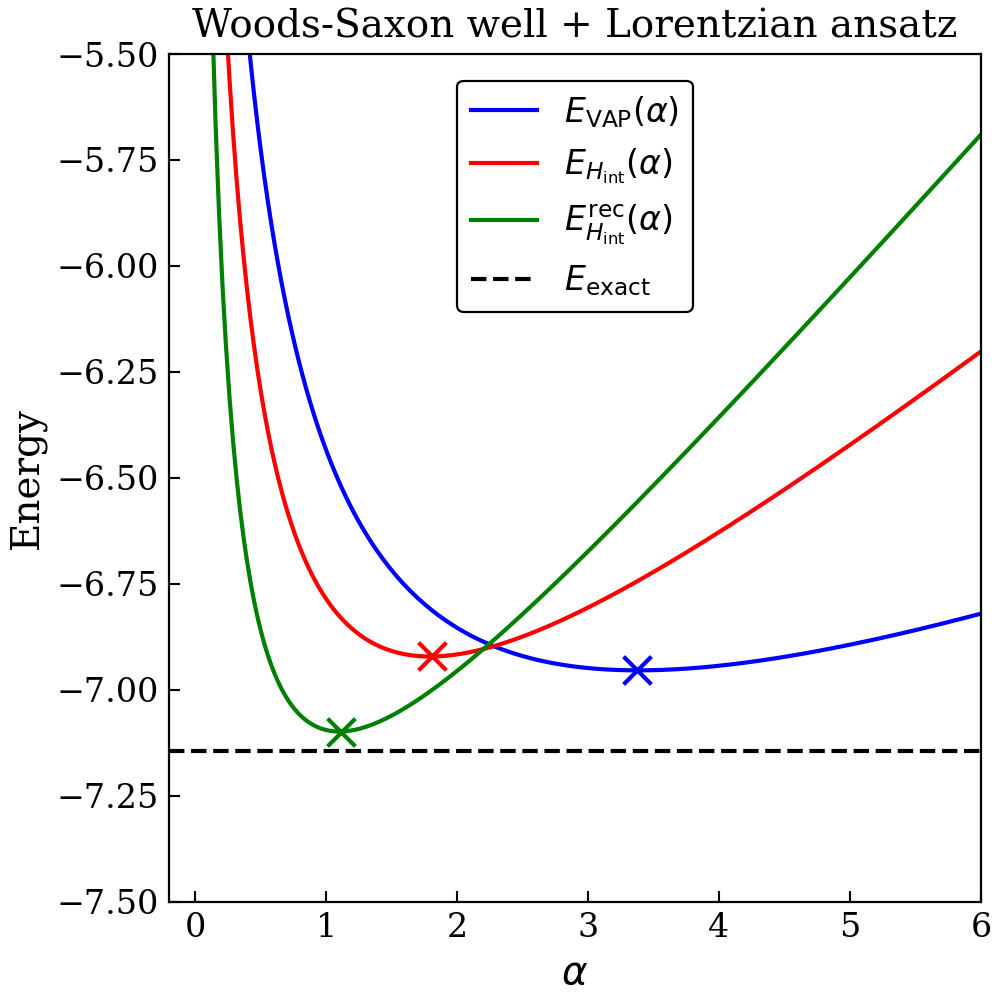}
    \end{subfigure}
    \hfill
    \begin{subfigure}[t]{0.45\textwidth}
        \centering
        \includegraphics[width=0.99\textwidth]{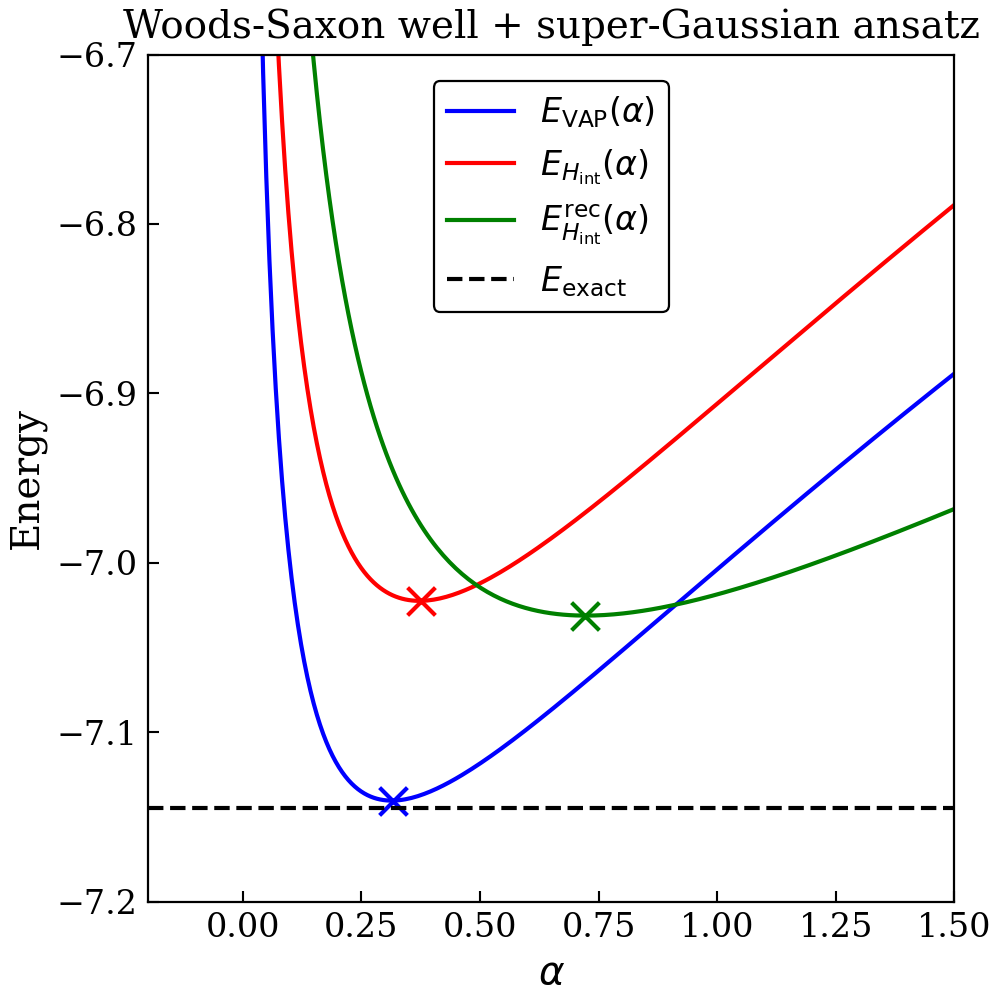}                
    \end{subfigure}%
    \caption{Energy as a function of variational parameter $\alpha$ for a Woods-Saxon potential well with (a) a Lorentzian product ansatz and (b) a super-Gaussian product ansatz. Results for VAP, $\Hint$, and the ansatz evaluated on a recentered gauge slice are compared to the exact result.}
    \label{fig:Var_E_Woods-Saxon_slice_exact}
\end{figure}

Let us see how this works out for our examples with $N=2$. 
After canceling the gauge factors, the variational ansatz with the recentering prescription reduces to integrations over $r \equiv x_2 - x_1$ of $\Hint$ and the wave function ansatz  
\beq
  \Phi_{\alphavec}(x_1, x_2) = 
  \phi(-r/2;\alphavec)\phi(+r/2;\alphavec).
\eeq
The variational results for the same four combinations of potential and ansatz used for illustration in Sec.~\ref{subsec:variational_examples} are shown in Figs.~\ref{fig:Var_E_gaussian-well_slice_exact} and \ref{fig:Var_E_Woods-Saxon_slice_exact}.
It is apparent from the $\alpha$ dependence and the minima that the recentered gauge-slice ansatz results are from a different variational space than the VAP and $\Hint$ results, even though the same single-particle ansatz is used.
With the Lorentzian ansatz, the recentered gauge-slice calculation has a better energy minimum than the VAP functional.
In contrast, with the super-Gaussian ansatz the recentered gauge-slice calculation does noticeably worse than VAP.

\end{document}